\documentclass{SciPost}
\pdfoutput=1
\usepackage{color,amsthm,amsmath,amsxtra,amsfonts,amssymb,dsfont,graphicx,bm,scalerel,bbm,graphicx}
\usepackage{tikz}

\usetikzlibrary{decorations.markings}
\usepackage{breqn} 
\usepackage{caption}
\usepackage{subcaption}
\usepackage{ytableau}

\newcommand{\fourbubble}[6]{
\begin{tikzpicture}[scale=#1, baseline={([yshift=-3pt]current bounding box.center)}]
    \draw[decoration={markings, mark=at position 0.25 with {\arrow>}},postaction={decorate}] (0,0) circle (0.4cm);
    \draw [fill] (-135:0.4) circle (0.05) node[left,scale=0.6*#1] {$#2$};
    \draw [fill] (-45:0.4) circle (0.05) node[right,scale=0.6*#1] {$#3$};
    \draw [fill] (45:0.4) circle (0.05) node[right,scale=0.6*#1] {$#4$};
    \draw [fill] (135:0.4) circle (0.05) node[left,scale=0.6*#1] {$#5$};
\end{tikzpicture}
}

\newcommand{\threebubble}[4]{
\begin{tikzpicture}[scale=#1, baseline={([yshift=-3pt]current bounding box.center)}]
    \draw[decoration={markings, mark=at position 0.25 with {\arrow>}},postaction={decorate}] (0,0) circle (0.3cm);
    \draw [fill] (-90:0.3) circle (0.05) node[below,scale=0.6*#1] {$#2$};
    \draw [fill] (30:0.3) circle (0.05) node[right,scale=0.6*#1] {$#3$};
    \draw [fill] (150:0.3) circle (0.05) node[left,scale=0.6*#1] {$#4$};
\end{tikzpicture}
}

\newcommand{\twobubble}[3]{
\begin{tikzpicture}[scale=#1, baseline={([yshift=-3pt]current bounding box.center)}]
    \draw[decoration={markings, mark=at position 0.55 with {\arrow>}},postaction={decorate}] (0.4,0) arc (0:180:0.4cm and 0.2cm);
    \draw[decoration={markings, mark=at position 0.55 with {\arrow>}},postaction={decorate}] (-0.4,0) arc (180:360:0.4cm and 0.2cm);
    \draw [fill] (0:0.4) circle (0.05) node[below, yshift=-2pt,scale=0.6] {$#2$};
    \draw [fill] (180:0.4) circle (0.05) node[below,yshift=-2pt,scale=0.6] {$#3$};
\end{tikzpicture}
}

\newcommand{\onebubble}[2]{
\begin{tikzpicture}[scale=#1, baseline={([yshift=-3pt]current bounding box.center)}]
    \draw[decoration={markings, mark=at position 0.55 with {\arrow>}},postaction={decorate}] (0.2,0) arc (0:180:0.2cm);
    \draw (0.2,0) arc (0:-60:0.4cm);
    \draw (-0.2,0) arc (180:240:0.4cm);
    \draw [fill] (-90:0.35) circle (0.05) node[right,scale=0.6] {$#2$};
\end{tikzpicture}
}

\newcommand{\nbubbleFour}[6]{
\begin{tikzpicture}[scale=#1, baseline={([yshift=-3pt]current bounding box.center)}]
    \draw[decoration={markings, mark=at position 0.25 with {\arrow>}},postaction={decorate}] (0,0) circle (0.5cm) node[scale=0.6*#1]{$#2$};
    \draw [fill] (-130:0.5) circle (0.05) node[left,scale=0.6*#1] {$#3$};
    \draw [fill] (-90:0.5) circle (0.05) node[below,scale=0.6*#1] {$#4$};
    \draw [fill] (-50:0.5) circle (0.05) node[right,scale=0.6*#1] {$#5$};
    \draw [fill] (-10:0.5) circle (0.05) node[right,scale=0.6*#1] {$#6$};
\end{tikzpicture}
}

\newcommand{\nbubbleThree}[5]{
\begin{tikzpicture}[scale=#1, baseline={([yshift=-3pt]current bounding box.center)}]
    \draw[decoration={markings, mark=at position 0.25 with {\arrow>}},postaction={decorate}] (0,0) circle (0.5cm) node[scale=0.6*#1]{$#2$};
    \draw [fill] (-130:0.5) circle (0.05) node[left,scale=0.6*#1] {$#3$};
    \draw [fill] (-90:0.5) circle (0.05) node[below,scale=0.6*#1] {$#4$};
    \draw [fill] (-50:0.5) circle (0.05) node[right,scale=0.6*#1] {$#5$};
\end{tikzpicture}
}

\newcommand{\nbubble}[2]{
\begin{tikzpicture}[scale=#1, baseline={([yshift=-3pt]current bounding box.center)}]
    \draw[decoration={markings, mark=at position 0.25 with {\arrow>}},postaction={decorate}] (0,0) circle (0.5cm) node[scale=0.9*#1]{$#2$};
\end{tikzpicture}
}

\usepackage{cancel}
\newcommand{\bc}{\begin{center}}
\newcommand{\ec}{\end{center}}
\newcommand{\beq}{\begin{equation}}
\newcommand{\eeq}{\end{equation}}
\newcommand{\beqa}{\begin{eqnarray}}
\newcommand{\eeqa}{\end{eqnarray}}
\newcommand{\beqs}{\begin{eqnarray*}}
\newcommand{\eeqs}{\end{eqnarray*}}

\newcommand{\bi}{\begin{itemize}}
\newcommand{\ei}{\end{itemize}}

\def\bra#1{\langle#1|}
\def\ket#1{|#1\rangle}

\newcommand{\Tr}{{\rm Tr}}

\def\bra#1{\mathinner{\langle{#1}|}}
\def\ket#1{\mathinner{|{#1}\rangle}}

\newcommand{\ZZ}{\mathbb{Z}}

\newcommand{\bE}{\mathbb{E}}

\def\Tr{{\,{\rm tr}}}
\def\Tr{{\,{\rm Tr}}}

\def\one{\mathbbm{1}}

\def\be{\begin{equation}}
\def\ee{\end{equation}}

\newcommand{\bea}{\begin{eqnarray}}
\newcommand{\eea}{\end{eqnarray}}
\def\nn{\nonumber\\}

\def\fr#1{(\ref{#1})}

\def\cvec{\boldsymbol{\mathfrak{c}}}
\def\sfix#1{\texorpdfstring{#1}{Lg}}

\begin{document}

\begin{center}{\Large \textbf{
Dynamics of Fluctuations in Quantum Simple Exclusion Processes}}
\end{center}


\begin{center}
Denis Bernard${}^{1}$, Fabian H.L. Essler${}^{2}$, Ludwig Hruza${}^{1}$, Marko Medenjak${}^{1,4}$, \end{center}

\noindent

\begin{center}
${}^1$ Laboratoire de Physique de l'\'Ecole Normale Sup\'erieure, CNRS, ENS \& PSL University, Sorbonne Universit\'e, Universit\'e de Paris, 75005 Paris, France.\\
${}^2$ The Rudolf Peierls Centre for Theoretical Physics, Oxford University, Parks Road, Oxford OX1 3PU, UK\\
${}^4$ Institut de Physique Th\'eorique Philippe Meyer, \'Ecole Normale Sup\'erieure, 
PSL University, Sorbonne Universit\'es, CNRS, 75005 Paris, France\\

\end{center}

\vskip 0.5 truecm

\begin{center}
\today
\end{center}

\vskip 1.0 truecm

\pagestyle{plain}

\section*{Abstract} 
We consider the dynamics of fluctuations in the quantum asymmetric simple exclusion process (Q-ASEP) with periodic boundary conditions.
The Q-ASEP describes a chain of spinless fermions with random hoppings that are induced by a Markovian environment. We show that fluctuations of the fermionic degrees of freedom obey evolution equations of Lindblad type, and derive the corresponding Lindbladians. We identify the underlying algebraic structure by mapping them to non-Hermitian spin chains and demonstrate that the operator space fragments into exponentially many (in system size) sectors that are invariant under time evolution. At the level of quadratic fluctuations we consider the Lindbladian on the sectors that determine the late time dynamics
for the particular case of the quantum symmetric simple exclusion process (Q-SSEP). We show that the corresponding blocks in some cases correspond to known Yang-Baxter integrable models and investigate
the level-spacing statistics in others. We carry out a detailed analysis of the steady states and slow modes that govern the late time behaviour and show that the dynamics of fluctuations of observables is described in terms of closed sets of coupled linear differential-difference equations. The behaviour of the solutions to these equations is essentially diffusive but with relevant deviations, that at sufficiently late times and large distances can be described in terms of a continuum scaling limit which we construct. We numerically check the validity of this scaling limit over a significant range of time and space scales. These results are then applied to the study of operator spreading at large scales, focusing on out-of-time ordered correlators and operator entanglement.

\vspace{10pt}
\noindent\rule{\textwidth}{1pt}
\tableofcontents\thispagestyle{fancy}
\noindent\rule{\textwidth}{1pt}
\vspace{10pt}
\vskip 1.5 truecm

\section{Introduction}
\label{sec:intro}

Environmental effects on many-particle quantum dynamics continue to attract a great deal of interest for experimental \cite{Johnson2018long,Tang2018thermalization} as well as fundamental theoretical reasons \cite{Bernard2018Transport,Skinner2019measurement}. In many cases of interest the environment is approximately Markovian, which allows for a description of the average system dynamics by means of the Lindblad formalism \cite{gorini1976completely,lindblad1976generators,breuer2002theory}. In the many-particle context Lindblad equations are generally quite difficult to solve and a natural question that arises is whether there exist paradigmatic models for which exact results can be obtained. The simplest such class of Lindblad equations can be mapped to imaginary-time Schr{\"o}dinger equations with
non-Hermitian ``Hamiltonians'' that are quadratic in creation/annihilation operators~\cite{prosen2008third,Horstmann2013noise,vernier2020mixing,alba2020spreading}. More recently it has been shown that there exist Lindblad equations whose evolution operators are related to interacting integrable quantum spin chains
\cite{Medvedyeva2016Exact,Rowlands2018noisy,Naoyuki2019dissipative,Naoyuki2019dissipativespin,ziolkowska2020yang,nakagawa2020exact,buca2020dissipative,Ribeiro2019integrable,lerma2020trigonometric,yuan2020solving,Essler2020Integrability,robertson2021exact,de_Leeuw_2021}. This opens the door to obtaining exact results on the average dissipative dynamics of many-particle Lindblad equations by employing methods from quantum integrability.

While historically the focus of attention has mainly been on the average system dynamics, there are interesting questions pertaining to fluctuations of system degrees of freedom induced by their coupling to the environment. While classical stochastic dynamics has been a well studied domain for many decades,  stochastic dynamical quantum extended systems remained largely unexplored, to the best of our knowledge, until the very recent developments in random quantum circuit models \cite{brown2015decoupling,brandao2016local,Nahum2017Quantum,nahum2018operator,Chan2018Solution}. These advances proceeded in parallel, but independently, to work on quantum exclusion processes and more generally stochastic dynamics in quantum many-body systems such as spin chains coupled to Markovian "noise" \cite{Bauer2017Stochastic}. These model systems provide quantum extensions of discrete versions of the classical fluctuating hydrodynamics whose impact on our understanding of transport in low dimensional systems cannot be underestimated \cite{Spohn2014Nonlinear,Spohn1991Large,Kipnis1989Hydrodynamics}. 
Moreover, the average late time dynamics in such models typically reduces to classical stochastic processes, which have been intensively analyzed over the last several decades \cite{Spitzer1970Interaction,Liggett1985Interacting,Derrida1998An,Schutz2000Exactly}. A key motivation for studying fluctuating quantum dynamics, as in \cite{Bauer2017Stochastic}, is the search for a quantum extension of the classical macroscopic fluctuation theory (MFT) \cite{bertini2005current,Bertini2005MFT}, see \cite{BernardInPreparationQMFT}. The latter is a general setting, whose formulation relies on stochastic hydrodynamic equations and gives rise to a precise description of the fluctuations in classical, diffusive, out-of-equilibrium systems. Quantum simple exclusion processes are paradigmatic models for fluctuating diffusive quantum dynamics. The closed quantum symmetric simple exclusion process (Q-SSEP), was introduced  in \cite{Bauer2019Equilibrium} (under another name), its open version in \cite{Bernard2019Open}, and its asymmetric analogue, the quantum asymmetric simple exclusion process (Q-ASEP) was proposed in \cite{Jin2020Stochastic}. The average dynamics of both the Q-ASEP and the Q-SSEP was recently shown to be integrable \cite{Essler2020Integrability} and it reduces to their classical analogues (which have been studied in great detail, see e.g. \cite{Gwa1992six,Gwa1992bethe,Derrida1993Exact,Schutz1993Phase,Kim1995Bethe,Sasamoto1999One,deGier2005Bethe,Blythe2007Nonequilibrium,deGierEssler11,Lazarescu2011Exact,mallick2011some,crampe2011matrix}) at late times. However, fluctuations in the Q-ASEP and Q-SSEP are not classical and reveal patterns which are possibly generic for mesoscopic diffusive quantum systems. In a recent work, fluctuations in the steady state were analysed in the framework of a continuum limit of the open Q-SSEP \cite{Bernard2021Solution} by constructing the associated invariant probability measure. 

In this work we carry out a detailed analysis of the (finite time) \emph{dynamics} of fluctuations in Q-ASEP and Q-SSEP. 

In section \ref{sec:qssep} we introduce the Q-ASEP and Q-SSEP and formulate the dynamics of fluctuations in terms of Lindblad equations. The corresponding Lindbladians are shown to take the form of non-Hermitian $gl(2R)$ quantum spin chains. In section \ref{sec:qssep-sector} we establish that the Lindbladians fragment \cite{Essler2020Integrability,robertson2021exact} into an exponential (in system size) number of sectors characterized by local conserved charges. The remaining part of the paper focuses on the case of the Q-SSEP. 
In section \ref{sec:qssep-integral} we focus on the five sectors that describe the late time dynamics of quadratic fluctuations in the Q-SSEP and show that two of them are Yang-Baxter integrable and two others are trivial. We then investigate the level-spacing statistics of the Lindbladian eigenvalues in the remaining sector (as well as the two integrable ones) in order to ascertain whether it may be integrable as well.
In section \ref{sec:qssep-steady} we identify the algebraic structure of all steady states and low-lying "magnon" excitations in terms of the underlying $gl(2R)$ algebra. In section \ref{sec:qssep-scaling} we turn to a detailed analysis of the dynamics of correlation functions on the lattice, and show that there is a scaling regime in which correlators are described by hierarchies of partial differential equations that take the form of diffusion equations with source terms that couple the different levels of the hierarchy. We then apply these findings to the large scale dynamics of operator spreading in section \ref{sec:op-entangled}. Section \ref{sec:conclusions} contains our conclusions. A number of technical details are discussed in six appendices.

\section{Q-ASEP \sfix{$\&$} Q-SSEP dynamics as spin chain dynamics}
\label{sec:qssep}

\subsection{Dynamics of fluctuations in the Q-ASEP \sfix{$\&$} Q-SSEP}
\label{sec:qssep-dyn}

The quantum asymmetric (Q-ASEP) or symmetric (Q-SSEP) simple exclusion processes \cite{Bauer2019Equilibrium,Bernard2019Open} are models of quantum many-body stochastic dynamics describing fermions hopping stochastically along a one dimensional chain. The chain can be open, with injection/extraction of fermions at the two ends of the chain, or closed, without any external injection/extraction processes. 
We shall deal with the latter case, in which the state converges to a steady state dressed by sub-leading fluctuations at late time \cite{Bernard2019Open,Bernard2021Solution}. The purpose of the following analysis is to provide the description for the dynamics of those fluctuations in terms of Lindbladian time evolution.

In the closed setup, the Q-SSEP $\&$ Q-ASEP dynamics is unitary but noisy. This means that the system density matrix $\rho_t$ evolves unitarily, $\rho_{t+dt} = e^{-idH_t}\,\rho_t\, e^{idH_t}$, with the Hamiltonian increment between times $t$ and $t+dt$ $dH_t$, which depends on the external noise\footnote{A better notation would be to label the noise by a variable, say $\omega$, taking values in some adapted probability space, and in denoting the Hamiltonian increments as $dH_t(\omega)$ to indicate its noise dependence. As a consequence, the density matrix, which also noise dependent, should be denoted as $\rho_t(\omega)$. We lighten the notation making the $\omega$-dependence implicit.}. For Q-SSEP $\&$ Q-ASEP, on the chain of length $L$ and periodic boundary conditions, the Hamiltonian increment reads \cite{Bauer2019Equilibrium,Bernard2019Open,Jin2020Stochastic}
\beq \label{eq:def-dH}
 dH_t := \sum_{j=1}^{L} \big( c^\dag_{j+1}c_{j}\, dW_t^j + c^\dag_{j}c_{j+1}\, d\overline{W}_t^j \big) ~,
 \eeq
where $c_j$ and $c_j^\dag$ are canonical fermion operators at site $j$ of the chain,
\beqa \label{CACR-0}
\{c_j,c_k^\dag\}=\delta_{j;k} ~,
\eeqa
while $W_t^j$ and $\overline{W}_t^j$ are classical complex Brownian motions in the Q-SSEP case and complex \emph{quantum} Brownian motions in the case of the Q-ASEP. Specifically, they correspond to Gaussian processes with zero mean $\mathbb{E}[dW^j_t]=0$, $\mathbb{E}[d\overline{W}^j_t]=0$ and covariances 
\beq \label{eq:def-dW}
\mathbb{E}[dW^j_t\, d\overline{W}^k_t]= pJ\,\delta^{j;k}\,dt ~,\quad 
\mathbb{E}[d\overline{W}^j_t\, dW^k_t]= qJ\,\delta^{j;k}\,dt ~,
\eeq
and independent increments.
 Here $p,\, q$ are real positive dimensionless parameters, $J$ is a dimensionfull rate parameter, and $\mathbb{E}$ denotes the average over the Brownian motions. In the symmetric case (Q-SSEP), $p=q=1$, the processes $W^j_t$ and $\overline{W}^k_t$ commute, while in the asymmetric case $p\neq q$ (Q-ASEP) the commutator is nontrivial,
\beq
[dW^j_t,d\overline{W}^k_t]= (p-q)J\, \delta^{j;k}\,dt ~.
\eeq

Periodic boundary conditions are assumed throughout.

The mean density matrix $\bar\rho_t:=\mathbb{E}[\rho_t]$, which is obtained by averaging over all possible realizations of the Brownian motion, satisfies a dissipative Lindblad dynamics \cite{Bauer2017Stochastic}
\be
\partial_t \bar \rho_t = \mathcal{L}(\bar \rho_t) ~,
\label{Lindblad}
\ee
where the Lindbladian $\mathcal{L}=\sum_{j=1}^L \mathcal{L}_{j;j+1}$ is the sum of local terms $\mathcal{L}_{j;j+1}$, which can be decomposed as 
\begin{equation}
  \mathcal{L}_{j;j+1}= J\,\big( p\overrightarrow{\mathcal{L}}_{j;j+1} +  q \overleftarrow{\mathcal{L}}_{j;j+1}\big)\ ,
\end{equation}
where  $\overrightarrow{\mathcal{L}}_{j;j+1}$ corresponds to the hopping from site $j$ to site $j+1$ and $\overleftarrow{\mathcal{L}}_{j;j+1}$ the other way around:
\begin{subequations} \label{eq:L-edge}
\begin{align}
 \overrightarrow{\mathcal{L}}_{j;j+1}(\rho)  &:= \ell^-_{j}\, \rho\, \ell^+_{j} -\frac{1}{2}(\ell^+_{j}\ell^-_{j}\, \rho + \rho\,\ell^+_{j}\ell^-_{j}) ~,\\
 \overleftarrow{\mathcal{L}}_{j;j+1}(\rho) &:= \ell^+_{j}\, \rho\, \ell^-_{j} -\frac{1}{2}(\ell^-_{j}\ell^+_{j}\, \rho + \rho\,\ell^-_{j}\ell^+_{j}) ~.
\end{align}
\end{subequations}
Here the jump operators are given by $\ell^+_{j} :=  c^\dag_{j+1}c_{j}$ and $\ell^-_{j} := c^\dag_{j}c_{j+1}$. As a consequence, the total Lindbladian can be decomposed as
\beqa
\mathcal{L}= J\,\big( p \overrightarrow{\mathcal{L}} + q \overleftarrow{\mathcal{L}} \big) ~,
\eeqa
where the Lindbladian $\overrightarrow{\mathcal{L}}$ (resp. $\overleftarrow{\mathcal{L}}$) generates fermionic hopping to the right (resp. left).
The cases where $p=0$ or $q=0$ correspond to the totally asymmetric processes (Q-TASEP).

Let $O$ be an operator acting on the system degrees of freedom only. We denote the associated Heisenberg picture operator acting on the full Hilbert space, comprised of the system and noise, by $O_t$, and its average over the noise by $\bar O_t:=\mathbb{E}[O_t]$. 
The dynamics of $\bar{O}_t$  follows from duality $\Tr(\bar \rho_t\, O)=\Tr(\rho_0\, \bar O_t)$, and is of the Lindblad type as well, 
\beqa
\partial_t\bar O_t= \mathcal{L}^*(O)_t ~,
\eeqa
where the dual Lindbladian $\mathcal{L}^*= J\,\big( p\overrightarrow{\mathcal{L}}^* +  q \overleftarrow{\mathcal{L}}^*\big)$, is also the sum of local contributions, $\overrightarrow{\mathcal{L}}^*=\sum_j\overrightarrow{\mathcal{L}}^*_{j;j+1}$ and $\overleftarrow{\mathcal{L}}^*=\sum_j\overleftarrow{\mathcal{L}}^*_{j;j+1}$. We have
\begin{subequations} \label{eq:Ldual-edge}
\begin{align}
 \overrightarrow{\mathcal{L}}^*_{j;j+1}(O)  &:=  \ell^+_{j}\, O\, \ell^-_{j} -\frac{1}{2}(\ell^+_{j}\ell^-_{j}\, O + O\,\ell^+_{j}\ell^-_{j}) ~,\\
 \overleftarrow{\mathcal{L}}^*_{j;j+1}(O) &:= \ell^-_{j}\, O\, \ell^+_{j} -\frac{1}{2}(\ell^-_{j}\ell^+_{j}\, O +  O\,\ell^-_{j}\ell^+_{j}) ~.
\end{align}
\end{subequations}
Although $\overrightarrow{\mathcal{L}}^*_{j;j+1}$ (resp. $\overleftarrow{\mathcal{L}}^*_{j;j+1}$) is by construction the dual of $\overrightarrow{\mathcal{L}}_{j;j+1}$ (resp. $\overleftarrow{\mathcal{L}}_{j;j+1}$), it is different from $\overleftarrow{\mathcal{L}}_{j;j+1}$ (resp. $\overrightarrow{\mathcal{L}}_{j;j+1}$) since we have $\overrightarrow{\mathcal{L}}^*_{j;j+1}(O)-\overleftarrow{\mathcal{L}}_{j;j+1}(O)= \frac{1}{2}\{O,n_j-n_{j+1}\}$. However, summing over all sites of the chain, with periodic boundary conditions, implies that
\be
\overrightarrow{\mathcal{L}}^*=\overleftarrow{\mathcal{L}} ~,\quad
\overleftarrow{\mathcal{L}}^*=\overrightarrow{\mathcal{L}} ~.
\label{duallind}
\ee
In particular, exchanging $p$ and $q$, by reversing the orientation of the chain, amounts to exchange the total Linbladian with its dual. 

The above Lindbladians describe the time evolution of averages $\mathbb{E}[\Tr(\rho_tO)]$ of operators $O$ acting on the system degrees of freedom but not of their fluctuations. Quadratic fluctuations associated with two operators $O_1$ and $O_2$ are given by
$\mathbb{E}[\Tr(\rho_tO_1)\Tr(\rho_tO_2)]$, and higher moments are defined analogously. The dynamics of the quadratic fluctuations $\Tr(\rho_tO_1)\Tr(\rho_tO_2)=\Tr\left[(\rho_t\otimes\rho_t)(O_1\otimes O_2)\right]$ can be obtained from a \emph{replicated} system
with density matrix $\rho_t^{(2)}:=\mathbb{E}[\rho_t\otimes\rho_t]$. Here the two
replicas of the system are coupled to the same realisation of the
noise, which is then averaged over. Physically this is equivalent to a
three-leg ladder, where the outer legs corresponding to the fermion
degrees of freedom are coupled in an identical fashion to the central
leg, which describes the noise degrees of freedom.

As a consequence, the dynamics of the quadratic fluctuations is still of Lindblad type,
\beq \label{eq:rho2}
\partial_t \rho^{(2)}_t = \mathcal{L}(\rho_t^{(2)}) ~,
\eeq
where $\mathcal{L}=J\,\big(p \overrightarrow{\mathcal{L}} +  q \overleftarrow{\mathcal{L}}\big)$ and $\overrightarrow{\mathcal{L}}=\sum_j \overrightarrow{\mathcal{L}}_{j;j+1}$ and $\overleftarrow{\mathcal{L}}=\sum_j \overleftarrow{\mathcal{L}}_{j;j+1}$ as in eqn \eqref{eq:L-edge}, but with altered jump operators\footnote{To be precise, we should have added to the Lindbladians a label referring to the number of replicas, say $\mathcal{L}^{(2)}$. We made it implicit to lighten the writing.}
\beq \label{jump2} 
\ell^\pm_{j} \to \ell^\pm_{j}\otimes \mathbb{I} + \mathbb{I}\otimes \ell^\pm_{j} = \ell^\pm_{1;j} + \ell^\pm_{2;j} \ ,
\eeq
which correspond to the sum of jumps on both replicas.
Here the indices $1,2$ refer to the labelling of the replicas. A similar structure applies for higher moments, and hence higher number of replicas
\be
\ell^\pm_{j} = \sum_{a=1}^R \ell^\pm_{a;j}\ ,
\ee
where the index $a$ labels the different replicas.

By duality, the dynamics of replicated observables is also of Lindblad type.
Quadratic fluctuations are encoded in $\mathbb{E}[O_{1}\otimes O_{2}]$, where as before the averaging is over the different realizations of the Brownian motions. Then we have
\beq
\partial_t \mathbb{E}[O_{1}\otimes O_{2}]_t= \mathcal{L}^*\big(\mathbb{E}[O_{1}\otimes O_{2}]\big)_t ~,
\eeq
where the dual Lindbladian $\mathcal{L}^*$ has the same structure as in eqn \eqref{eq:Ldual-edge} but with the replacements $\ell^\pm_{j} \to \ell^\pm_{j}\otimes \mathbb{I} + \mathbb{I}\otimes \ell^\pm_{j} = \ell^\pm_{1;j} + \ell^\pm_{2;j}$.  Similar remarks apply to higher moments.

\subsection{Q-ASEP \sfix{$\&$} Q-SSEP as non-hermitian \sfix{$gl(2R)$} spin chains}
\label{sec:qssep-chain}
The aim of this section is to represent the Q-SSEP $\&$ Q-ASEP replicated dynamics as non-Hermitian quantum spin chains. Viewing the replicas as different internal "color" degrees of freedom, the dynamics will include intra- and inter- replica interactions. As a consequence, the Q-SSEP $\&$ Q-ASEP dual Lindbladian for two replicas, i.e. for quadratic fluctuations, can be decomposed as 
\[
 \overrightarrow{\mathcal{L}}^* = \overrightarrow{\mathcal{L}}^*_1 + \overrightarrow{\mathcal{L}}^*_2 + \overrightarrow{\mathcal{L}}^*_{12} ~,\quad
  \overleftarrow{\mathcal{L}}^* = \overleftarrow{\mathcal{L}}^*_1 + \overleftarrow{\mathcal{L}}^*_2 + \overleftarrow{\mathcal{L}}^*_{12} ~,
\]
where $\overrightarrow{\mathcal{L}}^*_a$ ($\overleftarrow{\mathcal{L}}^*_a$), $a=1,2$ and $\overrightarrow{\mathcal{L}}^*_{12}$ ($\overleftarrow{\mathcal{L}}^*_{12}$) 
represent respectively intra and inter-replica Lindbladians.
Each of them is the sum of local Lindbladians acting on edges connecting two sites, say $\overrightarrow{\mathcal{L}}^*_1=\sum_j \overrightarrow{\mathcal{L}}^*_{1;j,j+1}$ and similarly for the other terms. For a higher number $R$ of replicas, the dual Lindbladians can be decomposed as
\begin{equation}
 \overrightarrow{\mathcal{L}}^* = \sum_a \overrightarrow{\mathcal{L}}^*_a + \sum_{a<b} \overrightarrow{\mathcal{L}}^*_{ab} ~,\quad
  \overleftarrow{\mathcal{L}}^* = \sum_a \overleftarrow{\mathcal{L}}^*_a + \sum_{a<b} \overleftarrow{\mathcal{L}}^*_{ab} ~,  
\end{equation}
where $\overrightarrow{\mathcal{L}}^*_{ab}$ ($\overleftarrow{\mathcal{L}}^*_{ab}$) are the inter-replica Lindbladians and $\overrightarrow{\mathcal{L}}^*_a$ ($\overleftarrow{\mathcal{L}}^*_a$), $a=1,2$, the intra-replica ones.
In the following we will focus on the two replica case but mention in passing what are the appropriate generalisations to an arbitrary number of replicas.

As we will now describe, these Linbladians can be written in terms of super-operators, i.e. linear maps acting on operators defined over the system Hilbert space, which fulfil the commutation relations of the $gl(4)$ algebra, or the $gl(2R)$ algebra in case of $R$ replicas.
\begin{itemize}
\item{} First, there are local generators $C_{a;j}$, $j=1,\cdots, L$ and $a=1,\cdots, R$ of $u(1)$ charges counting the fermionic degrees of the operators at site $j$. Their action on an arbitrary operator $X$ is defined by 
\be
C_{a;j}\,X := [n_{a;j},X]\ ,
\ee
where $n_{a;j}$ are the local number operators in the $a$-th replica.
\item{} Second, on each replica there are generators $J^z_{a;j}$, $J^\pm_{a;j}$, $j=1,\cdots, L$, and $a=1,\cdots,R$ forming an $sl(2)$ algebra of super-operators
\be
[J^z_{a;j},J^\pm_{a;k}]= \pm2\,\delta_{j;k}\,J^\pm_{a;j} ~,\quad [J^+_{a;j},J^-_{a;k}]= \delta_{j;k}\,J^z_{a;k} ~,
\ee
and commuting with the $u(1)$ charges defined above, $[C_{a;j},J^\star_{a;k}]=0$\ . These super-operators are defined below in eqn \eqref{eq:def-J*}.\\
The intra-replica Lindbladians are expressed in terms of these super-operators as :
\begin{align}
\label{eq:L1replica}
\overrightarrow{\mathcal{L}}^*_{a;j,j+1} &= \Big( {J}^+_{a;j+1} {J}^-_{a;j} + \frac{1}{4}( {J}^z_{a;j+1} +1)({J}^z_{a;j}-1)+\frac{1}{4} C_{a;j+1}C_{a;j}\Big),\nn
\overleftarrow{\mathcal{L}}^*_{a;j,j+1} &= \Big({J}^-_{a;j+1} {J}^+_{a;j} + \frac{1}{4}( {J}^z_{a;j+1} -1)({J}^z_{a;j}+1)+\frac{1}{4} C_{a;j+1}C_{a;j}\Big) ~.
\end{align}
In particular, in the symmetric case the sum $\overrightarrow{\mathcal{L}}^*_a + \overleftarrow{\mathcal{L}}^*_a = \sum_{j=1}^L\big({J}^+_{a;j+1} {J}^-_{a;j} + {J}^-_{a;j+1} {J}^+_{a;j}+ \frac{1}{2}{J}^z_{a;j+1}{J}^z_{a;j} + \frac{1}{2} (C_{a;j+1}C_{a;j}-1)\big)$ is the isotropic Heisenberg Hamiltonian. This formula was also observed in \cite{frassek2020duality} in the one replica case.
\item{} Finally, we define in eqn \eqref{eq:def-GAB} a set of super-operators $G^{AB}_{j}$, $A,B = (1r),\, (1l),\, (2r),\, (2l)$, satisfying the $gl(4)$ commutation relations,
\beq
 \big[G_j^{AB},G_k^{CD}\big]=\delta_{j,k}\big(\delta^{BC}G_j^{AD}-\delta^{DA}G_j^{CB}\big) ~,
 \eeq
such that (for two replicas)
\begin{subequations} \label{eq:L12replica}
\begin{align}
\overrightarrow{\mathcal{L}}^*_{12;j,j+1} &= \Big(G^{1l;2r}_{j+1}G^{2r;1l}_j + G^{2l;1r}_{j+1}G^{1r;2l}_j \\
&~~~~ +\frac{1}{2}\big(  G^{1l;2l}_{j+1}G^{2l;1l}_j  + G^{2l;1l}_{j+1}G^{1l;2l}_j 
+  G^{1r;2r}_{j+1}G^{2r;1r}_j +  G^{2r;1r}_{j+1}G^{1r;2r}_{j} \big)\Big) ~, \nonumber\\
\overleftarrow{\mathcal{L}}^*_{12;j,j+1} &= \Big( G^{2r;1l}_{j+1}G^{1l;2r}_j + G^{1r;2l}_{j+1}G^{2l;1r}_j \\
&~~~~ +\frac{1}{2}\big(  G^{1l;2l}_{j+1}G^{2l;1l}_j  + G^{2l;1l}_{j+1}G^{1l;2l}_j 
+  G^{1r;2r}_{j+1}G^{2r;1r}_j +  G^{2r;1r}_{j+1}G^{1r;2r}_{j} \big) \Big) ~. \nonumber
\end{align}
\end{subequations}
The relation between the above $gl(2)$ generators and the $gl(4)$ generators reads $J^z_{a;j}=G^{al;al}_j-G^{ar;ar}_j$, $J^+_{a;j}=(-)^{R_a^0} G^{al;ar}_j$, $J^-_{a;j}=G^{ar;al}_j (-)^{R_a^0}$, with $(-)^{R_a^0}$ some Wigner-Klein factors to be defined below, and $C_{a;j}=G^{al;al}_j + G^{ar;ar}_j -1$. 
\end{itemize}
The above remarks extend to the case of an arbitrary number $R$ of replicas. In that case, the generators $G^{AB}_{j}$ form a representation of $gl(2R)$ with index $A,B$ running over the pairs $(ar)$ and $(al)$ with $a=1,\cdots,R$ referring to the replica index and $r/l$ to right/left action. The dual Lindbladians $\overrightarrow{\mathcal{L}}^*_a$ ($\overleftarrow{\mathcal{L}}^*_a$) and $\overrightarrow{\mathcal{L}}^*_{ab}$ ($\overleftarrow{\mathcal{L}}^*_{ab}$) are simply given by the previous expressions but replacing the replica indices $1,2$ by $a,b$.

In the symmetric Q-SSEP case, with $R$ replicas, the total Lindbladian $\mathcal{L}_{\mathrm{ssep}} = J\,\big( \overrightarrow{\mathcal{L}}^*+ \overrightarrow{\mathcal{L}}^*\big)$ can written as (for Q-SSEP $\mathcal{L}^*_{\mathrm{ssep}} =\mathcal{L}_{\mathrm{ssep}}$)
\beq \label{eq:L-tensorC-GL}
\mathcal{L}_{\mathrm{ssep}}  = J \sum_j \Big( \sum_{A,B} G_{j+1}^{AB}G_j^{BA} - \frac{1}{2}(C_{j+1}+C_j) - R \Big) ~,
\eeq 
with 
\beqa
C_j:=\sum_a C_{a;j}=\sum_AG_j^{AA}-R ~,
\eeqa
the total $u(1)$ charge over the different replicas at site $j$. This expression makes the $gl(2R)$ symmetry explicit. Alternatively, one can decompose the $gl(2R)$ algebra as the sum of a $sl(2R)$ algebra plus a $u(1)$ and introduce the $sl(2R)$ generators $J^{AB}_j:=G^{AB}_j - \frac{\delta^{AB}}{2R}(C_j+R)$ with $\sum_A J_j^{AA}=0$. Then
\beq \label{eq:L-tensorC-SL}
 \mathcal{L}_{\mathrm{ssep}}  = J\sum_j\Big( \sum_{A,B} J_{j+1}^{AB}J_j^{BA}+ \frac{1}{2R}\,C_{j+1}C_j - \frac{R}{2} \Big) ~.
\eeq

We shall present the derivation of these results using two complementary approaches: one defining the appropriate super-operators by their actions on operators via specific left/right multiplications, the other one introducing explicit basis on the space of operators and defining the super-operators via their actions on this basis. They have different advantages depending on the question raised.

Since starting from Section \ref{sec:qssep-integral} this manuscript deals with Q-SSEP only we shall then drop the label ``ssep'' attached to the Lindbladian.

\subsubsection{Super-operator formalism}
When dealing with many replicas, we have as many fermion species as there are replicas. We denote them by $c_{a,j}$, $c^\dag_{a;j}$, with $a$ the replica index and $j$ the site index. They satisfy:
\begin{align}
\label{CACR}
\{c^\dag_{a;j},c_{a;k}\}&=\delta_{j;k}\ ,\quad\{c_{a;j},c_{a;k}\}=0\ ,\quad
\{c^\dagger_{a;j},c^\dagger_{a;k}\}=0\ ,\nn
\lbrack c^\dag_{a;j},c_{b;k}\rbrack &=0\ ,\quad\lbrack c_{a;j},c_{b;k}\rbrack=0\ ,\quad \lbrack c^\dagger_{a;j},c^\dagger_{b;k}\rbrack=0\ ,\quad \ b\not=a ~,
\end{align}
 i.e. fermion operators anti-commute when in the same replica but commute when in different replicas.

 We then define super-operators by left/right multiplications with $c_{a,j}$ or $c^\dag_{a;j}$. That is, for any operator $O$ on the physical system Hilbert space, we set:
\begin{subequations}
\begin{align}
 R^+_{a;j}\, O := O\, c^\dag_{a;j} &,\quad R^-_{a;j}\, O := O\, c_{a;j} ~,\\
  L^+_{a;j}\, O := c^\dag_{a;j}\, O &,\quad L^-_{a;j}\, O := c_{a;j}\, O ~.
\end{align}
\end{subequations}
For instance, the local $u(1)$ charge $C_{a;j}$, whose action is specified by $C_{a;j}\,O:=[n_{a;j},O]$, can be represented as 
\beqa
C_{a;j}= L^+_{a;j}L^-_{a;j} - R^-_{a;j}R^+_{a;j} = L^+_{a;j}L^-_{a;j} + R^+_{a;j}R^-_{a;j} -1 ~.
\eeqa
We recall that the total charge is defined as $C_j=\sum_a C_{a;j}$.
The operators defined in this way satisfy mixed commutation/anti-commutation relations (for instance left/right multiplication commute). It is convenient to transform them into a set of fermionic super-operators, anti-commuting for all replica indices and all sites by introducing Klein factors, which we implement through a Wigner like transformation on the replica space (it is not intrinsic, we could have made different choices). We set (for two replicas, but the generalisation to any number of replicas is straightforward)
\beqa 
\hat R^\pm_{1;j} := R^\pm_{1;j} &,&\ \hat L^\pm_{1;j} := (-)^{R_1^0}\, L^\pm_{1;j} ~, \nonumber\\
\hat R^\pm_{2;j} := (-)^{R_1^0+L_1^0}\,R^\pm_{2;j} &,&\ \hat L^\pm_{2;j} := (-)^{R_1^0+L_1^0+R_2^0}\, L^\pm_{2;j} ~, \nonumber
\eeqa
where $R^0_a:=\sum_j R^+_{a;j}R^-_{a;j}$ and $L_a^0:=\sum_j L^+_{a;j}L^-_{a;j}$ are the (total) number $R/L$ operators. By construction we then have the anti-commutation relations 
\beqa
\{\hat R^s_{a;j},\hat R^{s'}_{b;k}\} &=& \delta_{j,k}\delta_{a,b}\delta^{s+s';0} ~, \nonumber\\
 \{\hat R^s_{a;j},\hat L^{s'}_{b;k}\} &=& 0 ~,\\
  \{\hat L^s_{a;j},\hat L^{s'}_{b;k}\} &=& \delta_{j,k}\delta_{a,b}\delta^{s+s';0} ~. \nonumber
\eeqa

We denote these fermions by $f^{A\,\dag}_{j},\, f^{A}_{j}$ with  $A=(1r),\, (1l),\, (2r),\, (2l)$, where the labelling $r/l$ refers to right/left multiplication (the generalization to $R$ replicas is $A=(ar)$ or $A=(al)$ with $a=1,\cdots,  R$), with
\beqa
\{ f^{A\,\dag}_j , f^{B\,\dag}_k \}=0~,\quad \{ f^{A\,\dag}_j , f^{B}_k \}= \delta_{j,k}\delta^{AB}~,\quad \{ f^{A}_j , f^{B}_k \}=0 ~.
\label{ffermions}
\eeqa
There are $2R$ such effective fermions (or super-fermions as they are linear maps acting on operators).

The $gl(2R)$ generators are then defined by
\beqa \label{eq:def-GAB}
G_j^{AB}:= f^{A\,\dag}_{j} f^{B}_{j} ~.
\eeqa
By construction, they satisfy the $gl(2R)$ commutation relations. In particular, they include as many $sl(2)$  sub-algebras as the number of replicas which are generated by the super-operators $J^\pm_{a;j}$ and $J^z_{a;j}$ defined by 
\beqa \label{eq:def-J*}
J^+_{a;j}O=c^\dag_{a;j}Oc_{a,j},\quad J^-_{a;j}O=c_{a;j}Oc^\dag_{a,j},\quad J^z_{a;j}O=n_aO+On_a-O ~,
\eeqa
for any operator $O$. We have $J^z_{a;j}=G^{al;al}_j-G^{ar;ar}_j$, $J^+_{a;j}=(-)^{R_a^0} G^{al;ar}_j$, $J^-_{a;j}=G^{ar;al}_j (-)^{R_a^0}$.

The proof of eqns \eqref{eq:L1replica} and \eqref{eq:L12replica} is then just a matter of expanding the expressions \eqref{eq:Ldual-edge} of the dual Lindbladian in terms of the jump operators $\ell^\pm_j$ and  expressing them in terms of the effective fermions $f^{A\,\dag}_{j}$ and $f^{A}_{j}$. We shall not reproduce all computations here, as they are slightly lengthy, but only reproduce two examples. Let us first look at one intra-replica contribution to $\mathcal{L}^*(O)$ for some operator $O$. One such contribution is $\ell_{a;j}^+ O \ell_{a;j}^-$, for instance. We have:
\beqs
\ell_{a;j}^+ O \ell_{a;j}^- &=&  c^\dag_{a;j+1}c_{a,j}Oc^\dag_{a;j}c_{a,j+1} \\
&=& L^+_{a;j+1}R^-_{a;j+1}L^-_{a;j}R^+_{a;j}\, O = G_{j+1}^{al;ar}G^{ar;al}_j\, O ~.
\eeqs
An inter-replica contribution is for instance $\ell_{a;j}^+\ell_{b;j}^-O$. We have:
\beqs
\ell_{a;j}^+\ell_{b;j}^-O &=&  c^\dag_{a;j+1}c_{a,j}c^\dag_{b;j}c_{b,j+1} O \\
&=& L^+_{a;j+1}L^-_{b;j+1}L^+_{b;j}L^-_{a;j}\, O = G_{j+1}^{al;bl}G^{bl;al}_j\, O ~.
\eeqs
It is clear that all terms contributing to the (dual) Lindbladians are quadratic in the $gl(2R)$ generators.
Gathering all terms proves eqns \eqref{eq:L1replica} and \eqref{eq:L12replica}.

\subsubsection{Hilbert space doubling formalism}
A basis of quantum states over site $j$ of our one-dimensional lattice is
\be
|\mathfrak{0}\rangle_j\ ,\qquad |\mathfrak{1}\rangle_j=c^\dagger_j|\mathfrak{0}\rangle_j\ .
\ee
A basis of the Hilbert space of states ${\cal H}$ is thus
\be
|\boldsymbol{\sigma}\rangle=\otimes_{j=1}^L |\sigma_j\rangle_j\ ,\quad \sigma_j\in\{\mathfrak{0},\mathfrak{1}\}.
\ee
In the one-replica case the noise-averaged density matrix can be expanded in this
basis as
\be
\bar{\rho}=\sum_{\boldsymbol{\alpha},\boldsymbol{\beta}}\rho_{\boldsymbol{\alpha},\boldsymbol{\beta}}\ |\boldsymbol{\alpha}\rangle\langle\boldsymbol{\beta}|\ .
\ee
In the Hilbert-space doubling formalism this is considered as a vector
on the linear vector space of operators on ${\cal H}$
\be
|\!|\bar{\rho}\rangle=\sum_{\boldsymbol{\alpha},\boldsymbol{\beta}}\rho_{\boldsymbol{\alpha},\boldsymbol{\beta}}\ |\boldsymbol{\alpha}\rangle\otimes|\boldsymbol{\beta}\rangle\!\rangle\ .
\ee
A basis of the 4-dimensional space of states associated with site $j$
is given by
\begin{align}
|\!|1\rangle_j&=|\mathfrak{1}\rangle_j\otimes|\mathfrak{1}\rangle\!\rangle_j\ ,\qquad
|\!|2\rangle_j=|\mathfrak{0}\rangle_j\otimes|\mathfrak{1}\rangle\!\rangle_j\ ,\nn
|\!|3\rangle_j&=|\mathfrak{1}\rangle_j\otimes|\mathfrak{0}\rangle\!\rangle_j\ ,\qquad
|\!|4\rangle_j=|\mathfrak{0}\rangle_j\otimes|\mathfrak{0}\rangle\!\rangle_j\ .
\label{states_1rep}
\end{align}
Introducing notations such that
\be
{\cal O}={\cal O}\otimes\mathds{1}\ ,\qquad
\tilde{\cal O}=\mathds{1}\otimes{\cal O}\ ,
\ee
and using that the jump operators $\ell^+_{j} :=  c^\dag_{j+1}c_{j}$ fulfil
$(\ell_j^\pm)^T=\ell_j^\mp$, the Lindblad equation
\fr{Lindblad}, \fr{eq:L-edge} is then recast in the form
\be
\frac{d}{dt}|\!|\bar\rho_t\rangle={\cal L}|\!|\bar\rho_t\rangle\ ,
\ee
where
\be
{\cal L}=J\sum_jp\left[
\ell_j^-\tilde{\ell}_j^--\frac{1}{2}\left(\ell_j^+\ell_j^-+
\tilde{\ell}_j^+\tilde{\ell}_j^-\right)\right]
+q\left[
\ell_j^+\tilde{\ell}_j^+-\frac{1}{2}\left(\ell_j^-\ell_j^++
\tilde{\ell}_j^-\tilde{\ell}_j^+\right)\right]\ .
\label{Liou}
\ee
In order to represent ${\cal L}$ in the basis \fr{states_1rep} we
define Hubbard operators in the usual way by 
\be
E_j^{ab}\equiv |\!|a\rangle_j\ {}_j\langle b|\!|\ .
\label{Hubbardops}
\ee
Keeping track of minus signs arising from fermionic anticommutation
relations we have
\begin{align}
\ell^-_{j}&=(E_j^{12}+E_j^{34})(E_{j+1}^{21}+E_{j+1}^{43})\ ,\nn
\tilde{\ell}^-_{j}&=(E_j^{13}-E_j^{24})(E_{j+1}^{31}-E_{j+1}^{42}).
\end{align}
This results in
\begin{align}
{\cal L}&=J\sum_jp\left[J^+_jJ^-_{j+1}+\frac{1}{4}\big(
J^z_jJ^z_{j+1}-1\big)\right]
+q\left[J^-_jJ^+_{j+1}+\frac{1}{4}\big(J^z_jJ^z_{j+1}-1\big)\right]
+\frac{p+q}{4}C_jC_{j+1}\ ,
\label{Lasep}
\end{align}
where we have defined
\begin{align}
J^+_j&=E_j^{14}\ ,\quad
J^-_j=E_j^{41}\ ,\quad
J^z_j=E_j^{11}-E_j^{44}\ ,\quad
C_j=E_j^{22}-E_j^{33}\ .
\label{JandC}
\end{align}
Quantities of physical interest are expressed in this formalism as follows
\be
{\rm Tr}\left[\bar\rho_t {\cal O}\right]=\langle \mathds{1}|\!|{\cal O}|\!|\bar\rho_t\rangle\ ,
\ee
where
\be
\langle \mathds{1}|\!|=\otimes_{j=1}^L\Big[{}_j\langle \mathfrak{1}|\!|+{}_j\langle \mathfrak{4}|\!|\Big]\ .
\ee
In the two-replica case we proceed analogously. Our starting point are
two sets of fermion creation and annihilation operators corresponding
to the two replicas that commute with one another, \emph{cf.}
\fr{CACR}. To proceed we define new fermions that fulfil pure
anticommutation relations 
\be
\bar{c}_{1;j}=c_{1;j}\ ,\quad
\bar{c}_{2;j}=c_{2;j}(-1)^{N_1}\ ,\quad N_1=\sum_{j=1}^Lc_{1;j}\ .
\ee
In the Hilbert space doubling formalism the Lindblad equation
\fr{eq:rho2} for $\rho^{(2)}_t$ then takes the form 
\be
\frac{d}{dt} |\!|\rho^{(2)}_t\rangle=\hat{\cal L}_2|\!|\rho^{(2)}_t\rangle\ .
\ee
Now the space associated with site $j$ is 16-dimensional and a
convenient basis is given by the graded tensor product of the states
\fr{states_1rep} 
\be
|\alpha\rangle_j\otimes|\beta\rangle_j\ ,\quad \alpha,\beta\in\{1,2,3,4\}.
\ee
The jump operators \fr{jump2} in the Lindblad equation \fr{eq:rho2} for
$\rho^{(2)}_t$ are
\be
\ell^-_{a;j}=(\ell^+_{a;j})^\dagger=\bar{c}^\dagger_{a;j}\bar{c}_{a;j+1}\ ,\qquad
a=1,2.
\ee
The Lindbladian is given by
\be
\hat{\cal L}_2=J\sum_{j=1}^L\sum_{a=1}^2\left({\cal L}_{a,j}+{\cal L}^{\rm int}_{a,j}\right) ~,
\ee
with
\begin{align}
{\cal L}_{a,j}=&p\left[\ell_{a,j}\tilde{\ell}_{a,j}-\frac{1}{2}\left(
  \ell_{a,j}^\dagger\ell_{a,j}+\tilde{\ell}_{a,j}^\dagger\tilde{\ell}_{a,j}\right)\right]
+q\left[\ell^\dagger_{a,j}\tilde{\ell}^\dagger_{a,j}-\frac{1}{2}\left(
\ell_{a,j}\ell^\dagger_{a,j}+\tilde{\ell}_{a,j}\tilde{\ell}^\dagger_{a,j}\right)\right]\ ,\nn
{\cal L}^{\rm int}_{a,j}=&
p\left[\ell_{a,j}\tilde{\ell}_{\bar{a},j}-\frac{1}{2}\left(\ell^\dagger_{a,j}\ell_{\bar{a},j}
+\tilde{\ell}_{a,j}\tilde\ell^\dagger_{\bar{a},j}\right)\right]
+q\left[\ell_{a,j}^\dagger\tilde{\ell}_{\bar{a},j}^\dagger
-\frac{1}{2}\left(\ell_{a,j}\ell_{\bar{a},j}^\dagger+\tilde{\ell}^\dagger_{a,j}\tilde\ell_{\bar{a},j}\right)\right], 
\end{align}
where we defined $\bar{1}=2$, $\bar{2}=1$. We now construct explicit
matrix representations of $\hat{\cal   L}_2$. We start by decomposing
the 16 basis states on site $j$ according to
\begin{align}
  |\!|0\rangle_j&=|3\rangle_j\otimes|2\rangle_j\ ,\quad
  |\!|1\rangle_j=|1\rangle_j\otimes|1\rangle_j\ ,\quad
  |\!|2\rangle_j=|1\rangle_j\otimes|4\rangle_j\ ,\\
  |\!|3\rangle_j&=|4\rangle_j\otimes|1\rangle_j\ ,\quad
  |\!|4\rangle_j=|4\rangle_j\otimes|4\rangle_j\ ,\quad
  |\!|5\rangle_j=|2\rangle_j\otimes|3\rangle_j\ ,
&\nn
  |\!|6\rangle_j&=|1\rangle_j\otimes|2\rangle_j\ ,\quad
  |\!|7\rangle_j=|2\rangle_j\otimes|1\rangle_j\ ,\quad
  |\!|8\rangle_j=|4\rangle_j\otimes|2\rangle_j\ ,\quad
  |\!|9\rangle_j=|2\rangle_j\otimes|4\rangle_j\ ,
&\nn
  |\!|10\rangle_j&=|1\rangle_j\otimes|3\rangle_j\ ,\quad
  |\!|11\rangle_j=|3\rangle_j\otimes|1\rangle_j\ ,\quad
  |\!|12\rangle_j=|4\rangle_j\otimes|3\rangle_j\ ,\quad
  |\!|13\rangle_j=|3\rangle_j\otimes|4\rangle_j\ ,
&\nn
  |\!|14\rangle_j&=|2\rangle_j\otimes|2\rangle_j\ ,\quad
  |\!|15\rangle_j=|3\rangle_j\otimes|3\rangle_j\ .
\label{states_2rep}
\end{align}
We then define Hubbard operators by
\be
{\cal E}_j^{a,b}=|\!|a\rangle_j\ {}_j\langle b|\!|,
\label{Hubbardops2}
\ee
and operators $(G^{ab}_j)^\dagger=G^{ba}_j$
\begin{align}
G^{12}_j&={\cal E}_j^{1,3}+{\cal E}_j^{2,4}+{\cal E}_j^{6,8}+{\cal
  E}_j^{10,12}\ ,\nn
G^{13}_j&={\cal E}_j^{0,4}+{\cal E}_j^{1,5}-{\cal E}_j^{6,9}-{\cal E}_j^{11,12}\ ,\nn
G^{14}_j&={\cal E}_j^{0,3}-{\cal E}_j^{2,5}-{\cal E}_j^{6,7}+{\cal E}_j^{13,12}\ ,\nn
G^{23}_j&=-{\cal E}_j^{0,2}+{\cal E}_j^{3,5}-{\cal E}_j^{8,9}+{\cal E}_j^{11,10}\ ,\nn
G^{24}_j&=-{\cal E}_j^{0,1}-{\cal E}_j^{4,5}-{\cal E}_j^{8,7}-{\cal E}_j^{13,10}\ ,\nn
G^{34}_j&={\cal E}_j^{2,1}+{\cal E}_j^{4,3}+{\cal E}_j^{9,7}+{\cal
  E}_j^{13,11}\ ,\nn
G^{11}_j&={\cal E}^{0,0}_j+{\cal E}^{1,1}_j+{\cal E}^{2,2}_j+{\cal
  E}^{6,6}_j+{\cal  E}^{10,10}_j+{\cal  E}^{11,11}_j+{\cal  E}^{13,13}_j
+{\cal  E}^{15,15}_j\ ,\nn
G^{22}_j&={\cal E}^{0,0}_j+{\cal E}^{3,3}_j+{\cal E}^{4,4}_j+{\cal  E}^{8,8}_j+{\cal E}^{11,11}_j+{\cal  E}^{12,12}_j+{\cal  E}^{13,13}_j+{\cal  E}^{15,15}_j\ ,\nn
G^{33}_j&={\cal E}^{2,2}_j+{\cal E}^{4,4}_j+{\cal E}^{5,5}_j+{\cal  E}^{7,7}_j+{\cal  E}^{10,10}_j+{\cal  E}^{11,11}_j+{\cal  E}^{12,12}_j+{\cal  E}^{15,15}_j\ ,\nn
G^{44}_j&={\cal E}^{1,1}_j+{\cal E}^{3,3}_j+{\cal E}^{5,5}_j+{\cal
  E}^{9,9}_j+{\cal E}^{10,10}_j+{\cal  E}^{12,12}_j+{\cal
  E}^{13,13}_j+{\cal  E}^{15,15}_j\ .
\label{notationsfermions}
\end{align}
The operators $G_j^{ab}$ fulfil the $gl(4)$ commutation relations
\be
[G^{ab}_j,G^{cd}_j]=\delta_{b,c}G_j^{ad}-\delta_{a,d}G_j^{ca}\ .
\ee
In terms of \fr{notationsfermions} the two-replica Lindbladian becomes
\begin{align}
\hat{\cal L}_2=J\sum_{j=1}^Lp\overleftarrow{\cal L}_{j,j+1}
+q\overrightarrow{\cal L}_{j,j+1}\equiv Jp\overleftarrow{\cal L}_2+Jq\overrightarrow{\cal L}_2\ ,
\end{align}
where $\overleftarrow{\cal L}_2=\big(\overrightarrow{\cal
  L}_{2}\big)^\dagger$ and
\begin{align}
\overleftarrow{\cal L}_{j,j+1}=&
G^{12}_jG^{21}_{j+1}+G^{43}_jG^{34}_{j+1}+G^{13}_jG^{31}_{j+1}+G^{42}_jG^{24}_{j+1}
+\frac{1}{2}\left[G^{41}_jG^{14}_{j+1}+G^{32}_jG^{23}_{j+1}+{\rm h.c.}\right]\nn
&+\frac{1}{2}\sum_{a=1}^4G^{aa}_jG^{aa}_{j+1}-\frac{1}{4}(C_j+C_{j+1}+4).
\label{Lfermi}
\end{align}
Here we have defined $C_j=\sum_{a=1}^4 G^{aa}_j-2$, as above. This is in complete agreement with \fr{eq:L12replica} once we account
for the relation between the Lindbladian and its dual \fr{duallind}.

We note that the states \fr{states_2rep} transform under $gl(4)$ as follows:
\begin{itemize}
\item{} $\{|\!|a\rangle_j|a=0,\dots,5\}$ form the six-dimensional representation [6] of $gl(4)$ and have $C_j=0$;
\item{} $\{|\!|a\rangle_j|a=6,\dots,9\}$ form the fundamental representation [4] of $gl(4)$ and have $C_j=1$;
\item{} $\{|\!|a\rangle_j|a=10,\dots,13\}$ form the fundamental representation [$\bar{4}$] of $gl(4)$ and have $C_j=-1$;
\item{} $\{|\!|14\rangle_j\}$ and $\{|\!|15\rangle_j\}$ form one-dimensional representations of $gl(4)$ and have $C_j=\mp 2$.
\end{itemize}

The quantities of physical interest are expressed as follows
\be
{\rm Tr}\left[\bar\rho^{(2)}_t {\cal O}_1\otimes{\cal O}_2\right]=\langle \mathds{1}_2|\!|{\cal O}_1\otimes{\cal O}_2|\!|\bar\rho^{(2)}_t\rangle\ ,
\ee
where
\be
\langle \mathds{1}_2|\!|=\otimes_{j=1}^L\Big[{}_j\langle 1|\!|+{}_j\langle2|\!|+{}_j\langle3|\!|+{}_j\langle4|\!|\Big]\ .
\label{id2}
\ee

\section{\sfix{$U(1)$} symmetries, sectors and fragmentation}
\label{sec:qssep-sector}

\subsection{Symmetries}

The Q-ASEP $\&$ Q-SSEP models possess $u(1)$ symmetries, one for each site, which echo the fact that two pairs of complex Brownian motions $(W_t,\overline{W}_t)$ and $(\tilde W_t,\tilde{\overline{W}}_t)$ differing by a phase, $\tilde W_t=e^{i\theta}\, W_t$, have identical distributions. 

For any realization of the Brownian motions $W^j_t$, let us consider the representation of $U(1)^L$  on the physical Hilbert space generated by the local particle numbers $n_j$. That is: elements of $U(1)^L$ are represented by $e^{i\sum_j\theta_j n_j}$ with arbitrary parameter $\theta_j$. The two Hamiltonian increments,  $dH_t$ from eqn \eqref{eq:def-dH}, and $d\tilde H_t$, obtained by conjugation with these $U(1)$'s, with
\[ d\tilde H_t := e^{-i\sum_j \theta_j n_j}\, dH_t\, e^{i\sum_j\theta_j n_j} ,\]
have identical distributions. This holds because $d\tilde H_t$ coincides with eqn \eqref{eq:def-dH} but with Brownian increments $d\tilde W^j_t=e^{i(\theta_{j}-\theta_{j+1})}\, dW_t^j$ whose distribution is identical to that of $dW_t^j$. As a consequence, the two density matrices $\rho_t$ and $\tilde\rho_t=e^{-i\sum_j \theta_j n_j}\,\rho_t\, e^{i\sum_j\theta_j n_j}$, time evolved with the Hamiltonian increments $dH_t$ and $d\tilde H_t$, respectively, have identical distributions. This property holds for Q-ASEP and Q-SSEP as well.

This implies that, for any $R$ replicas, acting with the Lindbladian or conjugating with the total local particle numbers are commuting operations:
\beq
[\mathcal{L}\,,\,C_j]=0 ~,\quad [\mathcal{L}^*\,,\,C_j]=0,
\eeq
for all $j=1,\cdots, L$ with $C_j=\sum_a C_{a;j}$. Recall that the $C_{a;j}$'s act on any operators as $C_{a;j}\,O =[n_{a;j},O]$ with $n_{a;j}=c^\dag_{a;j}c_{a;j}$ the local number operator in the $a$-th replica. They count the number of fermion insertions in the operators $O$. Since the latter are made of product of the local fermionic operators, the spectrum of $C_j$ runs from $-R$ to $R$:
\[ C_j= -R, -R+1,\cdots,0,\cdots, R-1,R ~.\]
For instance, the operator $\prod_a c^\dag_{a;j}$ (resp. $\prod_a c_{a;j}$) is only the operator with $C_j$-charge $+R$ (resp. $-R$).

\subsection{Sectors and fragmentation}

Since the $C_j$'s commute with the Lindbladian, their eigenvalues are good quantum
numbers preserved by the dynamics. Phrased differently, the
Q-ASEP $\&$ Q-SSEP dynamics only involves mixing between operators with
identical $C$-charges. Note that one $C$-charge is assigned per
site. We denote the set of $C$-charges by
  $\{\mathfrak{c}_j|j=1,\dots,L\}$ by $\cvec$.   

We call the subspace of the operators of given $C$-charges \emph{a sector}. Each sector is labelled by the corresponding set $\cvec$. The
Q-ASEP $\&$ Q-SSEP dynamics takes place a within given 
sector and does not induce transitions between different sectors. In
particular, since the extremal $C$-charges $\pm R$ are non-degenerate,
the dynamics inside those sectors is trivial in the sense that operators with extremal $C$ charge do not move. The splitting of the
Q-ASEP $\&$ Q-SSEP dynamics into sectors has be been termed
``fragmentation'' in Ref. \cite{Essler2020Integrability}.  

Let us analyze the operator content of the different sectors
depending on the number of replicas. For a single replica, the
different values of the $C$-charge are $\pm 1$ and $0$ with the
following content, at each given site $j$ (with $n_j=c_j^\dag c_j$): 
\beqs
C^{[R=1]}_j=0 &:& \mathbb{I}_j,\ n_j , \leadsto [2] ~,\\
C^{[R=1]}_j= \pm 1 &:& c_j^\dag\ (c_j) , \leadsto [1],\ ([\bar 1]).
\eeqs

In order to describe time evolution in a given sector it is useful to define projection operators onto the different subspaces at a given site
\be
P_j^{[\alpha]}\ ,\quad \alpha=0,\pm 1\ .
\ee
In terms of the explicit representation \fr{states_1rep} these read
\be
P_j^{[0]}=|\mathfrak{1}\rangle\langle\mathfrak{1}|+|\mathfrak{4}\rangle\langle\mathfrak{4}|\ ,\quad
P_j^{[1]}=|\mathfrak{2}\rangle\langle\mathfrak{2}|\ ,\quad
P_j^{[-1]}=|\mathfrak{3}\rangle\langle\mathfrak{3}|\ .
\ee
The Lindbladian in a given sector $\cvec$ is then given by
\be
{\cal L}_{\cvec}=\Big(\prod_{j=1}^LP_j^{[c_j]}\Big){\cal L}\Big(\prod_{k=1}^LP_j^{[c_k]}\Big) .
\ee
The time evolution of the single-particle Green's function ${\rm Tr}\left[\rho(t) c^\dagger_jc_k\right]$, with $k>j+1$, occurs in sector
\be
\cvec_1=(0,\dots,0,\underbrace{1}_j,0,\dots,0,\underbrace{-1}_k,0,\dots,0)\ .
\ee
The corresponding Lindbladian is (with periodic boundary conditions)
\be
{\cal L}_{\cvec_1}=P^{[1]}_j+{\cal L}_{[j+1,k-1]}+P^{[-1]}_k+{\cal
    L}_{[k+1,j-1]}\ ,
\label{LGF}
\ee
where \cite{Essler2020Integrability}
\begin{align}
{\cal L}_{[1,\ell]}=&\sum_{j=1}^{\ell-1} p\left[E_j^{14}E_{j+1}^{41}-E_j^{44}E_{j+1}^{11}\right]
+q\left[E_j^{41}E_{j+1}^{14}-E_j^{11}E_{j+1}^{44}\right]\nn
&-\frac{q}{2}E_{\ell}^{11}-\frac{p}{2}E_{\ell}^{44}
-\frac{p}{2}E_{1}^{11}-\frac{q}{2}E_{1}^{44}\ .
\label{boundaryL}
\end{align}
The dynamics described by \fr{LGF} is that of two stationary
``defects'' at sites $j$ and $k$, which affect the dynamics in the
spatial regions between them through the boundary terms in
\fr{boundaryL}. The dynamics is integrable \cite{Essler2020Integrability}, but so far has
only been solved for the case of the SSEP.

For two replicas, the possible $C$-charges are $\pm 2$, $\pm 1$ and $0$ with contents (with the convention, $c^+\equiv c^\dag,\ c^-\equiv c,\ n=c^\dag c$):
\beqs
C^{[R=2]}_j=0 &:& \mathbb{I}_j\otimes\mathbb{I}_j,\ \mathbb{I}_j\otimes n_j,\ n_j\otimes\mathbb{I}_j,\ n_j\otimes n_j,\ c_j^+\otimes c_j^-,\ c_j^-\otimes c_j^+ \leadsto [6]\\
C^{[R=2]}_j=\pm1 &:& \mathbb{I}_j\otimes c_j^\pm,\ c_j^\pm\otimes n_j,\ n_j\otimes c_j^\pm, c_j^\pm\otimes\mathbb{I}_j, \leadsto [4],\ ([\bar 4]) \\
C^{[R=2]}_j=\pm 2 &:& c_j^\pm\otimes c_j^\pm , \leadsto  [1],\ ([\bar 1]).
\eeqs
The operators in the $\cvec_0=(0,\dots,0)$ sector form an algebra (because the product of two such operators still has $C$ charge zero). There is an action of this algebra on the other sectors, or more precisely the other sectors form modules of the $\cvec_0$ algebra under multiplication, because the $C$-charge is additive under products.

The above numbers $[\cdot]$ refer to the multiplicity of the corresponding sectors. They have a group theoretical interpretation. The space of local operators at site $j$ is isomorphic to that of the Fock space of the $2R$ effective fermions $f_j^{A\,\dag}$. Its dimension $2^{2R}= 2^R\cdot 2^R$ indeed coincides with that of the space of endomorphisms on the local physical Hilbert space (generated by the $R$ physical fermions $c_{a;j}^\dag$ at site $j$). As explained above, there is an action of $gl(2R)$ on the operators localised at site $j$ whose generators are the super-operators $G^{AB}_j=f_j^{A\,\dag} f_j^B$, see eqn \eqref{eq:def-GAB}. The algebra $gl(2R)$ has a central $u(1)$, generated by the $C$-charge $C_j=\sum_A f_j^{A\,\dag} f_j^A-R$, and $gl(2R)=sl(2R) \oplus u(1)$. In particular, the $C_j$ charge commutes with the $sl(2R)$ action and hence each sector of given $C$-charge form a $sl(2R)$ representation. Since the space $\mathrm{Vec}[O_j]$ of operators localised at site $j$ is the Fock space of $2R$ fermions, it decomposes as the sum of all fundamental representations of $sl(2R)$:
\[
\mathrm{Vec}[O_j] \equiv [\mu_0]\oplus[\mu_1]\oplus\cdots\oplus[\mu_{2R-1}]\oplus[\mu_{2R}] ~,
\]
where $[\mu_k]$ denotes the representation of $sl(2R)$ made of antisymmetric rank $k$ tensors. Their dimensions are $\frac{(2R)!}{k!(2R-k)!}$ and their $C$-charges are $k-R$. In particular, the $C=0$ sector corresponds to the representation $\mu_R$ with dimension ${(2R)!}/{(R!)^2}$.

For one replica, $R=1$, the operator content is $[1]\oplus[2]\oplus[\bar1]$ with respect to $sl(2)$. 

For two replicas, $R=2$, the operator content is 
\[
\ytableausetup{smalltableaux,centertableaux}
[1]\oplus [4] \oplus [6] \oplus [\bar 4] \oplus [\bar 1] \equiv 
\bullet \oplus \ydiagram{1} \oplus \ydiagram{1,1} \oplus \overline{\ydiagram{1}} \oplus \overline \bullet ,
\]
where we made explicit the description of the $sl(2R)$ representations in terms of Young tableaux. The representation $[4]$ is the vector representation of $sl(4)$, and $[\bar 4]$ its complex conjugated, the representation $[6]$ is that of rank $2$ antisymmetric tensor. Since $sl(4)\equiv so(6)$, as Lie algebras, the representation $[6]$ is also the vector representation of $so(6)$, and $[4]$ and $[\bar 4]$ the two chiral/anti-chiral spin representations of $so(6)$.

The effect of the fragmentation on the two-replica dynamics
is quite different from the single-replica case. We again introduce projection operators
\begin{align}
P^{[2,0]}_j&=\sum_{a=0}^5{\cal E}_j^{a,a}\ ,\quad
P^{[2,1]}_j=\sum_{a=6}^9{\cal E}_j^{a,a}   \ ,\quad
P^{[2,-1]}_j=\sum_{a=10}^{13}{\cal E}_j^{a,a}    \ ,\nn
P^{[2,-2]}_j&={\cal E}_j^{14,14}   \ ,\quad P^{[2,2]}_j={\cal E}_j^{15,15} .   
\end{align}
The Lindbladian in a given sector $\cvec$ then takes the form
\be
\hat{\cal L}_{2,\cvec}=\left(\prod_{j=1}^L P^{[2,c_j]_j}\right)\hat{\cal L}_2\left(\prod_{j=1}^L P^{[2,c_j]_j}\right)
\label{L2frag}
\ee
In contrast to the one-replica case the sites $j$ and $k$ now generally retain non-trivial dynamics and can be viewed as positions of “impurities” with internal structure that changes under time evolution. In the following we will focus on the particular sectors
\be
\cvec_0\equiv (0,\dots,0)\ ,\quad
\cvec_{\pm1}\equiv (\pm1,\dots,\pm 1)\ ,\quad
\cvec_{\pm2}\equiv (\pm2,\dots,\pm 2)\ .
\ee
As we shall point out below, these three sectors are the only gapless ones. In all other non-homogeneous sectors, the Lindbladian exhibits a finite spectral gap. Hence the operators belonging to these sectors decay exponentially fast in time, even in the large system size limit.

Thus, the two replica Q-ASEP $\&$ Q-SSEP dynamics in the sector $\cvec_0$ is equivalent to a $sl(4)$ spin chain in the anti-symmetric rank two representation. It can alternatively be viewed as $so(6)$ vector spin chain. The dynamics on the $\cvec_{\pm1}$ sector is that of $sl(4)$ vector spin chain. As we shall discuss below, the latter is integrable but, to our knowledge, the
former cannot be mapped to known integrable model although it might be integrable. Recall that, while the $U(1)$'s are symmetries of both the Q-ASEP and Q-SSEP dynamics, the $sl(2R)$ is a dynamical symmetry
only for Q-SSEP. 

\section{Integrability and non-integrability}
\label{sec:qssep-integral}

\subsection{Spin chain identification}

Thanks to eqn \eqref{eq:L-tensorC-GL} or eqn \eqref{eq:L-tensorC-SL}, $\mathcal{L}  \propto \sum_j\Big( \sum_{A,B} J_{j+1}^{AB}J_j^{BA}+ \frac{1}{2R}\,C_{j+1}C_j - \frac{R}{2} \Big)$, the Q-SSEP Lindblad dynamics has been re-written as a spin chain dynamics. In the one replicas case $R=1$, the spin chain is equivalent to the isotropic XXX Heisenberg model and thus integrable \cite{Korepin1993Quantum}. The purpose of this section is to determine whether the $R=2$ replica dynamics is integrable. The structure of the Lindbladian density \fr{L2frag} is different from that of integrable spin chains obtained by varying the representations of the symmetry algebra acting on a given site, see. e.g. \cite{Andrei1984Heisenberg,Lee1988Integrable,deVega1992New,Bedurftig1996Integrable}
in that the latter constructions generically lead to three-"spin" interactions, while in our case the Lindbladian density only involves two sites. 
This suggests that the dynamics in a general sector $\cvec$ is not integrable in a standard Yang-Baxter fashion. We therefore focus on the $\cvec_{\pm 1}$ and $\cvec_0$ sectors.

$\bullet$ {\it Integrability in the $\cvec_{\pm 1}$ sector.}

The quantity $\sum_{A,B} J_{j+1}^{AB}J_j^{BA}$ is the tensor Casimir acting on the tensor product of the $sl(2R)$ representations located at site $j$ and $j+1$. By $sl(2R)$ invariance, its diagonalisation is obtained by decomposing this tensor product into irreducible $sl(2R)$ sub-representations. As a consequence, $\sum_{A,B} J_{j+1}^{AB}J_j^{BA}$ can be written as a linear combination of the projectors on these irreducible sub-representations.

In the $\cvec_{\pm 1}$ sector, for two replicas, the representation on each site is the $sl(4)$ vector representation $\square=[4]$. Its tensor product with itself, $\square\otimes\square$,  decomposes  into two irreducible sub-representations, the symmetric and anti-symmetric tensors. The associated projectors are $\frac{1}{2}(1\pm P_{j;j+1})$ with $P_{j;j+1}$ the permutation operator. As a consequence, the Q-SSEP Lindbladian in the $\cvec_0$ sector reads
\[ \mathcal{L}^{[C=\pm1,\, R=2]}  \propto \sum_j P_{j;j+1} + \mathrm{const.} ~.\]
It is the $sl(2R)$ version of the isotropic Heisenberg spin chain, and it is known to be integrable \cite{Sutherland1975model}. In other words, the dynamics of Q-SSEP quadratic fluctuations in the $\cvec{\pm1}$ sector is integrable.

$\bullet$ {\it Integrability in the $\cvec_0$ sector?}

The formula eqn \eqref{eq:L-tensorC-SL} also holds in the $\cvec_0$ sector. In this sector and for two replicas, the on-site representation is the $[6]$ representation of $sl(4)$ or $so(6)$, that of rank $4$ anti-symmetric tensors of $sl(4)$ or that of $6$-dimensional vectors of $so(6)$. We have the decomposition rule, $[6]\otimes[6]=[1]+[15]+[20]$ as $sl(4)$ or $so(6)$ modules, with $[15]$ the rank two antisymmetric $so(6)$ tensors and $[20]$ the rank two traceless symmetric $so(6)$ tensors.
The projectors on these representations are (with $d=6$)
\[ P_A=\frac{1}{2}(1-P),\ P_S=\frac{1}{2}(1+P)-\frac{1}{d}Q,\ P_\bullet= \frac{1}{d}Q,\]
with $P$ the permutation operator and $Q$ the so-called trace operator. Thus $P$, $Q$ and the identity are the only $sl(4)\equiv so(6)$ intertwiners in $[6]\otimes[6]$. As a consequence,  in the $\cvec_0$ sector or for two replicas, $\sum_{A,B} J_{j+1}^{AB}J_j^{BA}$ is a linear combination of $P_{j;j+1}$, $Q_{j;j+1}$, up to an additive constant. Computing explicitly the tensor Casimir, see the end of Appendix \ref{Appendix:zero-mode}, we find
\[ \mathcal{L}^{[C=0,\, R=2]} \propto \sum_j (P_{j;j+1}-Q_{j;j+1}) + \mathrm{const.}\ .\]
The Hamiltonian of the known $sl(4)\equiv so(6)$ integrable model \cite{reshetikhin1985integrable} is $H= \sum_j ( 2 P_{j;j+1} - Q_{j;j+1} )$. Note the differences in the coefficients. 

Surprisingly, and despite this absence of direct connection to a known integrable models, we found numerical evidences that the dynamics in the $\cvec_0$ sector could be integrable. It therefore remains a challenge to analytically prove that the $\cvec_0$ dynamics is integrable and to decipher the underlying structures responsible for this property.

A similar analysis applies to higher number of replicas. In particular, because they associated to the $sl(2R)$ vector representations, the dynamics in the sectors with all $C$ equal to $\pm(R-1)$ are always integrable in the usual sense (they are mapped to the $sl(2R)$ analogues of the isotropic Heisenberg spin chain). Whether the other sectors are integrable is an open question.

\begin{figure}[ht]
     \centering
     \begin{subfigure}[b]{0.32\textwidth}
         \centering
         \includegraphics[width=\textwidth]{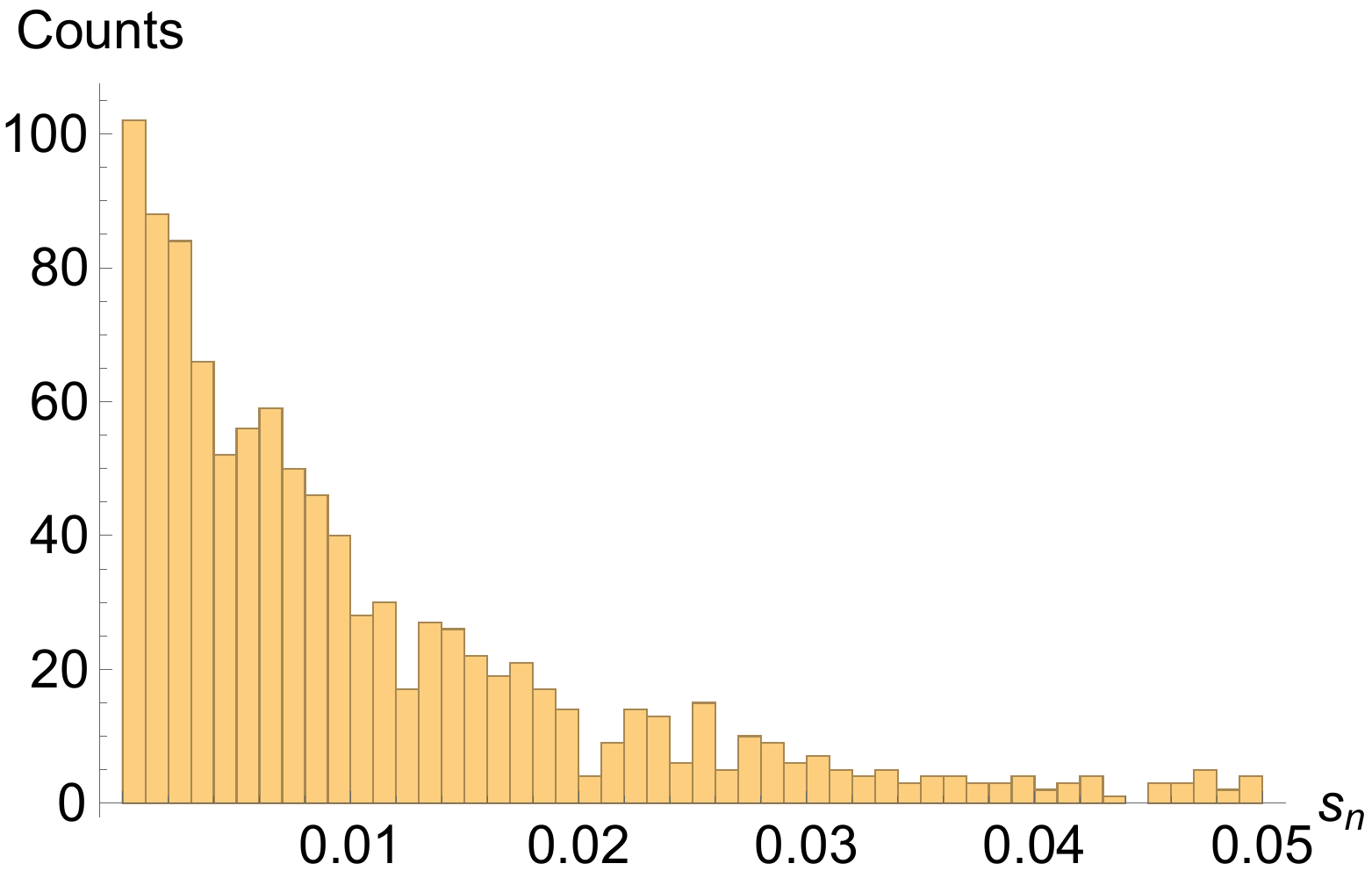}
         \caption{Spacings $s_n=|e_n-e_{n-1}|$ of adjacent eigenvalues}
     \end{subfigure}
     \hfill
     \begin{subfigure}[b]{0.32\textwidth}
         \centering
         \includegraphics[width=\textwidth]{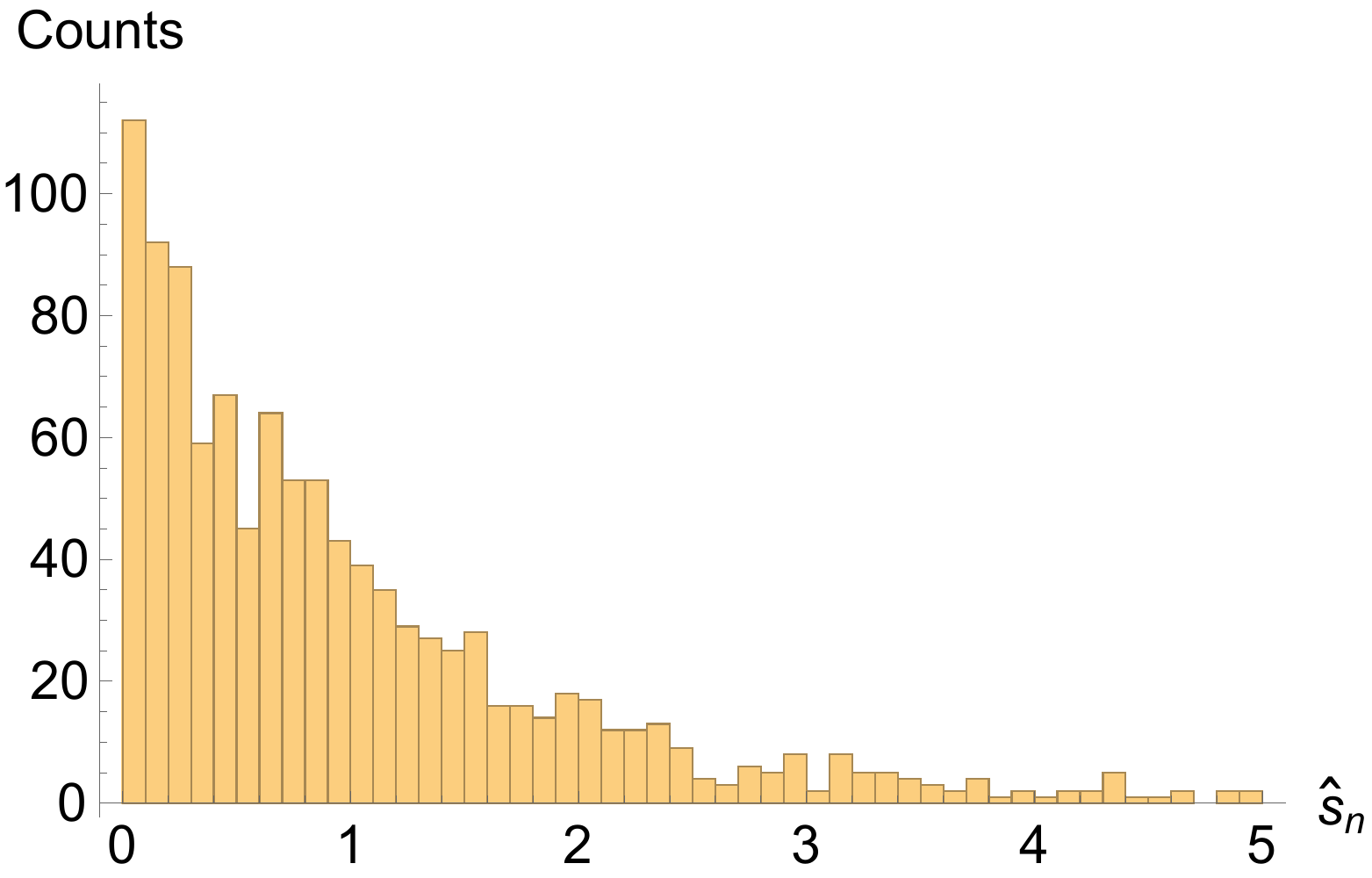}
         \caption{Spacings $\hat s_n=|\hat e_n-\hat e_{n-1}|$ of unfolded eigenvalues $\hat e_n$}
         \label{fig:c=1_unfolded}
     \end{subfigure}
     \hfill
     \begin{subfigure}[b]{0.32\textwidth}
         \centering
         \includegraphics[width=\textwidth]{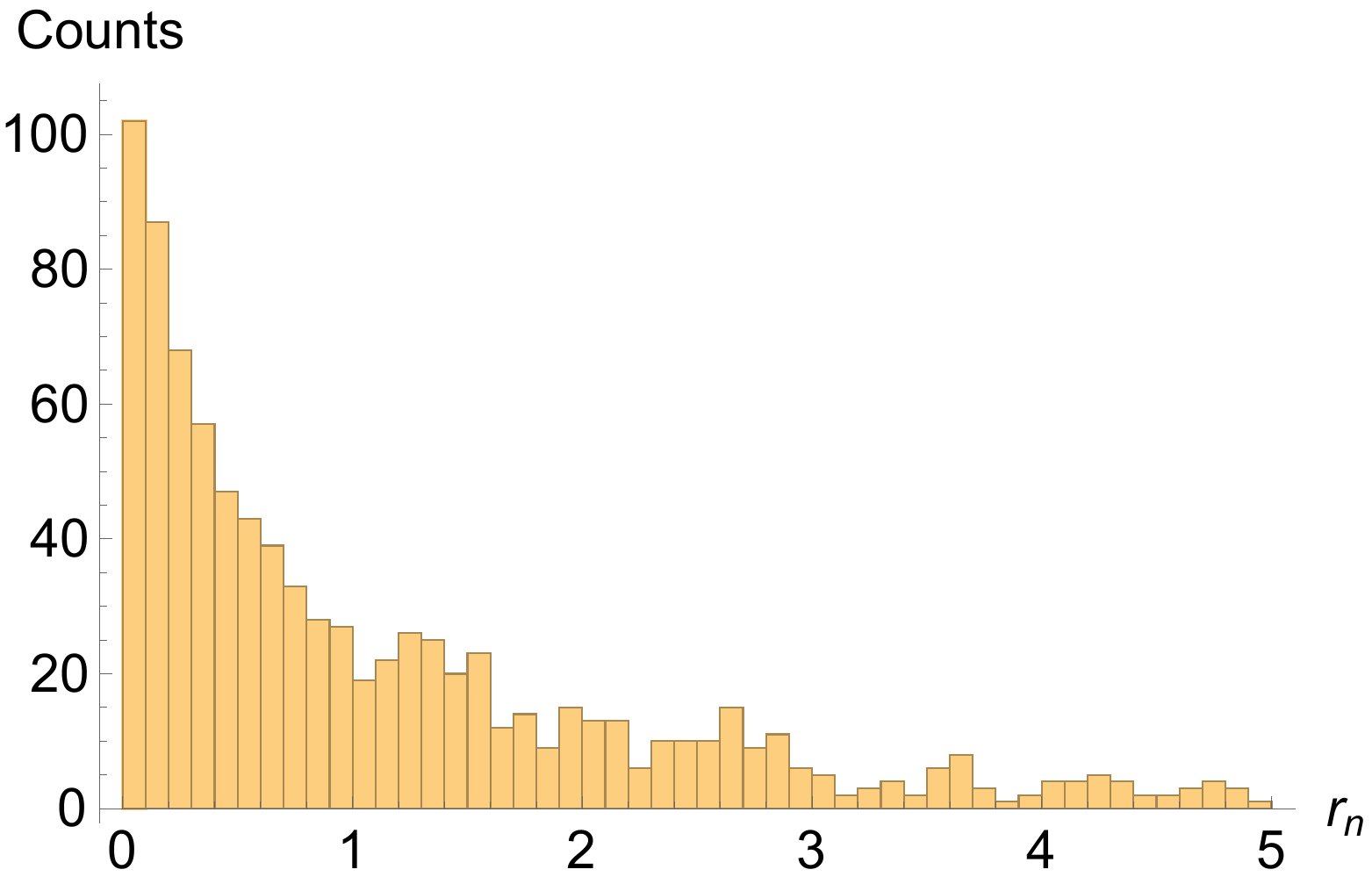}
         \caption{Ratio $r_n=s_n/s_{n-1}$ of adjacent spacings}
         \label{fig:c=1_ratio}
     \end{subfigure}
        \caption{Histograms for the spacings and ratios of the $N=1077$ eigenvalues of the $\cvec_1$ Lindbladian on $L=11$ sites after the degeneracies have been removed. The chosen symmetry sector is defined by $k=2\pi/11$ ($e^{ik}$ is the eigenvalue under translation by one site) and $(n_1,n_2,n_3,n_4)=(1,2,3,5)$ where $n_i$ is the number of times the local state $i=1,2,3,4$ appears in the tensor-product-state (the $\cvec_1$ sector has local dimension $4$ on each site). This corresponds to the Cartan charges $(J_1^z, J_2^z, \frac{C_1-C_2}{2})=(-2,-3,\frac{3}{2})$. The average ratio of consecutive spacings is $\langle \tilde r \rangle =0.3826$.}
        \label{fig:c=1}
\end{figure}

\begin{figure}[ht]
     \centering
     \begin{subfigure}[b]{0.32\textwidth}
         \centering
         \includegraphics[width=\textwidth]{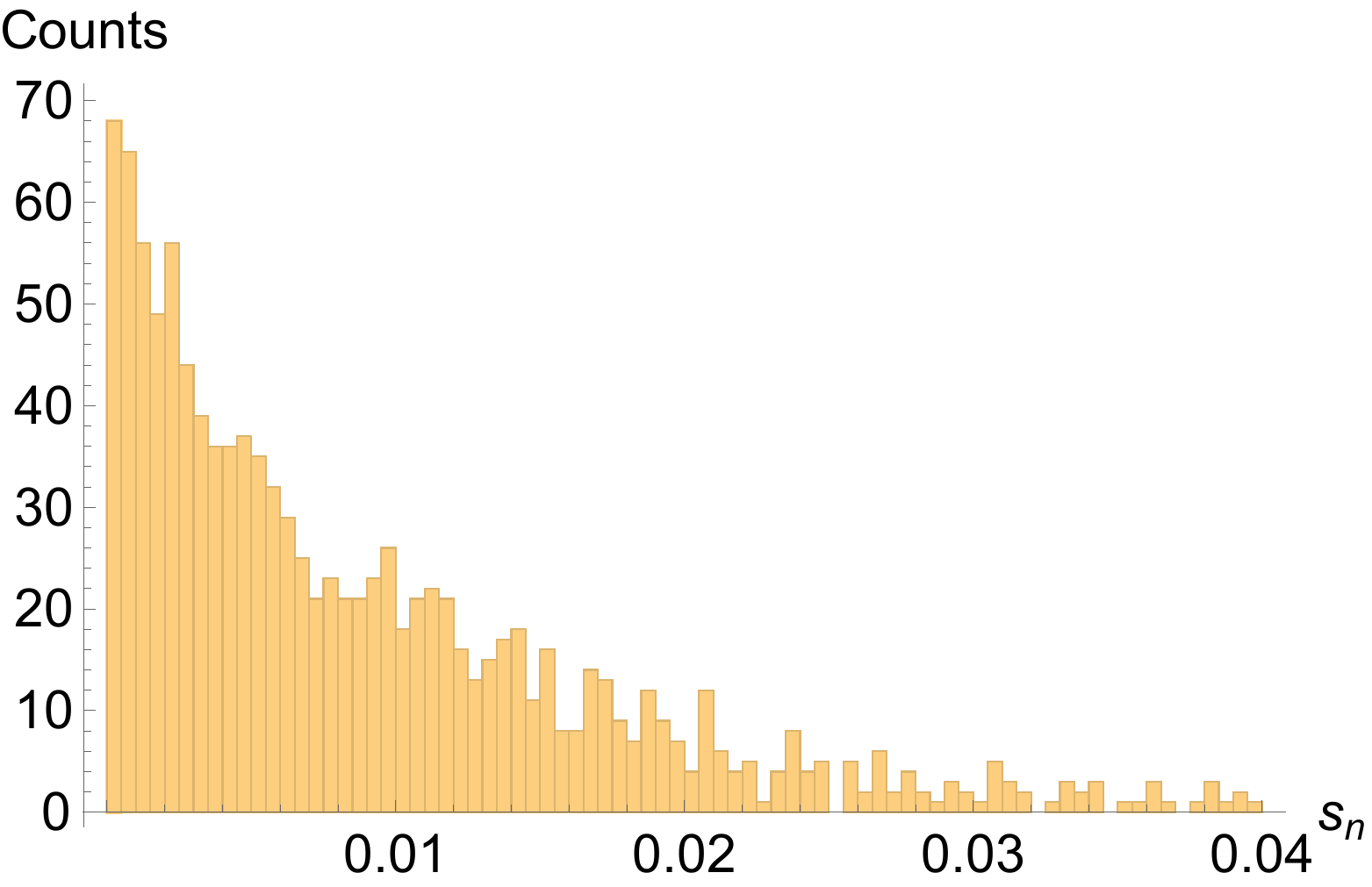}
     \end{subfigure}
     \hfill
     \begin{subfigure}[b]{0.32\textwidth}
         \centering
         \includegraphics[width=\textwidth]{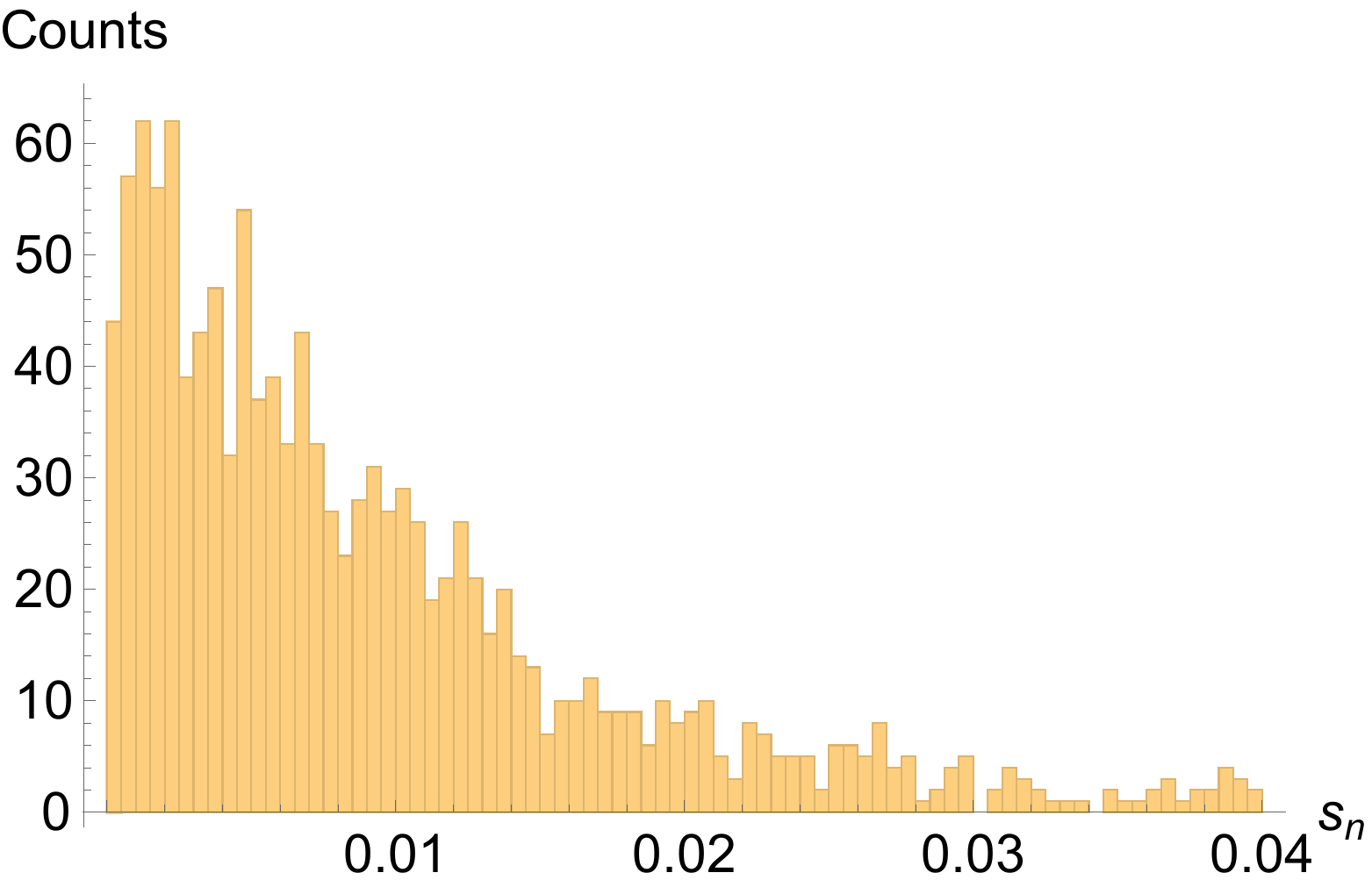}
     \end{subfigure}
     \hfill
     \begin{subfigure}[b]{0.32\textwidth}
         \centering
         \includegraphics[width=\textwidth]{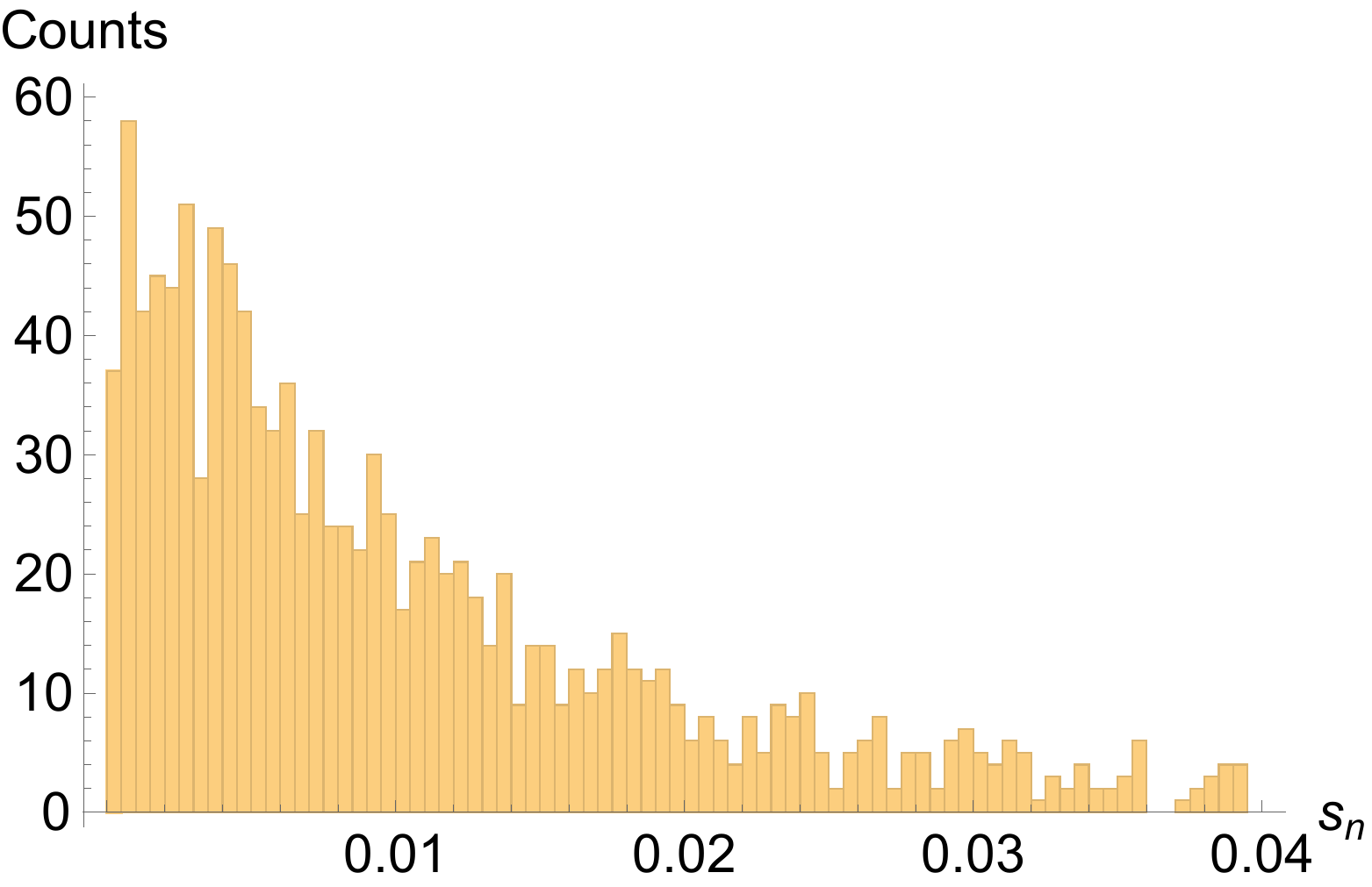}
         \end{subfigure}
     
     \begin{subfigure}[b]{0.32\textwidth}
         \centering
         \includegraphics[width=\textwidth]{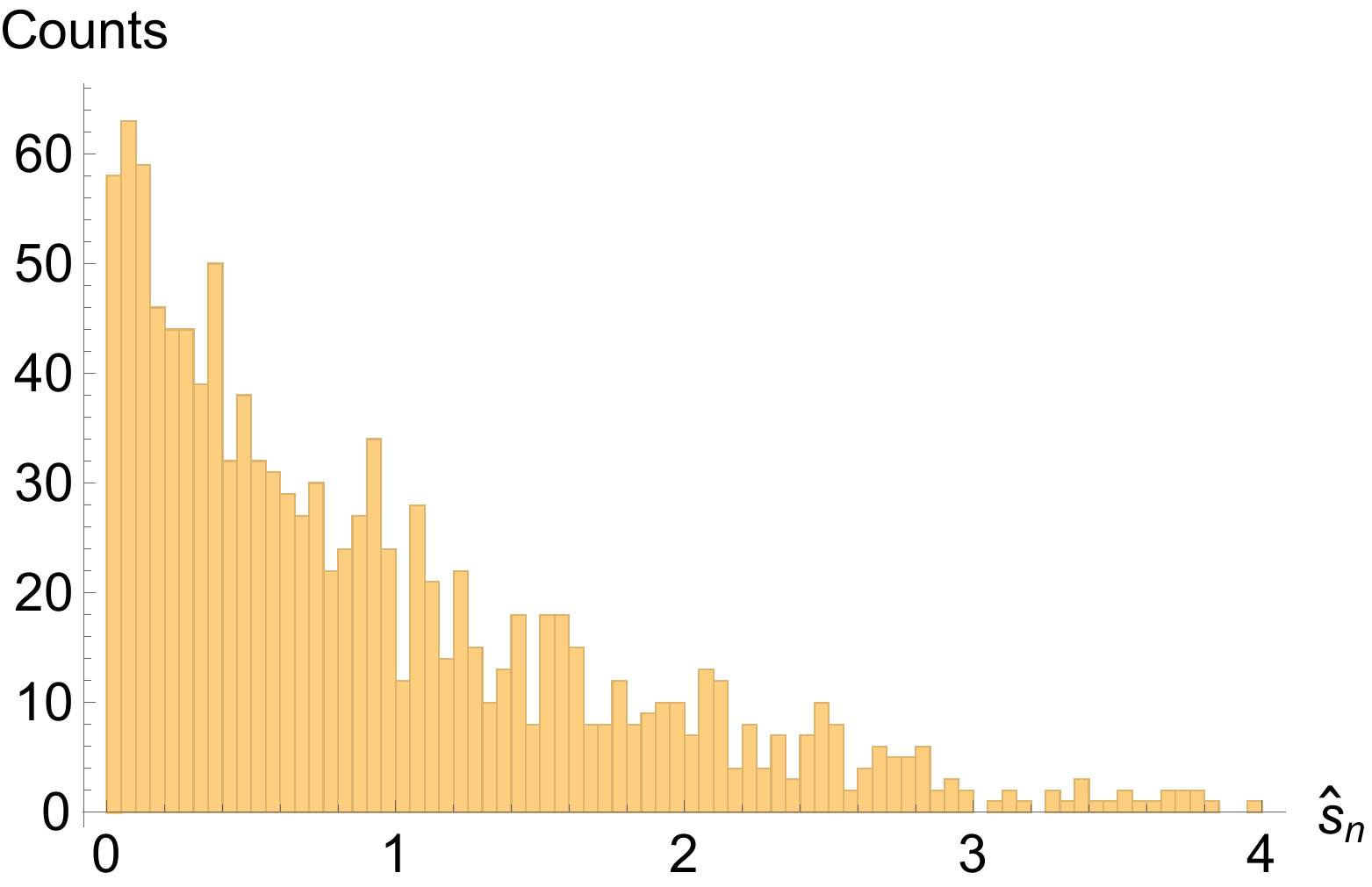}
         \centering\caption{Integrable case ($g=1/2$), $N=1205$, $\langle \tilde r \rangle =0.3817$}
     \end{subfigure}
     \hfill
     \begin{subfigure}[b]{0.32\textwidth}
         \centering
         \includegraphics[width=\textwidth]{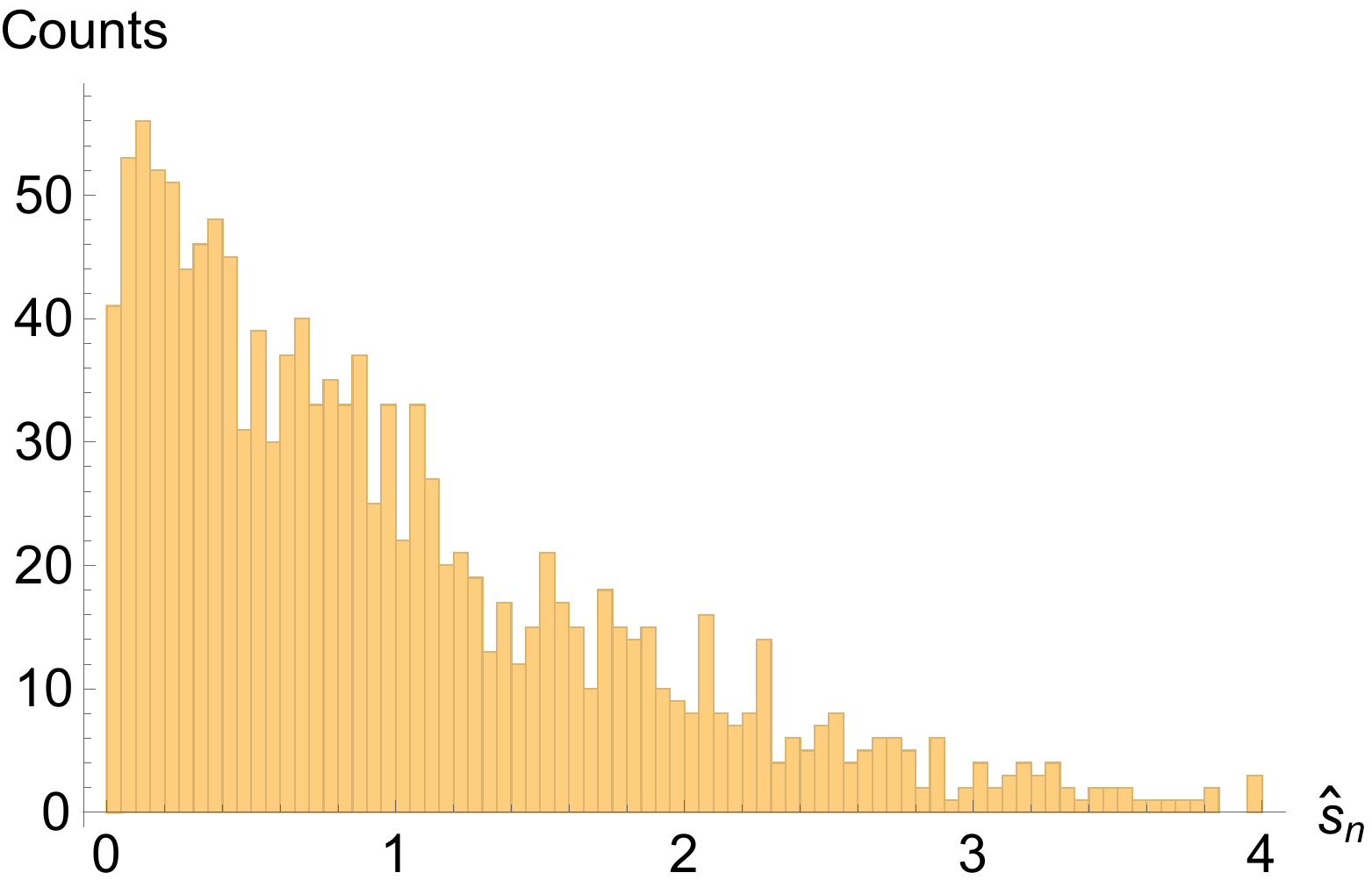}
         \centering\caption{Q-SSEP case ($g=1$), $N=1335$, $\langle \tilde r \rangle =0.3980$}
     \end{subfigure}
     \hfill
     \begin{subfigure}[b]{0.32\textwidth}
         \centering
         \includegraphics[width=\textwidth]{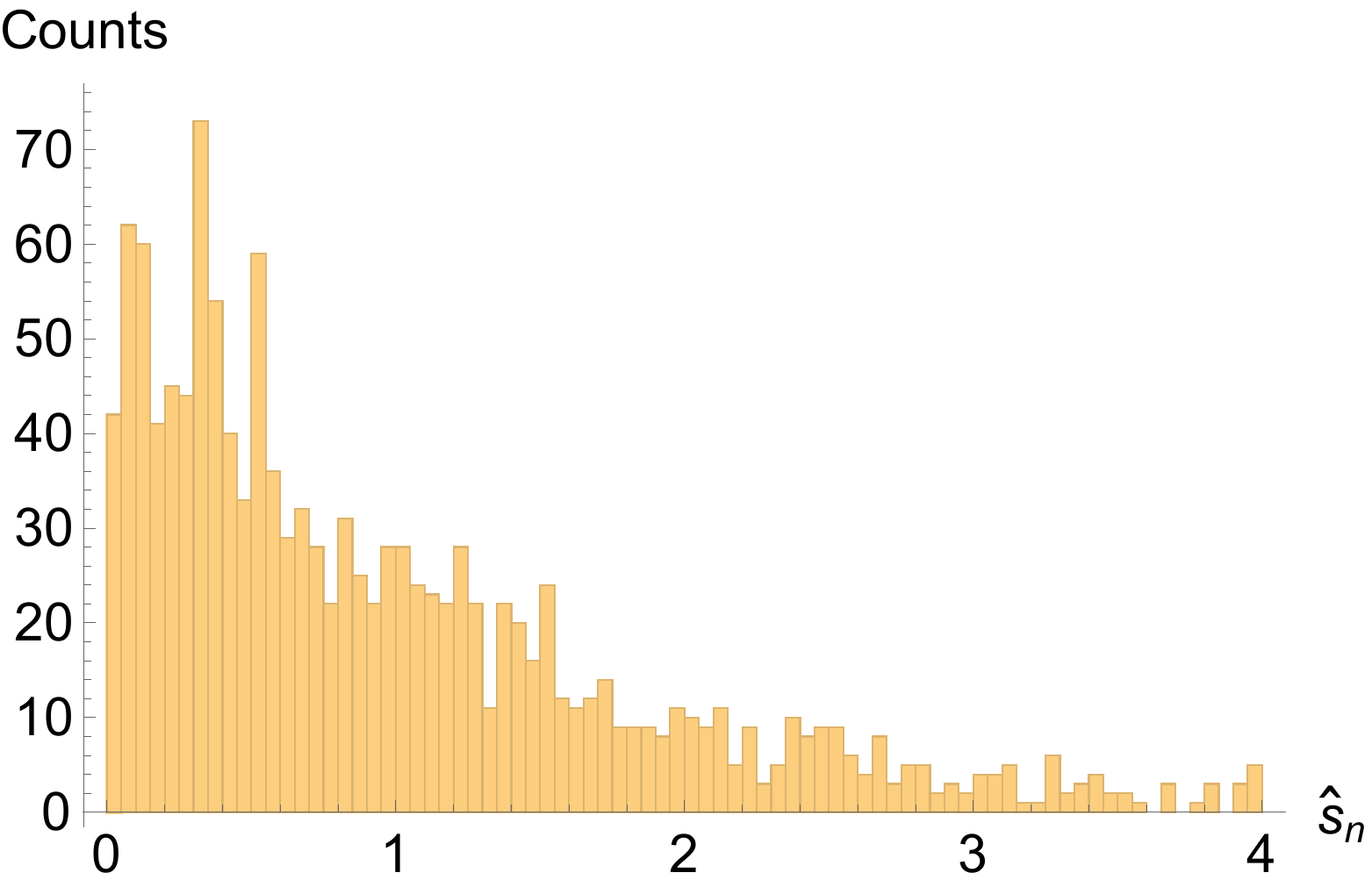}
         \centering\caption{A stronger perturbation with $g=2$, $N=1336$, $\langle \tilde r \rangle =0.3950$}
     \end{subfigure}
     \hfill
        \caption{Histograms of the spacings between consecutive raw eigenvalues (first row) and unfolded eigenvalues (second row) for a deformation of the Q-SSEP Linbladian for quadratic fluctuations on $L=10$ sites in the $\cvec_0$ sector. The deformations are parametrised by $g$ and defined as $\mathcal{L}(g) \propto \sum_j (P_{j;j+1}-gQ_{j;j+1}-1)$ such that $g=1$ corresponds to the original Q-SSEP case. The chosen symmetry sector is defined by $k=2\pi/10$ and Cartan charges $(J_1^z,J_2^z,\frac{C_1-C_2}{2})=(5,3,3)$. The value for $N$ provides the number of eigenvalues left after removing the degeneracies.}
        \label{fig:c=0_L=10}
\end{figure}

\subsection{Numerics}
Here we investigate the integrability of the two-replica Lindbladian numerically in sectors $\cvec_0$ and $\cvec_1$ by considering the level-spacing statistics of its eigenvalues. Most studies that have been conducted on the level-spacing statistics of integrable spin chains deal with the case where the global symmetry algebra is $u(1)$ or $sl(2)$. In contrast to this, the symmetry algebra of our spin chain model is of higher rank, namely $sl(4)$. This leads to a new source of degeneracies of eigenvalues, which makes the level statistics differ from the usual $sl(2)$ case at first sight. To our knowledge, this is the first time this problem has been addressed in the literature.

Our analysis of the level-spacing statistics suggests that the $\cvec_{\pm 1}$ sectors are indeed integrable, while for the $\cvec_0$ sector it provides weak signs of integrability breaking, which sometimes lack an overall consistency. A key problem in the analysis is that the effects of integrability breaking perturbations become visible only for sufficiently large system sizes \cite{szaszschagrin2021weak, Modak2014Universal}. We suspect that the maximal system size we could achieve in practice is simply too small to get a completely consistent picture. Moreover, the presence of a higher rank symmetry algebra may further obscure signs of integrability breaking in the level-spacing statistics.

Let us start by recalling some known results on the level-spacing statistics of integrable and non-integrable models. Integrable Hamiltonians possess the very particular property that their eigenvalues are i.i.d. random variables, as if the Hamiltonian was just a random diagonal matrix. This was first conjectured by Berry and Tabor \cite{berry1977level} and has been confirmed in many explicit examples such as the XXX Heisenberg chain \cite{Poilblanc1993Poisson,Kudo2005Level}. Importantly, the spacing $s_n=e_n-e_{n-1}$ between adjacent eigenvalues follows an exponential distribution
\begin{equation}\label{eq:poisson}
p(s)=e^{-s}.
\end{equation} 
To be precise, this holds only for the so-called "unfolded spectrum" of the Hamiltonian, where one performs a local change of variable on the eigenvalues $e_n$ such that the density of the new variables $\hat e_n$ is uniform (see Appendix \ref{app:unfolding}). Instead, as showed in \cite{atas2013distribution}, one can also consider the ratio of consecutive spacings $r_n=s_{n}/s_{n-1}$ whose distribution is independent of the local density of eigenvalues and is given by
\begin{equation}\label{eq:ratio}
p(r)=\frac{1}{1+r^2}\ .
\end{equation}

In contrast to integrable Hamiltonians the eigenvalues of a generic Hamiltonian - a random Matrix - tend to repel each other. The spacing between eigenvalues of a $2\times2$ random matrix in the GOE (Gaussian Orthogonal Ensemble) -- which would be the appropriate ensemble to deal with since the Q-SSEP Linbladian is symmetric and real -- has a probability distribution know as \textit{Wigners surmise}
\begin{equation}\label{wigner}
p(s)=\frac{\pi s}{2}e^{-\pi s^2/4}\ .
\end{equation}
This turns out to be good approximation also for the level-spacing of large GOE random matrices. In particular, there is a zero probability to find consecutive eigenvalues with spacing zero. The same is true if instead of the spacing one again considers the ratio of adjacent spacings $r$, which behaves as $p(r)\sim r^\beta$
for small $r$, where $\beta$ is the Dyson index of the matrix ensemble \cite{atas2013distribution}.

The ratio of adjacent spacings has the nice property, that the average of
\begin{equation}
\tilde r=\mathrm{min}(r,1/r)\in[0,1]
\end{equation}
over the given ensemble takes is a constant, and can therefore be used to classify the ensemble to which some numerical distribution might belongs. For Poisson random variables\footnote{A short derivation of this fact: Since the probability of consecutive spacings $s_1$ followed by $s_2$ is equivalent the inverse order, i.e. $P(s_1,s_2)=P(s_2,s_1)$, it follows that $r=s1/s2$ and $1/r=s_2/s_1$ have the same distributions. Therefore $R(\tilde r)=2p(r)\Theta(1-r)$, from which the average value $\langle \tilde r \rangle=2\ln2-1$ can be computed.}, $\langle \tilde r \rangle=2\ln2-1\approx0.3863$, while for the eigenvalues of a matrix in the GOE, $\langle \tilde r \rangle\approx0.5359$.

Before discussing the results, let us also comment on the reduction of the Lindbladian to its remaining symmetry sectors. The eigenvalues in each symmetry sector are statistically independent and therefore one should treat each sector independently. In practice, we bring the Lindbladian to block diagonal form with respect to all its mutually commuting symmetries $I_i$, $[\mathcal{L},I_i]=0$, $[I_i,I_j]=0$. After fixing to $\cvec_{\pm1}$ or $\cvec_0$, the maximally set of commuting symmetries are translation $T$, the three Cartan elements $(J_1^z=\sum_j J_{1,j}^z,J_2^z,\frac{C_1-C_2}{2})$ of $gl(4)$ (which are the analogues of the magnetization for $sl(2)$) and depending on the choice of the three Cartan elements, a permutation $F$ of the states (which is the generalization of a spin flip in the $m=0$ sector for $sl(2)$). However, once all these charges have been fixed, there is still a degeneracy in the eigenvalues of the Lindbladian left, which would lead to an artificial large peak at zero in the level-spacing statistic. This is because the Cartan subalgebra for $gl(4)$ consists of more than one element, hence there are more than one "lowering-operator" and therefore weight-spaces in an irreducible $gl(4)$ representation can be more than one-dimensional. But the $gl(4)$ symmetry of the Lindbladian ensures that all states in an irreducible $gl(4)$ representation have the same eigenvalue and hence, selecting a weight-space (i.e. fixing the Cartan charges) will not lift all degeneracies (as it would do for the $sl(2)$ spin chain). We therefore manually deleted all the degenerate copies of eigenvalues from the complete set of eigenvalues in a given symmetry sector and analysed the level-spacing and ratio statistics for the remaining eigenvalues. The procedure is not entirely correct, because there can also be degeneracies in the spectrum solely due to integrability, which would be neglected in our procedure. But it turns out that for the overall statistics this only plays a minor role.

Fig.~\ref{fig:c=1} shows the results for the $\cvec_1$ sector and both the shape of distribution and the value for $\langle\tilde r \rangle$ suggest that this model is integrable. The results for the $\cvec_0$ sectors are less clear. In some of the sectors with fixed Cartan elements there is no sign of integrability breaking. In others, such as in Fig.~\ref{fig:c=0_L=10} there are weak signs. Increasing the perturbation of the known integrable model with the trace operator $Q$, parametrised by the coupling strength $g$, one observes a gradual deviation from the exponential distribution, which is correctly reproduced for the integrable case ($g=1/2$). However, a real Wigner-like distribution is not visible, even for higher values of $g$, and we suspect this to be due to the small system size we are able to achieve in practice. Also note, that the value for $\langle\tilde r\rangle$ seems to suggest integrability breaking for the Q-SSEP case. However, in our example, it does not consistently increase with $g$ as one would expect: Value for $g=2$ is lower than that for $g=1$. Again we think this is due to the limited system size. Finally, it might be interesting to look at the value of $N$, which describes the number of eigenvalues left after removing the degeneracies. For the known integrable case (a), $N=1205$ is lower than for the other two cases (b) and (c) where the number almost coincides, $N=1335$ and $N=1336$, respectively\footnote{The slight difference could arise due to the numerical imprecision.} This hints, that in the integrable case (a), there were more degenerate eigenvalues than the $gl(4)$ symmetry with its higher dimensional weight spaces could explain. We think that these additional degeneracies are probably a result of integrability - or reversely, their absence (as in (b) and (c)) is a sign of integrability breaking. To sum up, all these observations suggest that the quadratic fluctuations for Q-SSEP are not integrable in the $\cvec_0$ sector.

\section{Steady states and slow modes}
\label{sec:qssep-steady}
We now look at the steady states of the Q-SSEP. The Q-SSEP invariant measure has been described in \cite{Bauer2019Equilibrium} using a mapping to random matrix theory. We here adopt an alternative approach which yields a  better access to their structure in terms of the underlying $gl(2R)$ symmetry. The case of the Q-ASEP steady states will be discussed elsewhere \cite{BernardInPreparationFluctuations}.

\subsection{Q-SSEP steady states}

The steady states of the Q-SSEP dynamics are the zero modes of the Q-SSEP Lindbladian $\mathcal{L}$. Since $\mathcal{L}$ commutes the $gl(2R)$ action, these modes form $gl(2R)$ multiplets. Recall that the Q-SSEP Lindbladian is self-dual, $\mathcal{L}=\mathcal{L}^*$. 

We first study the zero modes in the $\cvec_0$ sector (i.e. with all charges $C_j$ along the chain set to zero). The identity belongs to this sector and is an eigenstate with eigenvalue zero since $\mathcal{L}^*\, \mathbb{I}=0$ by definition and $\mathcal{L}=\mathcal{L}^*$.  The identity is a factorised operator $\mathbb{I}=\mathbb{I}_1\mathbb{I}_2\cdots\mathbb{I}_L$ with $\mathbb{I}_j$ the identity on the $j$-th site. As a consequence of the $sl(2R)$ symmetry, all factorised operators $x_1 x_2\cdots x_L$ with $x_j$ localised on site $j$ and with $x$ in the $sl(2R)$ orbit of the identity are also zero modes. This orbit is isomorphic to the middle $sl(2R)$ fundamental representation $[\mu_R]$ and the $sl(2R)$ action on this representation is faithful. As a consequence, all factorised operators of the form
\beq
 |\!|\Omega^0_x\rangle := x_1 x_2\cdots x_L,\quad x\in[\mu_R] ,
\eeq
are zero eigenstates of the Q-SSEP Lindbladian in the $\cvec_0$ sector: $\mathcal{L}|\!|\Omega_x\rangle=0$. Of course, linear combinations of such states are also steady states. They form an irreducible representation of $sl(2R)$ with highest weight $L\mu_R$ (with a slight abuse of notation consisting in denoting by $\mu_k$ the highest weight of the representation $[\mu_k]$). Indeed, the representation formed by the above operators $|\!|\Omega^0_x\rangle$, $x\in[\mu_R]$, is irreducible with the highest weight vector $|\!|\Omega^0_{\mu_R}\rangle$ equal to the tensor product of $L$ identical highest weight vectors, each of weight $\mu_R$. The Young tableau of the representation of highest weight $L\mu_R$ is a rectangular tableau with $R$ lines and $L$ columns. Its dimension can be computed using the so-called hook rule (see below for examples).

As detailed in the Appendix \ref{Appendix:zero-mode}, this remark holds true in any sector. If $|\mu_{R+k}\rangle$ is the highest weight vector of the representation $[\mu_{R+k}]$ corresponding to local operator with $C$ charge equals to  $k$, then $|\mu_{R+k}\rangle\otimes\cdots\otimes|\mu_{R+k}\rangle$ is a zero mode of the Q-SSEP Lindbladian. By $sl(2R)$ invariance, this implies that all states,
\beq \label{eq:Omega-x}
 |\!|\Omega^k_x\rangle := x_1 x_2\cdots x_L,\quad x\in[\mu_{R+k}] ,
 \eeq
are zero eigenstates of the Q-SSEP Lindbladian in the $\cvec_k$ sector. Linear combinations of such states are also zero modes. They form the $sl(2R)$ representation whose Young tableau is rectangular with $R+k$ lines and $L$ columns. 

Let us give examples for one or two replicas.
For $R=1$ replica, the structure of the zero modes as $gl(2)$ modules is : $\cvec_0\leadsto [L+1]$ (spin $L/2$) and $\cvec_{\pm 1}\leadsto [1]$ (spin $0$).

For $R=2$ replicas,  the structure of the zero modes as $gl(4)$ modules is : $\cvec_0\leadsto [L\mu_2]$ (Young tableau with $2$ lines of length $L$), $\cvec_{\pm 1}\leadsto [L\mu_1]$ (Young tableau with $1$ line of length $L$) and $\cvec_{\pm 2} \leadsto [L\mu_0]=[1]$ the trivial representation. The number of zero modes are:
\[ \cvec_0 \leadsto \frac{(L+1)(L+2)^2(L+3)}{12},\ \cvec_{\pm 1} \leadsto \frac{(L+1)(L+2)(L+3)}{6},\ \cvec_{\pm 2} \leadsto 1 .\]

We have checked that these zero modes form all possible Q-SSEP steady states for $L=1$ and $L=2$, and we conjecture that this is true for all $L$. It is known from \cite{Bauer2019Equilibrium} that the Q-SSEP steady distribution is $SL(L)$ invariant, so that there is a peculiar (nice) interplay between the $SL(L)$ and $GL(2R)$ perspectives on the zero modes. See Appendix \ref{Appendix:interplay} for details.

All Q-SSEP steady operators have homogeneous $C$-charge along the chain. In contrast, a domain wall operator connecting sectors with different $C$ charges $c_1$ and $c_2$, say of the form $|\mu_{R+c_1}\rangle^{\otimes^M}\otimes|\mu_{R+c_2}\rangle^{\otimes^{(L-M)}}$ and their $gl(2R)$ orbits, are eigenstates of the Q-SSEP Lindbladian with eigenvalues $-\frac{J}{2}|c_1-c_2|$. Hence, domain walls decay exponentially fast in time, cf. \cite{Essler2020Integrability}. More generally, for an inhomogeneous sector with domain walls at a series of edges $w_1,\cdots,w_d$ connecting sectors with charges $c^-_\alpha,\ c^+_\alpha$ ($c^-_\alpha\not=c^+_\alpha$), $\alpha=1,\cdots,d$, the gap is the sum of the gaps associated to each of these domain walls, i.e. $-\frac{J}{2}\sum_\alpha |c^-_\alpha- c^+_\alpha|$. A proof of this statement is given in Appendix \ref{Appendix:zero-mode}.

\subsection{Q-SSEP low lying magnon states}

Let us first consider the one replica case $(R=1)$. In the $\cvec_0$ sector the zero modes form a spin $L/2$ representation of $sl(2R)$. The other zero modes are $sl(2R)$ singlets. With respect to the $sl(2R)$ basis, $J^\pm=\sum_jJ^\pm_j$, $J^z=\sum_j J^z_j$, the highest weight vector of this spin $L/2$ representation is the operator $|\!|\Omega^1_n\rangle:=n_1n_2\cdots n_L$ with $n_j$ the number operator at site $j$. Since the one replica Linbladian has been identified with the XXX Heisenberg spin chain Hamiltonian, one may construct magnon excitations by reversing the polarization of this fully polarized state in a coherent way. Algebraically, this is done by acting with the super-operator $\sum_j e^{ipj}J_j^-$ depending on a momentum $p$. It yields the eigenstates
\beq
 \sum_j e^{ipj}\ n_1n_2\cdots (J_j^-n_j)\cdots n_L~,
 \eeq
with $J_j^-n_j=\mathbb{I}_j-n_j$. Their eigenvalues are $\omega_p:=J\,(\cos p-1)$. At small momenta, the dispersion relation is diffusive, $\omega_p\simeq - J\,p^2$, as expected. 

This observation generalizes to any number of replicas.
Since the zero modes $|\!|\Omega^k_x\rangle= x_1 x_2\cdots x_L$ with $x\in[\mu_{R+k}]$, given in eqn \eqref{eq:Omega-x} with $R$ the number of replicas and $k$ the $C$ charge, are also ferromagnetic like states, one can construct one-magnon eigenstates. The latter are obtained by considering the fully polarized state $|\!\ket{\Omega_{\mu_{R+k}}}$ with $x$ chosen to be the highest weight vector $\mu_{R+k}$ of the $sl(2R)$ representation $[\mu_{R+k}]$ and by acting on it with the appropriate $gl(2R)$ generator. Namely, one considers the operators
\beq \label{eq:1magnon-R}
 |\!|p;\gamma;\mu_{R+k}\rangle := \sum_j e^{ipj}\ G^{-\gamma}_j |\!|\Omega_{\mu_{R+k}}\rangle,
 \eeq
with $\gamma$ a positive root of $sl(2R)$ and $G^{-\gamma}$ the associated root generator. As shown in Appendix \ref{Appendix:zero-mode}, these states are eigenstates of the $R$ replica Lindbladian iff $\gamma$ is one of the simple roots of $sl(2R)$. There are $2R-1$ such simple roots but only one of them (dual to the weight $\mu_{R+k}$) yields a non vanishing state so that the labelling $\gamma$ is actually redundant since it is linked to $\mu_{R+k}$. The dispersion relation is also diffusive: 
\be
\omega_p = 2J\,(\cos p-1) \simeq - J\,p^2\ .
\label{magnondisp}
\ee
By $sl(2R)$ invariance, any state obtained by acting with $SL(2R)$ on $|\!|p;\gamma;\mu_{R+k}\rangle$ is an eigenstate. That is: all states $\otimes_j g_j |\!|p;\alpha;\mu_{R+k}\rangle$, with $g\in SL(2R)$ and $\alpha$ a simple root, are eigenstates. We have $\otimes_j g_j |\!|p;\alpha;\mu_n\rangle=|\!|p;g\alpha g^{-1};x\rangle$ with $x=g\,\mu_{R+k}$, they form an irreducible $sl(2R)$ representation isomorphic to $[\mu_{R+k}]$. 

\subsection{Cat state preparation in the non trivial integrable sector}

In order to demystify the meaning of different sectors and demonstrate how integrability could be used for addressing interesting physical questions, we discuss the dynamics of the cat states, which partially belong to the integrable $\cvec_{\pm 1}$ subsectors for two replicas. 

Imagine initially preparing the system in the pure state $\ket{\psi}$. Its initial $R=2$ replicated density matrix is $\rho^{(2)}_0=\rho_0\otimes\rho_0$, with $\rho_0= \ket{\psi}\bra{\psi}$. In order for $\rho^{(2)}$ to have a non zero component on the $\cvec_{\pm 1}$ sectors, one has to consider $\ket{\psi}$ to be a cat state, i.e. a particular linear combination of states with macroscopically different numbers of particles. The simplest state $\ket{\psi}$ to consider would be the sum $\ket{\psi_0} + \ket{\psi_M}$ with $\ket{\psi_0}$ (resp. $\ket{\psi_M}$) a state with $0$ (resp. $M$) particles. However, this is actually not enough for producing quadratic fluctuations in the $\cvec_{\pm 1}$ sector. One has to consider {\it three position cat states} $\ket{\psi}$ which are the sum of three states with different particle numbers of the form
\beq
\ket{\psi}= \ket{\psi_0} + \ket{\psi_M} + \ket{\psi_{L-M}}
\eeq
with $\ket{\psi_0}$, $\ket{\psi_M}$ and $\ket{\psi_{L-M}}$ having $0$, $M$ and $L-M$ particles respectively. We assume $0<M<L/2$. (We could also consider more general cat states but this is the simplest with the required properties).

When considering quadratic fluctuations, we have to consider the initial tensor product $\ket{\psi}\otimes\ket{\psi}$. To ensure that the quadratic fluctuation dynamics survives the large time limit, we have to decipher on which $SU(L)$ representations this product is decomposed onto in order to verify that it has a non trivial component on $\cvec_{\pm1}$ steady states. See Appendix \ref{Appendix:interplay} for details.

The state $\ket{\psi_0}$ belongs to the trivial $SU(L)$ representation $\Lambda_0$, while the states $\ket{\psi_M}$ (resp. $\ket{\psi_{L-M}}$) belong the fundamental representation $\Lambda_M$ (resp. $\Lambda_{L-M}$), with Young tableau with one column and $M$ (resp. $L-M$) boxes. The tensor product $\ket{\psi}\otimes\ket{\psi}$ thus contains $\ket{\psi_0}\otimes\ket{\psi_0}$ which belongs to the scalar representation $\Lambda_0\otimes\Lambda_0\equiv \Lambda_0$. It also contains the products $\ket{\psi_M}\otimes\ket{\psi_{L-M}}$ (resp. $\ket{\psi_{L-M}}\otimes\ket{\psi_M}$) which are elements of $\Lambda_M\otimes\Lambda_{L-M}$ (resp. $\Lambda_{L-M}\otimes\Lambda_{M}$). The decomposition of the latter tensor product of $SU(L)$ representations contains the fully antisymmetric representation $\Lambda_0^+$ which, as $SU(L)$ module, is isomorphic to $\Lambda^0$. However, $\Lambda_0^+$ and $\Lambda_0$ are isomorphic  as $SU(L)$ modules but not as $U(L)$ modules since they differ by their $C$ charges. 

As a consequence, the initial quadratic density matrix $\rho^{(2)}:=\ket{\psi}\bra{\psi}\otimes\ket{\psi}\bra{\psi}$ contains blocks intertwining $\Lambda_0\to \Lambda_0^+$ and $\Lambda_0^+\to\Lambda_0$. These blocks are in the $\cvec_{\pm 1}$ sectors. They are sub-blocks in the matrices $\ket{\psi_0}\bra{\psi_M}\otimes\ket{\psi_0}\bra{\psi_{L-M}}$ or $\ket{\psi_0}\bra{\psi_{L-M}}\otimes\ket{\psi_0}\bra{\psi_{M}}$ and their Hermitian conjugates.

To check that they have a non trivial projection on the $\cvec_{\pm 1}$ steady states, let us compute their overlaps with the $\cvec_{\pm 1}$ steady operator (that is: operators in the kernel of the dual Lindbladian in the sector $\cvec_{\pm 1}$). These operators, say with $\cvec_{-1}$, have exactly one fermionic annihilation operator per site in either of the two replicas. The simplest such operator is of the form
\[ \mathfrak{C}:=\prod_{m\in \mathfrak{M}}c_{1;m}\cdot \prod_{n\in \mathfrak{M}^c} c_{2;n} ~,\]
with $\mathfrak{M}$ a sub-set of $\{1,\cdots,L\}$ and $\mathfrak{M}^c$ its complement. Any other operator in the $\cvec_{\pm 1}$ sector is obtained by acting on the latter with $\cvec_0$ operators (which amounts to multiply $\mathfrak{C}$ by the number operators $n_{a;j}$ at different sites or replicas). Let us choose $\mathrm{Card}\,\mathfrak{M}=M$ and hence $\mathrm{Card}\,\mathfrak{M}=L-M$. Then,
\[ 
\Tr\big( \rho^{(2)}\, \mathfrak{C}) = \bra{\psi_0}\prod_{m\in \mathfrak{M}}c_{m}\ket{\psi_M}\bra{\psi_0}\prod_{n\in \mathfrak{M}^c} c_{n}\ket{\psi_{L-M}} \not= 0~.
\]
This is non zero if the state $\ket{\psi_M}$ (resp. $\ket{\psi_{L-M}}$) has particles inserted at the site selected by $\mathfrak{M}$ (resp. $\mathfrak{M}^c$). The same holds true if we multiply $\mathfrak{C}$ by an operator in the $\cvec_0$ sector.

Hence, the three position cat states, $\ket{\psi}= \ket{\psi_0} + \ket{\psi_M} + \ket{\psi_{L-M}}$, provide examples of physical states such that the dynamics of their quadratic fluctuations has a non trivial component on the integrable sector $\cvec_{\pm 1}$. Of course, part of the dynamics of those fluctuations is also in the non integable sector $\cvec_0$.

\section{Q-SSEP dynamics on the lattice and in the scaling limit}
\label{sec:qssep-scaling}
In this section we consider the dynamics of (quadratic) fluctuations in the Q-SSEP in some detail. We show that the dynamics is essentially diffusive, but at sufficiently late time and large distances the deviations from diffusive behaviour are captured by a continuum description. More precisely, we show that there is a scaling regime in which correlation functions are described by hierarchies of partial differential equations that take the form of diffusion equations with source terms that couple the different levels of the hierarchy. By numerically solving the equations of motion for some examples we show that the scaling limit gives an accurate description of an intermediate time regime of the lattice dynamics.

\subsection{Two point functions for two replicas}
\label{sec:qssep-2points}

In this section we investigate the equations of motion for averages involving two replicas, i.e. quantities of the form $\mathbb{E}[{\cal O}_1\otimes{\cal O}_2]_t$.
We first show that in the Q-SSEP the mean expectations evolve diffusively, \emph{cf.} eqn \eqref{eq:one-poin-diff}, while the quadratic fluctuations satisfy a set of linear equations encoding diffusion plus interactions, \emph{cf.} eqn \eqref{nndot}.

Averaging the Heisenberg equations of motion for system plus noise
over the latter gives
\begin{align}
\mathbb{E}\left[{\cal O}_1\otimes{\cal O}_2\right]_{t+\Delta t}&=
\mathbb{E}\left[{\cal O}_1\otimes{\cal O}_2\right]_t
-\frac{1}{2}\mathbb{E}\Big[[dH^{(2)}_t,[dH^{(2)}_t,{\cal O}_1\otimes{\cal O}_2]]\Big]_t
+\dots\ ,
\label{eomgeneral}
\end{align}
where the two-replica Hamiltonian increment reads
\be
dH^{(2)}_t =\sum_{j=1}^L\sum_{a=1}^2
c^\dagger_{a,j+1}c_{a,j}dW_t^j
+c^\dagger_{a,j}c_{a,j+1}d\overline{W}_t^j\ .
\ee
Here $dW_t^j$, $d\overline{W}_t^j$ are complex Brownian noises,
\emph{cf.} eqn \eqref{eq:def-dW}, and $c_{a,j}$ and $c^\dagger_{a,j}$ are the canonical fermion annihilation and creation operators defined in \fr{CACR} (we recall that the fermion operators of different species commute). As we will see, for the Q-SSEP the operators of interest are 
\be
n_{a,j}=c^\dagger_{a,j}c_{a,j}\ ,\quad
S_j^+=c^\dagger_{1,j}c_{2,j}=S_j^{-\,\dagger},\quad
S_j^-=c_{1,j}c^\dagger_{2,j}=(S_j^+)^\dagger \ .
\label{nspm}
\ee
Working out the relevant commutators and noise-averages in
\fr{eomgeneral} we obtain
\begin{align} \label{eq:one-poin-diff}
\frac{d}{dt}\mathbb{E}[n_{a,j}]_t&=J\Delta_{j,k}\mathbb{E}[n_{a,k}]_t\ ,\nn
\frac{d}{dt}\mathbb{E}[S^\pm_{j}]_t&=J\Delta_{j,k}\mathbb{E}[S^\pm_{k}]_t\ ,
\end{align}
where $\Delta_{j,k}=\delta_{j,k+1}+\delta_{j,k-1}-2\delta_{j,k}$ is
the lattice Laplacian and repeated indices are summed over. This establishes that one-point functions
exhibit purely diffusive dynamics. Assuming that the initial density matrix is invariant under the exchange of the two replicas, the average fermion density must be independent of the replica index $a$. For later convenience we define it as 
\be \label{eq:def=rho}
\rho_j(t)={\rm Tr}\left[\rho^{(2)}(0)\mathbb{E}[n_{a,j}]_t\right].
\ee
The fermion density fulfils a simple diffusion equation on the lattice 
\be
\frac{d}{dt}\rho_j(t)=J\Delta_{j,k}\rho_k(t)\ .
\label{ndot}
\ee 

The analogous calculation for
operators involving two operators $n_{a,j}$, $S^\pm_j$ gives
\begin{align}
\frac{d}{dt}\mathbb{E}[n_{1,j}n_{2,k}]_t&=J\big(\Delta_{j,n}+\Delta_{k,m}\big)
\mathbb{E}[n_{1,n}n_{2,m}]_t
-J\delta_{j,k}\big(\mathbb{E}[S^+_{j+1}S^-_j]_t+\mathbb{E}[S^+_{j-1}S^-_j]_t+{\rm h.c.}\big)\nn
&+J\big(\delta_{j,k-1}+\delta_{j,k+1}\big)\big(\mathbb{E}[S^-_jS^+_k]_t+\mathbb{E}[S^+_jS^-_k]_t\big)
\label{nndot}\\
\frac{d}{dt}\mathbb{E}[S^-_{j}S^+_{k}]_t&=
J\big(\Delta_{j,n}+\Delta_{k,m}\big)\mathbb{E}[S^-_{n}S^+_{m}]_t\nn
&+J\delta_{j,k-1}\ \mathbb{E}\big[\bar{n}_{1,j}n_{2,j+1}+n_{2,j}\bar{n}_{1,j+1}\big]_t
+J\delta_{j,k+1}\ \mathbb{E}[\bar{n}_{1,j}n_{2,j-1}+n_{2,j}\bar{n}_{1,j-1}\big]_t\nn
&-J\delta_{j,k}\ \mathbb{E}[\bar{n}_{1,j}(n_{2,j-1}+n_{2,j+1})+n_{2,j}(\bar{n}_{1,j-1}+\bar{n}_{1,j+1})]_t\ ,
\end{align}
where we have defined $\bar{n}_{a,j}=n_{a,j}-1$.

We now take the trace with an initial density matrix that is invariant under
exchange of the replicas and define the following averages of the (replicated)
system degrees of freedom
\begin{align}
g_+(j,k;t)&:={\rm Tr}\left[\rho^{(2)}(0)\mathbb{E}[n_{1,j}n_{2,k}]_t\right]\ ,\nn
g_-(j,k;t)&:=\begin{cases}
-{\rm Tr}\left[\rho^{(2)}(0)\mathbb{E}[S^-_{j}S^+_{k}]_t\right] & \text{ if }j\neq k\ ,\\
{\rm Tr}\left[\rho^{(2)}(0)\mathbb{E}[n_{1,j}n_{2,j}]_t\right] & \text{ if }j= k\ .
\end{cases}
\end{align}
Alternatively, we can express these as
\be 
g_+(j,k;t)=\mathbb{E}[G_{jj}G_{kk}]\ ,\qquad
g_-(j,k;t)=\mathbb{E}[G_{jk}G_{kj}]\ ,\qquad
G_{ij}=\Tr(\rho_t c^\dag_i c_j)\ .
\ee 
This shows that $g_+(j,k;t)$ and $g_-(j,k;t)$) encode respectively density and coherence correlation fluctuations.

Treating the case $j=k$ carefully, we obtain
\begin{align}
\frac{d}{dt}g_\sigma(j,k;t)&=J\big(\Delta_{j,n}g_\sigma(n,k;t)
+\Delta_{k,m}g_\sigma(j,m;t)\big)
+2J\delta_{j,k}\big[g_{-\sigma}(j,j+1;t)+g_{-\sigma}(j-1,j;t)\big]\nn
&-2J\big(\delta_{j,k-1}+\delta_{j,k+1}\big)g_{-\sigma}(j,k;t).
\label{EoMfinal}
\end{align}

In the following we will also consider the connected correlation
functions, 
\begin{align}
g_{+,c}(j,k;t)&\equiv g_+(j,k;t)-\rho_j(t)\rho_k(t)\ ,\nn
g_{-,c}(j,k;t)&\equiv g_-(j,k;t)-\delta_{j,k}\big(\rho_j(t)\big)^2\ ,
\end{align}
where $\rho_j(t)$ is the average fermion density in replica $a$ defined above in eqn \eqref{eq:def=rho} (it
is independent of $a$ as a result of our choice of initial conditions).

It is easy to see that \fr{EoMfinal} have time-independent sum rules
\be
A_\sigma=\sum_{j,k=1}^Lg_\sigma(j,k;t)\ ,\quad \frac{dA_\sigma}{dt}=0.
\label{sumrules}
\ee
These conservation laws are consequences of the fact that the dynamics is unitary for any given realization of the noise, and hence it preserves the spectrum of the system density matrix.
For the initial states we consider here the constants $A_\sigma$ have
regular expansions in inverse powers of the system size
\be
A_\sigma=a^{(0)}_\sigma+\frac{a^{(1)}_\sigma}{L}+\frac{a^{(2)}_\sigma}{L^2}+\dots
\ee


The calculation of correlation functions can of course also be formulated in the Hilbert space doubling approach. For example, we have
\be
g_+(j,k;t)=\langle\mathds{1}_2|\!|n_{1,j}n_{2,k}|\!|\rho^{(2)}(t)\rangle\ ,
\ee
where $\langle\mathds{1}_2|$ has been defined in \fr{id2}. The equation of motion thus reads
\be
\frac{d}{dt}g_+(j,k;t)=
\langle\mathds{1}_2|\!|n_{1,j}n_{2,k}\hat{\cal L}_{2}|\!|\rho^{(2)}(t)\rangle\ .
\label{SEq}
\ee
This, and the analogous equation for $g_-(j,k;t)$, can be cast in the form of an imaginary time two-particle Schr\"odinger equation for a non-Hermitian "Hamiltonian" obtained from $\hat{\cal L}_2$ by exploiting the existence of the two $gl(2)$ sub-algebras. This is sketched in Appendix \ref{app:2particlesector}, and a discussion of the spectrum of the Lindbladian in the two-particle sector is given in Appendix \ref{Appendix:two-particle}.

\subsection{Numerical solution}
The coupled equations \fr{EoMfinal} can be straightforwardly solved
numerically. We take the initial two-replica density matrix to encode
no correlatons between the replicas
\be
\mathbb{E}[\rho^{(2)}(t=0)]=\rho^{(1)}\otimes\rho^{(1)}\ .
\ee
\subsubsection{CDW initial state}
\begin{figure}[ht]
\begin{center}
\includegraphics[width=0.45\textwidth]{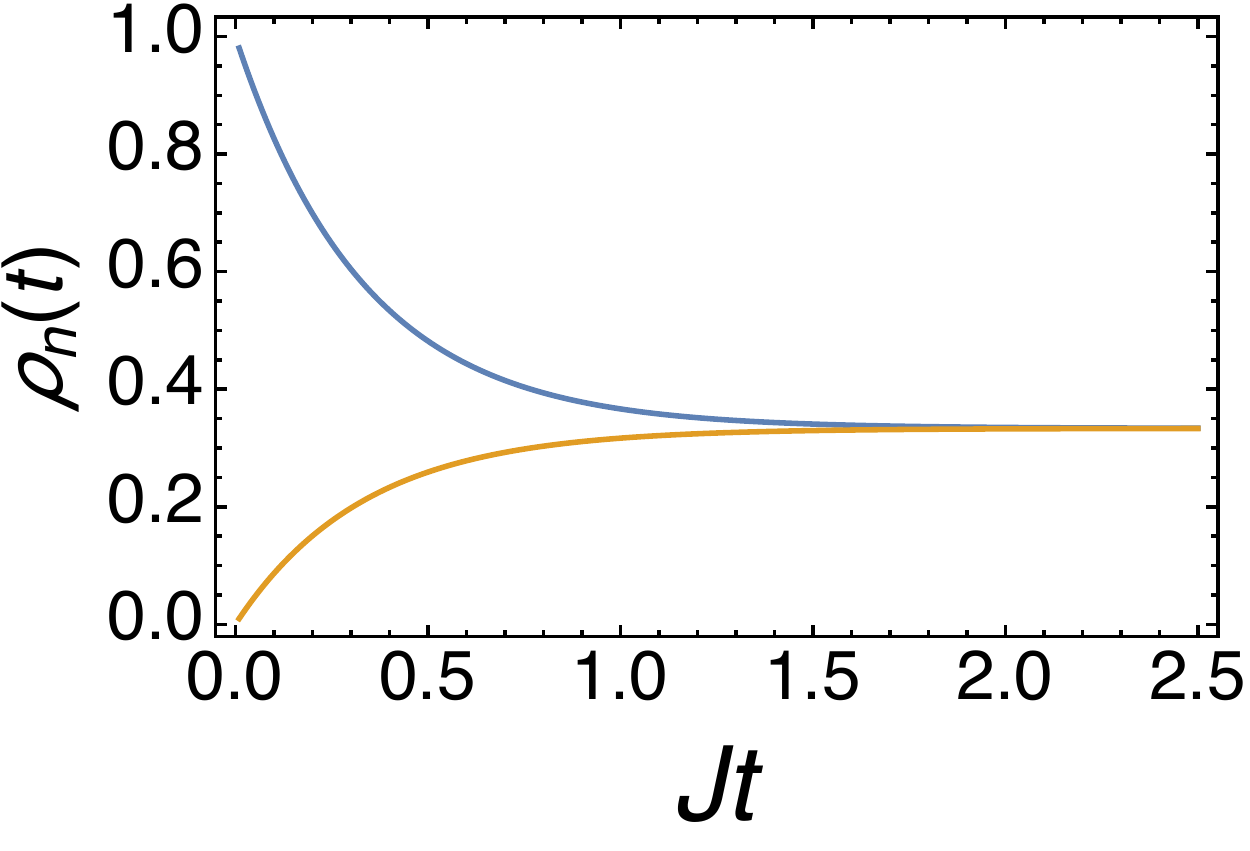}
\end{center}
\caption{Density $\rho_n(t)$ for the CDW initial state on the
sublattice $n=3k-2$ (blue) and the other two sublattices $n=3k-1$,
$n=3k$ (yellow) as functions of time. We observe that the density
relaxes quickly to its steady state value $1/3$.} 
\label{fig:density}
\end{figure}

\begin{figure}[ht]
\begin{center}
\includegraphics[width=0.42\textwidth]{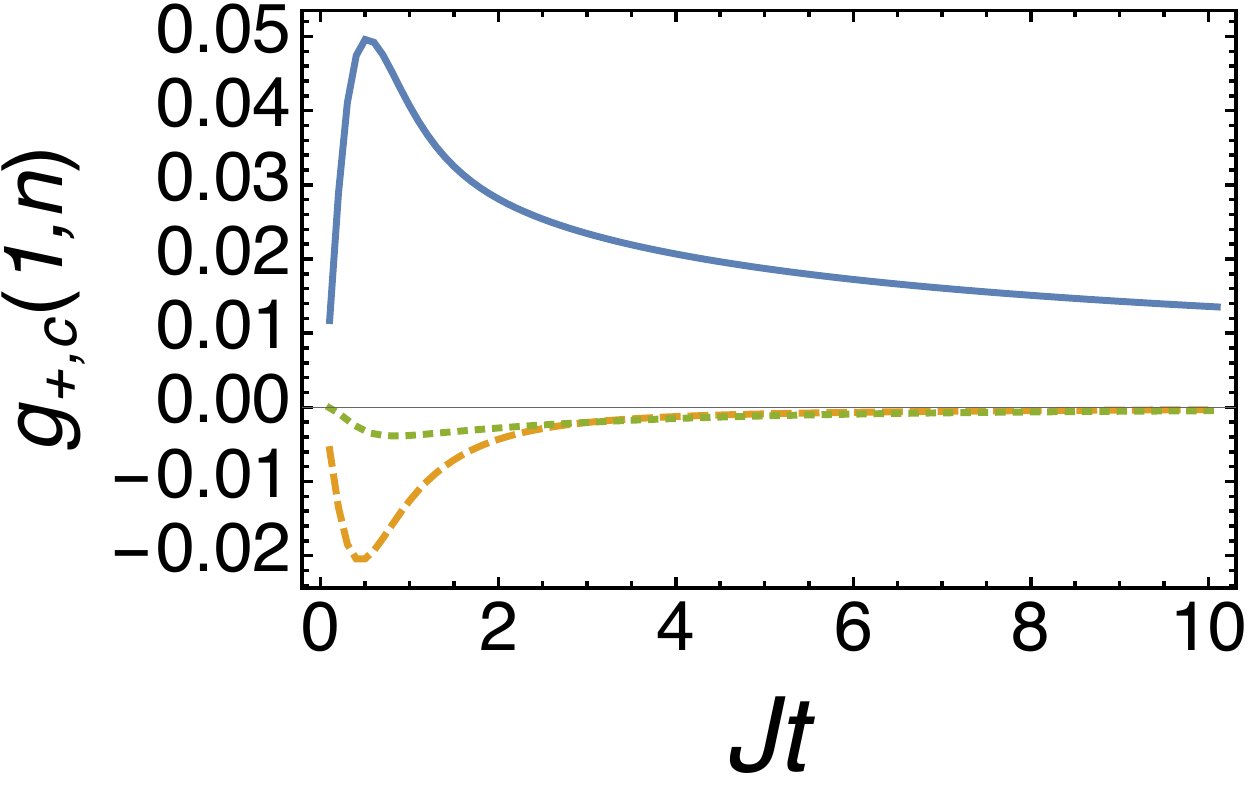}
\qquad
\includegraphics[width=0.42\textwidth]{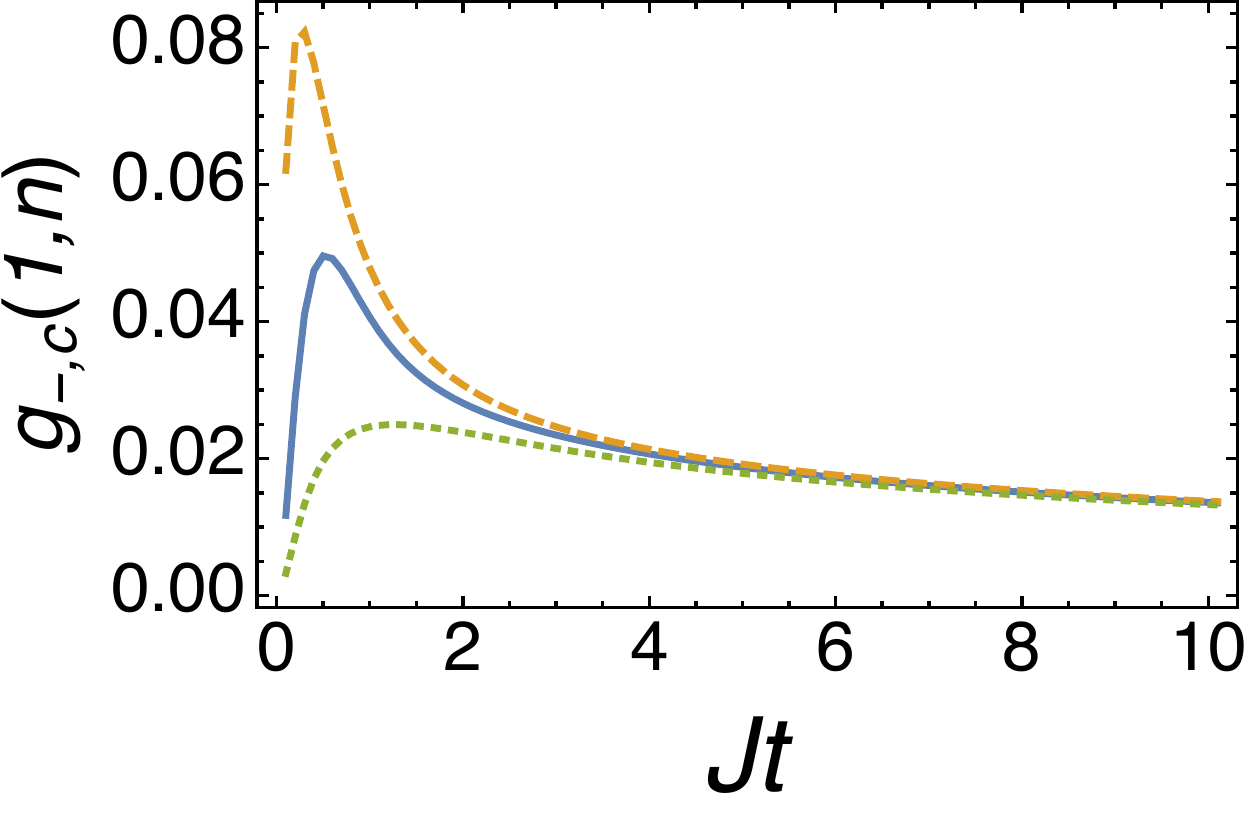}
\end{center}
\caption{Connected correlators $g_{\sigma,c}(1,n)$ as functions of
  time for $n=1$ (solid blue), $n=2$ (dashed yellow), $n=3$
  (dotted green) on a ring of $L=36$ sites. The correlations are very
  short-ranged.} 
\label{fig:nnc3D}
\end{figure}

As our first example we choose an initial product state that is
invariant under translations by three lattice sites
\be
\rho^{(1)}=|{\rm CDW}\rangle\langle{\rm CDW}|\ ,\quad
|{\rm CDW}\rangle=\prod_{j=1}^{L/3}c^\dagger_{3j-2}|0\rangle.
\ee
Here $L$ is taken to be a multiple of $3$ so that the charge-density
order is commensurate with the system size. This gives the initial
conditions  
\begin{align}
g_+(j,k,0)&=D_jD_k\ ,\quad D_j=\sum_{i=1}^{L/3}\delta_{j,3i-2}\ ,\nn
g_-(j,k,0)&=D_j^2\delta_{j,k}\ .
\end{align}
The fermion density in the steady state is $\rho_j(\infty)=1/3\equiv\rho$ and connected density and coherence fluctuations behave as
\begin{align}
g_{+,c}(j,k;\infty)&=\frac{\rho(1-\rho)}{L}\left[\delta_{j,k}-\frac{1}{L}\right]+{\cal
  O}(L^{-3}),\nn
g_{-,c}(j,k;\infty)&=\frac{\rho(1-\rho)}{L} \left[1-\frac{\delta_{j,k}}{L}\right]
+{\cal  O}(L^{-3}).
\label{SSCDW}
\end{align}
The results of numerical integration of the equations of motion is
shown in Figs~\ref{fig:density} and  \ref{fig:nnc3D}. The fermion
density is seen to relax quite quickly to its steady state value
$1/3$ as shown in Fig.~\ref{fig:density}.
The qualitative behaviour of the connected two-point functions is as
follows: 
\begin{itemize}
\item The connected correlations initially vanish, then build
up over a time scale $Jt\sim1$, and subsequently decay towards their steady state values \fr{SSCDW};
\item The decay of $g_+(j,j;t)$ towards its steady-state values is well-characterised by a power law $t^{-1/2}$ on times scales $Jt\ll L$;
\item The region in which connected correlations are non-negligible spreads diffusively, i.e. there is a parabolic ``envelope''.
\end{itemize}

\subsubsection{Domain wall initial state}
\label{sssec:DWS}
\begin{figure}[ht]
\begin{center}
(a)\includegraphics[width=0.5\textwidth]{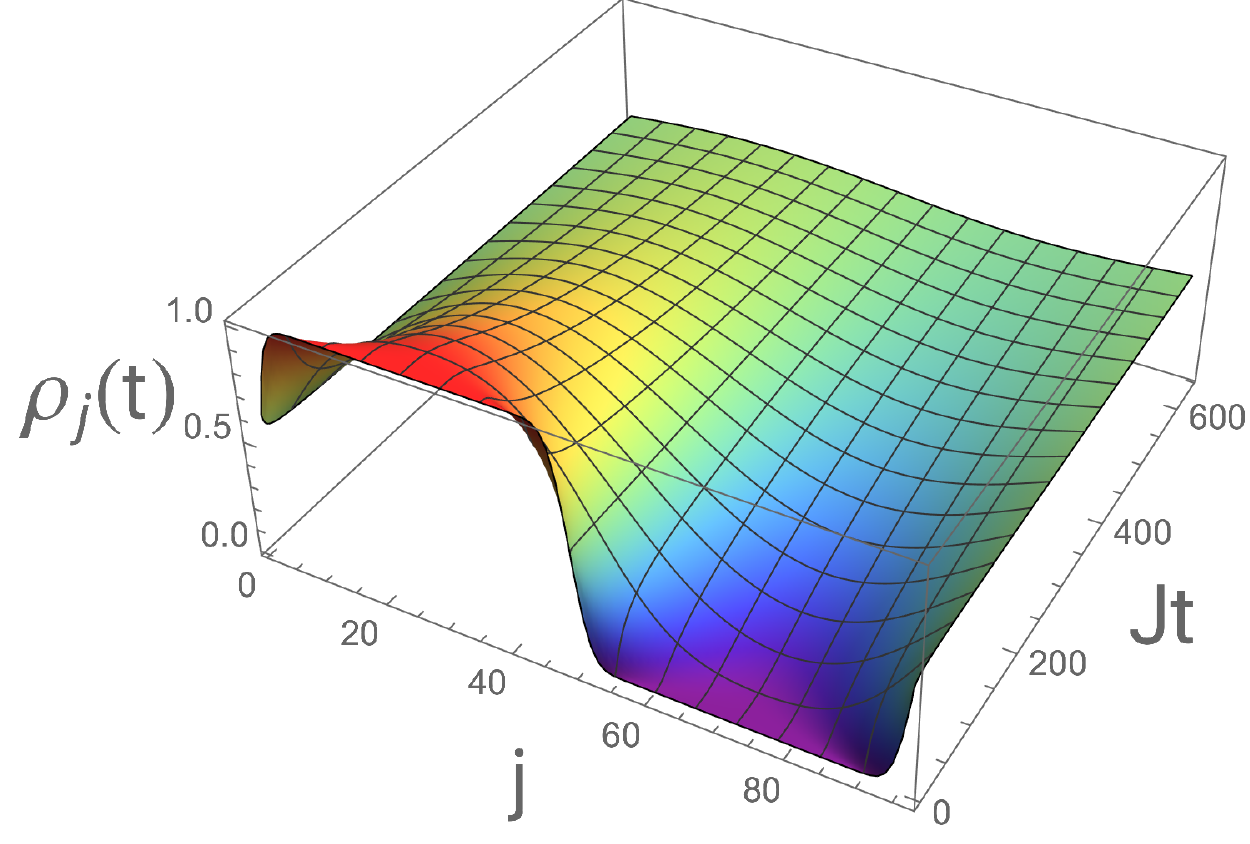}\qquad
(b)\includegraphics[width=0.38\textwidth]{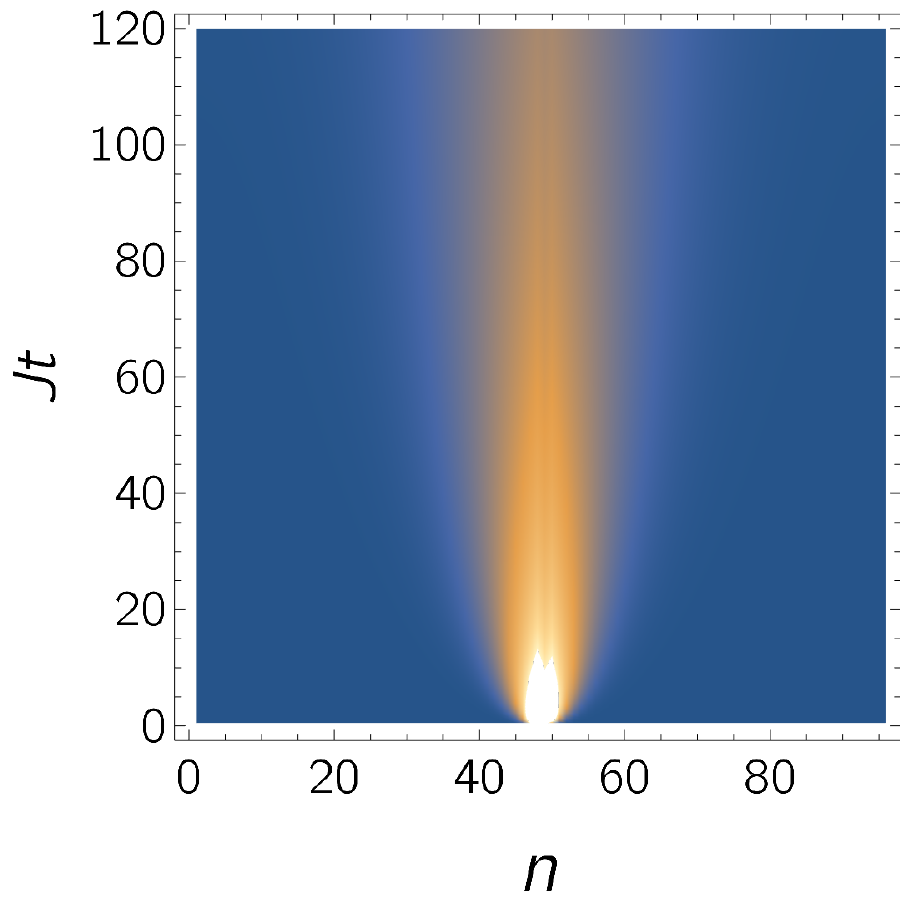}
\end{center}
\caption{(a) Density $\rho_j(t)$ for the uncorrelated domain-wall initial
  state \fr{DWS} as a function of $j$ and time on a ring of $L=96$
  sites. (b) Connected correlator $g_{-,c}(\frac{L}{2}+1,n,t)$ for the
same parameters.} 
\label{fig:rhoDW}
\end{figure}

The next example we consider is that of an uncorrelated domain wall state at half-filling, where all fermions are initially located on sites $[1,\frac{L}{2}]$, i.e.
\be
\rho^{(1)}=|{\rm DW}\rangle\langle{\rm DW}|\ ,\qquad
|{\rm DW}\rangle=\prod_{j=1}^{L/2}c^\dagger_j|0\rangle\ .
\label{DWS}
\ee
The initial conditions are thus
\begin{align}
  g_+(j,k,0)&=\begin{cases}
  1 & \text{ if } 1\leq j,k\leq \frac{L}{2}\ ,\\
  0 & \text{ else.}
  \end{cases}\nn
  g_-(j,k,0)&=\begin{cases}
  1 & \text{ if } 1\leq j=k\leq \frac{L}{2}\ .\\
  0 & \text{ else.}
  \end{cases}
\label{DWIC}
\end{align}
In Fig.~\ref{fig:rhoDW} we show the ensuing time evolution of the density and connected two-point function $g_{-,c}(\frac{L}{2}+1,n,t)$.
The initial state is unentangled, but connected correlations are seen to build up in time and are negligible outside a causal region that spreads diffusively in time, i.e. there is a parabolic "envelope".

\subsubsection{Correlated domain wall initial state}
\begin{figure}[ht]
\begin{center}
(a)\includegraphics[width=0.5\textwidth]{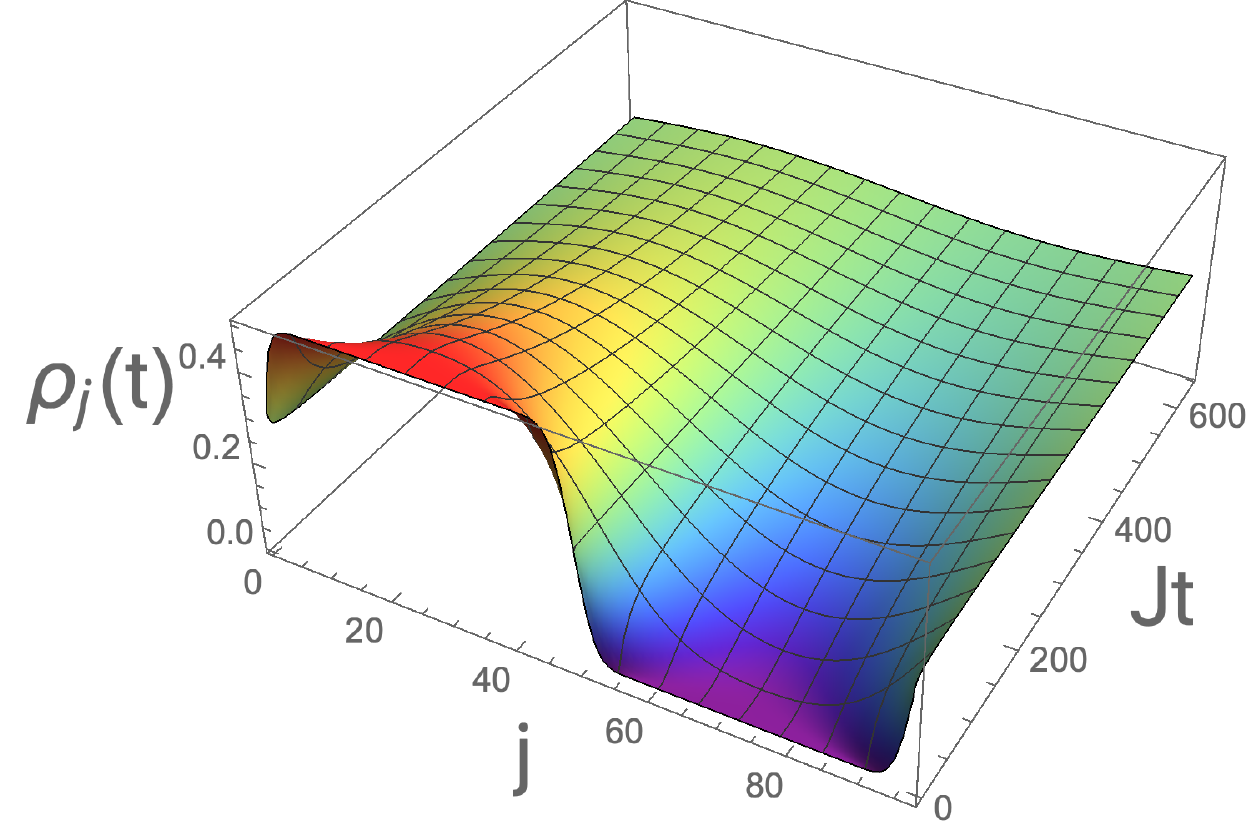}\qquad
(b)\includegraphics[width=0.38\textwidth]{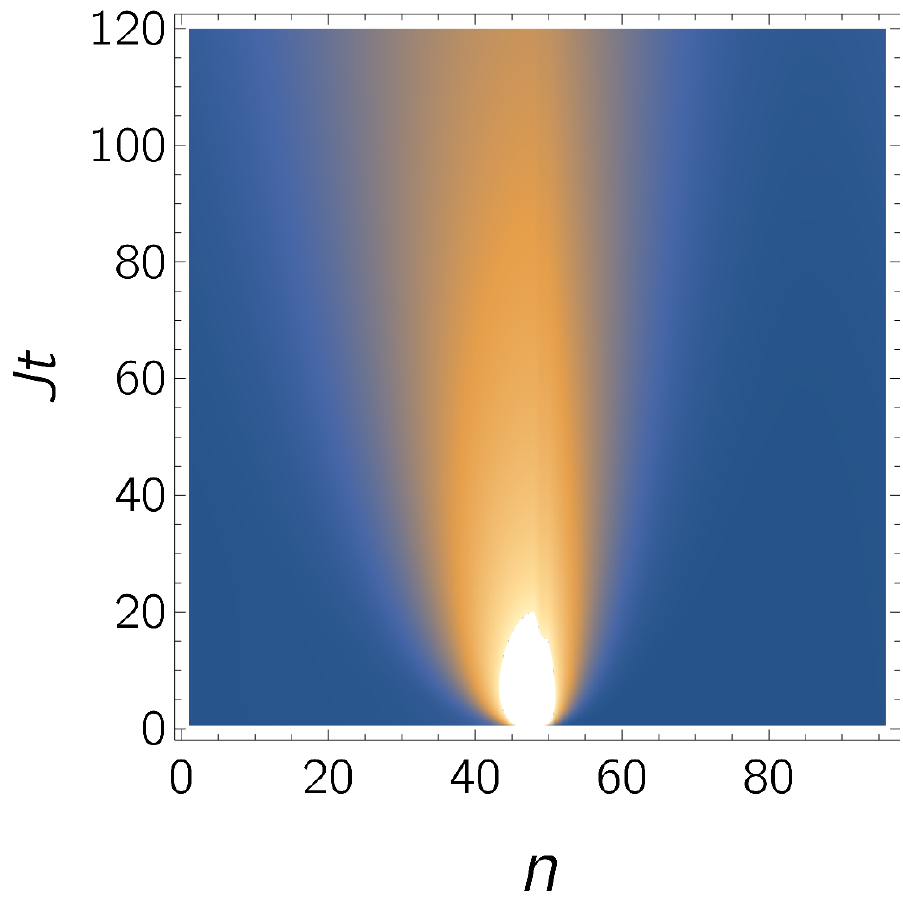}
\end{center}
\caption{(a) Density $\rho_j(t)$ for the correlated domain-wall initial
  state \fr{corrDW} as a function of $j$ and time on a ring of $L=96$
  sites. (b) Connected correlator $g_{-,c}(\frac{L}{2}+1,n,t)$ for the
same parameters.} 
\label{fig:rho2}
\end{figure}

In the two previous examples the initial states were unentangled. We
now consider the half-filled ground state of a tight-binding model
on half the system, i.e. sites $1\leq j\leq L/2$ (we assume $L$ to be even).
Defining
\begin{align}
  \tilde{c}(k_n)&=\sqrt{\frac{2}{L/2+1}}\sum_{j=1}^{L/2}c_j\ \sin(k_nj)\ ,\quad
k_n=\frac{\pi n}{\frac{L}{2}+1}\ ,\quad n=1,\dots,\frac{L}{2}\ ,
\end{align}
our initial state is
\be
|\Psi(0)\rangle=\prod_{n=1}^{L/4}\tilde{c}^\dagger(k_n)|0\rangle\ .
\label{corrDW}
\ee
The initial two-point functions are then easily calculated using
\be
c_j=\sqrt{\frac{2}{L/2+1}}\sum_{k_n}\tilde{c}(k_n)\ \sin(k_nj)\ .
\ee
Defining $2k_F=\frac{\pi}{2}$ we have for $j,k\in [1,L/2]$ 
\begin{align}
g_{j,k}&=\langle\Psi(0)|c^\dagger_j
c_k|\Psi(0)\rangle=f(j-k)-f(j+k)\ ,\quad j\neq k,\nn
f(n)&=\begin{cases}
\frac{\sin\big(2k_Fn\big)}{(L+2)\sin\big(\frac{\pi n}{L+2}\big)} &
\text{if } n\neq L+2\ ,\\
\frac{1}{2} & \text{if } n=L+2.
\end{cases}
\end{align}
The corresponding initial values for our one and two-point functions
are then
\begin{align}
\rho_j(0)&=\begin{cases}
\frac{1}{2} & \text{ if } 1\leq j\leq \frac{L}{2}\\
0 & \text{ if } \frac{L}{2}\leq j\leq L
\end{cases}\ ,\qquad
  g_+(j,k,0)=\rho_j(0)\rho_k(0)\ ,\nn
  g_-(j,k,0)&=\begin{cases}
  g_{j,k}g_{k,j} & \text{ if } 1\leq j\neq k\leq \frac{L}{2}\ ,\\
  \frac{1}{4} & \text{ if } 1\leq j=k\leq \frac{L}{2}\ .\\
  0 & \text{ else.}
  \end{cases}
\end{align}
The stationary behaviour at late times can again be determined
explicitly
\begin{align}
\lim_{t\to\infty}g_+(j,k,t)&=\frac{1}{16}\left[1-\frac{3}{L^2}+\frac{3}{L}\delta_{j,k}\right],\nn
\lim_{t\to\infty}g_-(j,k,t)&=\frac{1}{16}\left[\frac{3}{L}+\delta_{j,k}(1-\frac{3}{L^2})\right].
\end{align}
In Fig.~\ref{fig:rho2} we show the dynamics of the average density
$\rho_j(t)$ as a function of time.

The domain wall is seen to spread diffusively in a way that is qualitatively quite similar to the case of an uncorrelated domain wall initial state considered above. The connected correlations are non-vanishing in the initial state and as a result the region in which they are sizeable spreads in an asymmetric way.
\subsection{Scaling limit}
The aim of this section is to show that there exists a scaling limit in which a continuum description applies and to derive a set of partial differential equations describing the dynamics of quadratic fluctuations in this limit. We then establish the regime of applicability of results obtained in the scaling limit for the domain wall initial conditions considered above by comparing numerical result for the continuum and lattice descriptions. The scaling limit is defined as
\be
a_0\to 0,\quad J\to\infty\ ,\quad \mathfrak{D}=Ja_0^2\text{ fixed}\ ,
\label{SL}
\ee
where $a_0$ is the lattice spacing. The limit \fr{SL} has proved to
provide very useful insights into the stationary behaviour \cite{Bernard2021Solution}, and
we now show that this extends to the full dynamics. By introducing
co-ordinates 
\be
x=ja_0\ ,\quad y=ka_0\ ,
\ee
we can turn difference equations into differential equations for 
the functions $G_\sigma(x,y,t)=g_\sigma(j,k;t)$ and
$\rho(x,t)=\rho_j(t)$. The scaling limit of \fr{ndot} is simply
\be
\left[\frac{d}{dt}-\mathfrak{D}\frac{d^2}{dx^2}\right]\rho(x,t)=0\ .
\ee
To work out the scaling limits of (\ref{EoMfinal}) it is useful to rewrite
the lattice equations in the form 
\begin{align}
\frac{d}{dt}g_\sigma(j,k;t)=&J\sum_n\Delta_{j,n}g_\sigma(n,k;t)+\Delta_{k,n}g_\sigma(j,n;t)\nn
&+J\delta_{j,k}\sum_n\Delta_{j,n}g_{-\sigma}(n,k;t)+\Delta_{k,n}g_{-\sigma}(j,n;t)
-2JD_{j,k}g_{-\sigma}(j,k;t),
\end{align}
where $D_{j,k}=\delta_{j+1,k}+\delta_{j-1,k}-2\delta_{j,k}$. Taking
$h(n)$ to be a discretization of a test function we have
\begin{align}
\sum_k \delta_{j,k} h(k)&=h(j)\Rightarrow \delta_{j,k}\longrightarrow
a_0\delta(x-y)\ ,\nn
\sum_k D_{j,k} h(k)&=\Delta_{j,n}h(n)\Rightarrow D_{j,k}\longrightarrow a_0^3\delta''(x-y).
\end{align}
This then gives the following result for the scaling limit
\begin{align}
\frac{d}{dt}G_\sigma(x,y;t)&=\mathfrak{D}\nabla^2 G_\sigma(x,y;t)
+\mathfrak{D}a_0\left[\delta(x-y)\nabla^2-2\delta''(x-y)\right]G_{-\sigma}(x,y;t)\nn
& ~~~~ +{\cal O}(Ja_0^4).
\label{eomcont}
\end{align}
We observe that in the limit $a_0\to 0$ we obtain a simple diffusion
equation. This indicates that for large separations we do not have
connected correlations, in agreement with the numerical solution of
the lattice model. The sub-leading term in \fr{eomcont} should be
understood in the sense that we keep the overall physical length
$\mathfrak{L}=La_0$ of our system fixed while taking the lattice spacing $a_0$
to zero and the number of sites $L$ to infinity. Hence we have
\be
\mathfrak{L} = a_0\,L
\ee
The corrections to the diffusion equations are indeed proportional to
$1/L$ as expected, say from the known properties of the steady state \cite{Bauer2019Equilibrium, Bernard2019Open}. Eqns \fr{eomcont} are linear and therefore can be solved by
expanding 
\begin{align} \label{eq:expansion-G}
G_\sigma(x,y;t)=G^{(0)}_\sigma(x,y;t)+\frac{a_0}{\mathfrak{L}}G^{(1)}_\sigma(x,y;t)+\dots\ .
\end{align}
The functions ${G}^{(n)}_\sigma(x,y;t)$ then fulfil a hierarchy of partial differential equations
\begin{align}
\partial_t {G}^{(0)}_\sigma(x,y;t)&=\mathfrak{D}\nabla^2
G^{(0)}_\sigma(x,y;t)\ ,\nn
\partial_t {G}^{(1)}_\sigma(x,y;t)&=\mathfrak{D}\nabla^2
G^{(1)}_\sigma(x,y;t)+ h_{\sigma}(x,y;t)\ ,
\label{conteq}
\end{align}
where the inhomogeneities are given by
\begin{align}
  h_{\sigma}(x,y;t)&=\mathfrak{DL}\left[\delta(x-y)\nabla^2-2\delta''(x-y)\right]G^{(0)}_{-\sigma}(x,y;t).  
\label{hsigma}
\end{align}
Importantly, for generic initial conditions (assuming the absence of long-range order) the connected Green's functions on the lattice will decay to zero with distance, which implies that 
\be
G^{(0)}_+(x,y;0)=\rho(x,0)\rho(y,0)\ ,\qquad G^{(0)}_-(x,y;0)=0.
\ee
This results in 
\be
G^{(0)}_+(x,y;t)=\rho(x,t)\ \rho(y,t)\ ,\qquad G^{(0)}_-(x,y;t)=0.
\ee
The sub-leading contributions fulfil (driven) diffusion equations
\begin{align}
\left[\frac{d}{dt}-\mathfrak{D}\nabla^2\right]G^{(1)}_\sigma(x,y;t)&=
h_\sigma(x,y;t).
\label{gminus}
\end{align}
These are heat equations with a time-dependent external source acting along the line $x=y$. 

Let us recapitulate the main results obtained in this section.
\begin{itemize}
\item{}
The average fermion density $\rho(x,t)$ evolves diffusively
\be
 \partial_t \rho(x,t) = \mathfrak{D}\nabla_x^2 \rho(x,t)\ .
\ee
 \item{} The density correlations $G_+(x,y,t)$ and the coherence fluctuations $G_-(x,y;t)$ admit expansions in powers of $a_0/\mathfrak{L}$, \emph{cf.} eqn \eqref{eq:expansion-G}.
\item As a consequence of the absence of long range order in the initial state, the off-diagonal coherences vanish at leading order, i.e.
\beqa G^{(0)}_-(x,y;t)=0.\eeqa
\item Since the initial density correlations are factorized $G^{(0)}_+(x,y,0)=\rho(x,0)\rho(y,0)$ and then evolve diffusively, the density correlations remain uncorrelated at leading order in the expansion for all times $t$, i.e. 
\be 
G^{(0)}_+(x,y;t)=\rho(x,t)\rho(y,t)\ .
\ee
\item As a consequence of the triangular structure of the equations of motion, the coherence fluctuations, which are of order ${\cal O}(L^{-1})$, evolve diffusively but are subject to a source term depending on the spatial variation of the fermion density
\beqa
\partial_t G^{(1)}_-(x,y;t) = \mathfrak{D}(\nabla_x^2+\nabla_y^2) G^{(1)}_-(x,y;t) + h_{-}(x,y;t) ~,
\eeqa
where $ h_{-}(x,y,t)$ is given by \fr{hsigma} or equivalenty by $ h_{-}(x,y,t)= 2\,\mathfrak{D}\mathfrak{L}\, \nabla_x\nabla_y\big(\delta(x-y)\rho(x,t)\rho(y,t)\big)$. 

\item As the densities in the two replicas are by construction initially uncorrelated they remain uncorrelated at order $a_0/\mathfrak{L}$ at any time
\beqa 
G^{(1)}_+(x,y;t)=0 ~.
\eeqa
This means that the connected density correlations are of order ${\cal O}\big((a_0/\mathfrak{L})^2\big)$, in agreement with known properties of the steady state \cite{Bernard2021Solution}.
\end{itemize}

\subsubsection{Late time regime \sfix{$\mathfrak{D}t/\mathfrak{L}^2\gg 1$}}
In this regime we have
\be
\rho(x,t)=\rho(x,\infty)+{\cal O}(e^{-\gamma t})\ ,\quad
\gamma\geq \frac{4\pi^2\mathfrak{D}}{\mathfrak{L}^2}.
\ee
Here we have used the fact that there are low-lying magnon excitations with dispersion \fr{magnondisp} and the smallest non-zero momentum is $2\pi/L$.
At asymptotically late times the relaxation to the steady state is
exponential in time, and only the lowest ``excited'' mode contributes
to the dynamics. 
\subsubsection{Intermediate time regime \sfix{  $\mathfrak{D}t/\mathfrak{L}^2\ll 1\ll\mathfrak{D}t/a_0^2$}}\label{sec:intermediate_time_regime}
In this regime the continuum approximation applies as $\mathfrak{D}t\gg a_0^2$, but correlations have not yet spread throughout the entire volume and all low-lying excited modes (with momenta close to zero) contribute to the dynamics.
In order to work out the solution of our system of equations in this regime we
require the Green's function
\begin{align}
\left[\frac{d}{dt}-\mathfrak{D}\nabla^2\right]G_2(\boldsymbol{x},t;\boldsymbol{y},t')=
\delta^{(2)}\big(\frac{\boldsymbol{x-y}}{\mathfrak{L}}\big)\delta(t-t').
\end{align}
With $\boldsymbol{k}_{n,m}=\frac{2\pi}{\mathfrak{L}}(n,m)^T$, we have
\begin{align}
G_2(\boldsymbol{x},t;\boldsymbol{y},t')&=\theta(t-t')\sum_{n,m\in\mathbb{Z}}
e^{-i\boldsymbol{k}_{n,m}\cdot(\boldsymbol{x}
  -\boldsymbol{y})-\mathfrak{D}(t-t')\boldsymbol{k}_{n,m}^2}\nn
&=\frac{\theta(\tau)e^{-\frac{\boldsymbol{\xi}^2}{4\tau}}}{4\pi\tau}
\vartheta_3\big(i\frac{\xi_1}{4\tau},e^{-\frac{1}{4\tau}}\big)
\vartheta_3\big(i\frac{\xi_2}{4\tau},e^{-\frac{1}{4\tau}}\big),
\end{align}
where $\vartheta_3$ is an elliptic Theta function whose arguments were transformed by a modular transformation and we have defined
\be
\boldsymbol{\xi}=\frac{\boldsymbol{x-y}}{\mathfrak{L}}\ ,\quad
\tau=\frac{\mathfrak{D}(t-t')}{\mathfrak{L}^2}.
\ee
The correlation functions of interest can be decomposed as e.g.
\be
G^{(1)}_-(x,y;t)={\cal G}(x,y;t)+{\cal H}(x,y;t),
\ee
where ${\cal G}(x,y,t)$ is the solution of the homogeneous equation
with the appropriate initial conditions and ${\cal H}(x,y,t)$ is the
solution of the inhomogeneous equation with vanishing initial
conditions. With $\boldsymbol{x}=(x,y)$, the latter can be written as

\begin{align}
{\cal H}(x,y,t)&=\int_0^\infty dt'\int_0^{\mathfrak{L}}\int_0^{\mathfrak{L}} \frac{d^2\boldsymbol{z}}{\mathfrak{L}^2}
G_2(\boldsymbol{x},t;\boldsymbol{z},t')\ h_{-}(z_1,z_2,t')\nn
&=\int_0^t dt'\int_0^1 dz\
K(\zeta_1-z,\zeta_2-z,\tau)\rho^2(z\mathfrak{L},t') ~,
\label{hxyt}
\end{align}
where
\begin{align}
K(\zeta_1,\zeta_2,\tau)&=\frac{2\mathfrak{D}}{\pi\mathfrak{L}^2(4\tau)^3}
e^{-\frac{\zeta_1^2+\zeta_2^2}{4\tau}}g_1(\zeta_1,\tau)
g_1(\zeta_2,\tau),\nn
g_1(\zeta,\tau)&=2\zeta\vartheta_3\big(i\frac{\zeta}{4\tau},e^{-\frac{1}{4\tau}}\big)
-i\vartheta_3'\big(i\frac{\zeta}{4\tau},e^{-\frac{1}{4\tau}}\big) \ .
\end{align}
Here we have defined
\be
\tau(t')=\frac{\mathfrak{D}(t-t')}{\mathfrak{L}^2}\ ,\quad
\zeta_1=\frac{x}{\mathfrak{L}}\ ,\
\zeta_2=\frac{y}{\mathfrak{L}}.
\ee
The two equivalent expressions in eqn \eqref{hxyt} are related by integration by parts.


\subsubsection{Numerical test of the scaling limit}
\label{sec:numerical}
\begin{figure}[ht]
\begin{center}
\includegraphics[width=0.6\textwidth]{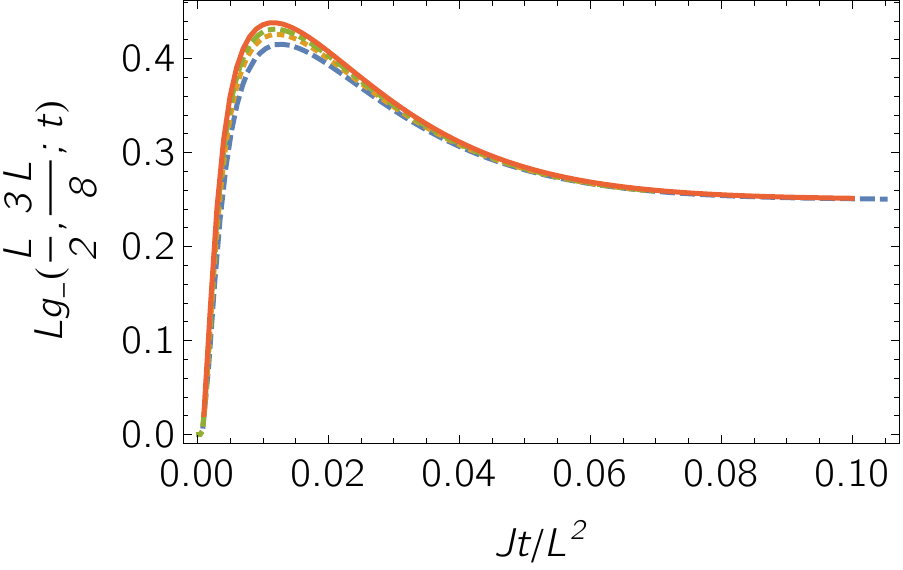}
\end{center}
\caption{Comparison between $g_{-}(L/2,3L/8;t)$ as a function of
$Jt/L^2$ on a ring of $L=48$ (blue dashed), $L=96$ (yellow dotted),
  $L=192$ (gree dot-dashed) sites with domain wall initial conditions
  and $G^{(1)}_-(x,y;t)$ (red solid). The continuum limit result is approached throughout the
depicted time range as $L$ increases.}
\label{fig:latt_vs_cont1}
\end{figure}

We now test the accuracy of the continuum approximation for the
domain-wall initial state \fr{sssec:DWS}.
In the continuum we replace the initial conditions \fr{DWIC} by
\begin{align}
G^{(0)}_+(x,y;0)=\theta\big(\frac{\mathfrak{L}}{2}-x\big)\theta\big(\frac{\mathfrak{L}}{2}-y\big)\ ,\nn
G^{(1)}_-(x,y;0)=\delta\big(\frac{x-y}{\mathfrak{L}}\big)\theta\big(\frac{\mathfrak{L}}{2}-x\big)\ .
\label{initialcond}
\end{align}
Importantly these correctly reproduce the sum rules
\fr{sumrules}. Strictly speaking we should smear out the
delta-function over some microscopic length scale associated with the
high-energy cutoff of the continuum theory. We have verified that this
essentially only modifies the short time behaviour. Using the initial
conditions \fr{initialcond} we find
\begin{align}
\rho(x,t)&=\frac{1}{2}+\frac{2}{\pi}\sum_{n=0}^\infty\frac{\sin(k_{2n+1}x)}{2n+1}e^{-\mathfrak{D}tk_{2n+1}^2}\ ,
\label{densityDW}
\\
{\cal G}(x,y;t)&=
\frac{\rho(\frac{x+y}{2},t/2)e^{\frac{-(\xi_1-\xi_2)^2}{8\tau}}}{\sqrt{8\pi\tau}}\vartheta_3\big(-\frac{\pi}{2}-i\frac{\xi_1-\xi_2}{8\tau}e^{-\frac{1}{8\tau}}\big)\nn
&+\frac{e^{\frac{-(\xi_1-\xi_2)^2}{8\tau}}}{4\sqrt{2\pi\tau}}
\vartheta_3\big(i\frac{\xi_1-\xi_2}{8\tau}e^{-\frac{1}{8\tau}}\big)
-\frac{e^{\frac{(\xi_1-\xi_2)^2}{8\tau}}}{4\sqrt{2\pi\tau}}
  \vartheta_3\big(-\frac{\pi}{2}-i\frac{\xi_1-\xi_2}{8\tau}e^{-\frac{1}{8\tau}}\big)\ .
\end{align}
The contribution ${\cal H}(x,y;t)$ is then obtained by numerically
integrating \fr{hxyt} with the density \fr{densityDW}. The result can
then be directly compared to the full lattice model. Recalling that
$x=ja_0$, $y=ka_0$ and the lattice spacing is $a_0=\mathfrak{L}/L$ we
expact that for sufficiently large $L$ we have  
\be
g_-(L\frac{x}{\mathfrak{L}},L\frac{y}{\mathfrak{L}};t)\approx
\frac{1}{L}G^{(1)}_-(x,y;t)+{\cal O}(L^{-2})\ .
\ee
In Fig.~\ref{fig:latt_vs_cont1} we show a comparison of the lattice
and continuum results for $(x/\mathcal{L},y/\mathcal{L})=(\frac{1}{2},\frac{3}{8})$ and $L=48,96,192$.

\begin{figure}[ht]
\begin{center}
\includegraphics[width=0.6\textwidth]{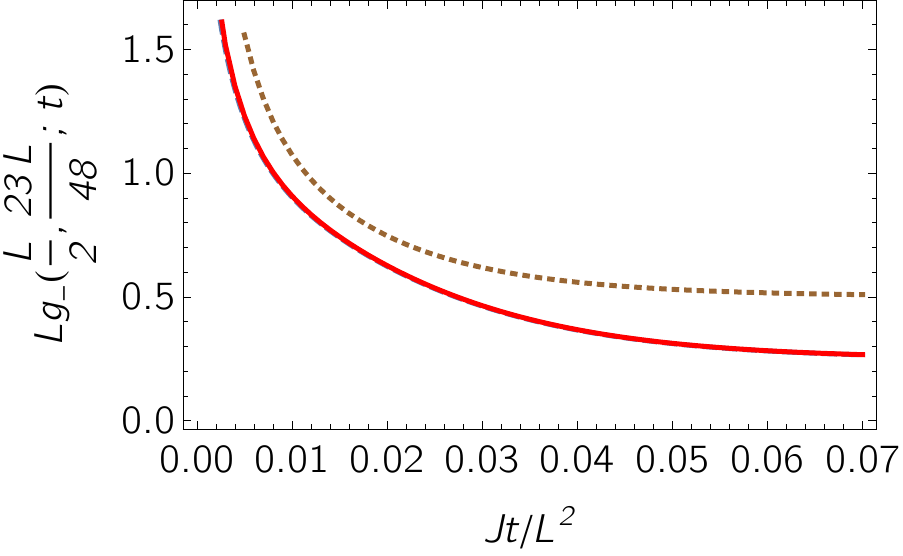}
\end{center}
\caption{Comparison between $g_{-}(L/2,23L/48;t)$ as a function of
$Jt/L^2$ on a ring of $L=96$ (blue dashed) sites with domain wall initial conditions
  and $G^{(1)}_-(x,y;t)$ (red solid). The dotted drown line is the
  result for ${\cal G}(x,y;t)$ only. 
} 
\label{fig:latt_vs_cont2}
\end{figure}
We observe that as $L$ is increased, the lattice result approaches the
continuum one, as expected. The agreement between lattice and
continuum is excellent throughout the intermediate time range
considered. In Fig.~\ref{fig:latt_vs_cont2} we show the
analogous comparison for
$(x/\mathfrak{L},y/\mathfrak{L})=(\frac{1}{2},\frac{23}{48})$ and
$L=96$. The lattice and continuum results match perfectly, and the
$L$-dependence is negligible in the sense that the results for
$L=48,192$ are indistinguishable on the scale of the plot. We also
show the result for ${\cal G}(x,y;t)$ only, which is clearly
different. This shows that the ``interaction term'' $h_-(x,y;t)$ in
the continuum equations \fr{conteq} is important.

\subsubsection{"Diagonal" correlations in the scaling limit}
So far we have considered the two-point functions $G_\sigma(x,y;t)$ for non-vanishing separations $x\neq y$. As we will see later on, the "diagonal" correlations $G_\sigma(\xi\sqrt{t},\xi\sqrt{t};t)$ are of interest in the study of operator spreading. We therefore turn to their analysis next.

\begin{figure}[ht]
	\begin{center}
		\vspace{0mm}
		\includegraphics[width=0.99\textwidth]{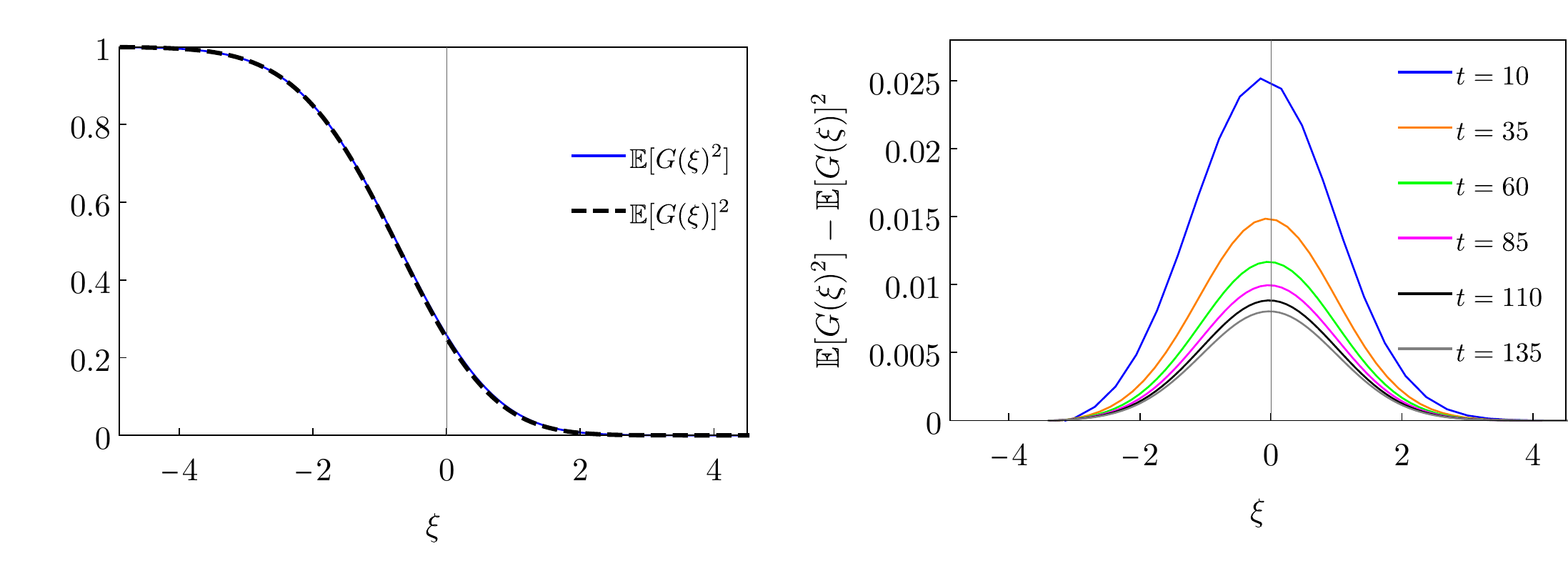}
		\caption{On the left we plot the comparison of $\bE[G(\xi)^2]$ at time $t=150$ (the blue line), and the square of expression \eqref{asy} (the black dashed line). On the right we plot the difference between the asymptotic profile, and finite time results. The system size is fixed to $L=120$.}
		\label{op_ent1}
	\end{center}
\end{figure}

\begin{figure}[ht]
	\begin{center}
		\vspace{0mm}
		\includegraphics[width=0.5\textwidth]{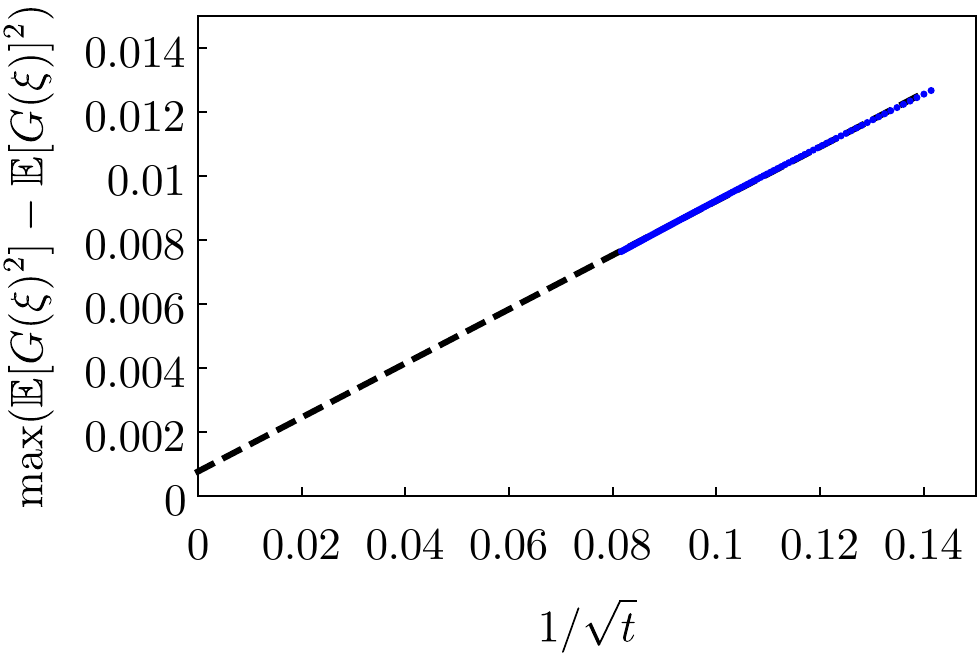}
		\caption{\label{appr} In this plot we show the scaling of the maximal difference, with respect to the coordinate $\xi$, between the average of square of $\bE[G(\xi)^2]$ for times between $50$ and $150$ and the square of expression \eqref{asy} (blue dots). We can see that the difference decays as $\frac{1}{\sqrt{t}}$. Indeed, the linear extrapolation (black line) yields the value at the origin $8\times 10^{-4}$.}
	\end{center}
\end{figure}

In the following we present a numerical check that as a function $\xi=\frac{x}{\sqrt{t}}$ the diagonal correlations $G_\sigma(\xi\sqrt{t},\xi\sqrt{t};t)$ factorize in the leading order in $\frac{1}{L}$
\begin{equation}
	\label{conjecture}
	\bE[G_\sigma(\xi\sqrt{t},\xi\sqrt{t};t)^\alpha]=\bE[G_\sigma(\xi\sqrt{t},\xi\sqrt{t};t)]^\alpha,
\end{equation}
in the limit $L\to\infty$, $t\to\infty$. In order to avoid finite size effects, in the study of operator spreading, we also take the limit $\mathfrak{L}\to\infty$.
In Fig.~\ref{op_ent1} we plot $\bE[G(\xi)^2]=G^{(0)}_+(\xi \times \sqrt{t},\xi\times \sqrt{t},t)$ at different times and compare it to the asymptotic distribution, which is approached as $\frac{1}{\sqrt{t}}$, as presented in Fig.\ref{appr}. 

\subsection{Evolution of higher point correlation functions}
\label{sec:qssep-all-points}
We now turn to multi-point correlation functions in the Q-SSEP. To ease notations we set $\mathfrak{L}=1=\mathfrak{D}$ (which implies $J=L^2$) in the remainder of this section. It is straightforward to reintroduce $\mathfrak{L},\mathfrak{D}$ in the various equations by dimensional analysis. The main point we make here is to show that higher point correlation functions satisfy equations of motion with a triangular structure similar to eqn \eqref{eomcont} in the scaling limit. 

We consider n-point correlation functions of the form
\begin{equation}
g_{n}(l_{1},...,l_{n})=\bE[G_{l_{1}l_{2}}G_{l_{2}l_{3}}...G_{l_{n}l_{1}}]=:\left[\nbubbleThree{1}{n}{l_{n}}{l_{1}}{...}\right]=:\left[\nbubble{1}{n}\right],
\end{equation}
which we call $n$-point bubbles due to its the diagrammatic representation. The discrete indices $l_{1},...,l_{n}$ take values in $\{1,...,L\}$, $L$ is the number of sites of the system, $G_{ij}=\Tr(\rho c_{j}^{\dagger}c_{i})$
and $\bE[\bullet]$ is the expectation value w.r.t. the noise. In the diagrammatic representation each index $l_a$ is a node and $G_{l_a l_b}$ is a directed edge from node $l_a$ to $l_b$. 

In the scaling limit ($J=L^2\to \infty$) the $n$-bubbles become functions of the continuous variables $x_{a}:=l_a/L\in[0,1]$ and they have an expansion in inverse powers of $L$,
\begin{equation}
g_n=g_n^{(0)}+L^{-1}g_n^{(1)}+L^{-2}g_n^{(2)}+....
\end{equation}
As shown in Appendix \ref{app:derivation_n_bubbles}, the first $n-1$ terms in this expansion are actually zero and the leading order of an $n$-bubble scales with $L^{-(n-1)}$. We furthermore show that any expectation value of products of bubbles will factorize at leading order into products of expectation value. Therefore, if interested in the leading order of any correlation function it is enough to study bubbles
\footnote{
Open diagrams such as $\bE[G_{xy} G_{yz}]$ decay exponentially in time and are of no interest. Pinched diagrams such as $\bE[G_{xy} G_{yx} G_{xx}]$ can be obtained by continuity from the three-bubble $\bE[G_{xy} G_{yz} G_{zx}]$ as the limit where $z\to x$ (to be precise, this is true for the connected part only). Therefore, the leading order of any correlation function is encoded into bubbles.
}.

Denoting the leading contribution of any term by the superscript $\#$, we find the time evolution of $n$-point bubbles to be
\begin{equation}\label{eq:evolution_continuous_n_bubble_lead}
\begin{split}
\frac{d}{dt} \left[\nbubble{1}{n}\right]^\#=
&\sum_{a=1}^L \Delta_{x_a} \left[\nbubble{1}{n}\right]^\# \\
&+ \sum_{\substack{a,b=1 \\ a<b}}^L \partial_{x_a}\partial_{x_b}
\left(
\left[\nbubbleThree{1}{b-a}{l_{b-1}}{l_{a}}{l_{a+1}}\right]^\#
\left[\nbubbleThree{1}{n-b+a}{l_{a-1}}{l_{b}}{l_{b+1}}\right]^\#
+(l_a\leftrightarrow l_b)
\right)
\end{split}.
\end{equation}
The equations suggests that the leading contribution of a $n$-point bubble diffuses and is sourced by the product of the leading contribution of smaller bubbles - hence the "triangular" structure. The source term arises if to indices $x_a$ and $x_b$ are in contact and its construction can be visualized as follows: The original $n$-bubbles is squeezed together such that the nodes $x_a$ and $x_b$ touch each other and the diagram forms an eight. Then the eight is split apart into two disconnected bubbles and we sum over the two possible ways how one can attribute $x_a$ and $x_b$ to the two nodes at the splitting junction.
To illustrate this equation, we give its explicit form in the case of 3-point bubbles.
\begin{equation}\label{eq:evolution_continuous_3_bubble_lead}
\begin{split}
\frac{d}{dt} \left[\threebubble{1}{x}{y}{z}\right]^\#=
&(\Delta_x+\Delta_y+\Delta_z) \left[\threebubble{1}{x}{y}{z}\right]^{\#} \\
&+ \partial_{x} \partial_{y} \left(\delta(x-y) \left[\onebubble{1}{x}\right]^{\#} \left[\twobubble{1}{z}{y}\right]^{\#}+(x\leftrightarrow y)\right) \\
&+ \partial_{y} \partial_{z} \left(\delta(y-z) \left[\onebubble{1}{y}\right]^{\#} \left[\twobubble{1}{x}{z}\right]^{\#}+(y\leftrightarrow z)\right) \\
&+ \partial_{z} \partial_{x} \left(\delta(z-x) \left[\onebubble{1}{z}\right]^{\#} \left[\twobubble{1}{y}{x}\right]^{\#}+(z\leftrightarrow x)\right)
\end{split}
\end{equation}

\subsection{Dictionary for 2-replica operators}

Since initially the aim was to find equations for the evolution of
any correlation function on two replicas, we will show here how Wick's
theorem can be used to find a correspondence between any connected
2-replica correlation function and the leading diagrams contributing
to these correlation functions - which are bubbles by the argument
in the last section.

Any 2-replica operator can be written in terms of $n_{a,i}=c_{a,i}^{\dagger}c_{a,i}$
(where $a=1,2$) and pairs of $S_{j}^{+}=c_{1,j}^{\dagger}c_{2,j}$
and $S_{k}^{-}=c_{1,k}c_{2,k}^{\dagger}$ (fermions on different replica
commute in our convention). Note that we need the same number of $S^{+}$
and $S^{-},$ otherwise the quantum expectation number is zero for
a Gaussian state. Wick's theorem
applies if the state of the system is described by a Gaussian density matrix
$\rho=e^{c^{\dagger}Mc}/\Tr(e^{c^{\dagger}Mc})$, which contains all possible quadratic terms in fermionic creation and annihilation operators.

As an example, the four-point correlation function $\bE[\langle S_x^+ S_y^- n_{1,a} n_{2,b} \rangle]$, where $\langle \cdot \rangle$ is the quantum expectation value, decomposes into
\begin{equation}
-\left[\onebubble{1}{a}\onebubble{1}{b}\twobubble{1}{x}{y}\right]
+\left[\onebubble{1}{a}\threebubble{1}{x}{b}{y}\right]
+\left[\onebubble{1}{b}\threebubble{1}{x}{y}{a}\right]
+\left[\fourbubble{1}{x}{b}{y}{a}\rbrack\right].
\end{equation}
We learned in the last section that products of bubbles factorize at leading order. Therefore, the correspondence two-replica correlation functions and bubbles (which we will in the following denote by $\sim$) is on the level of the connected part. To leading order in $L$ we have
\begin{equation}
    \bE[\langle S_x^+ S_y^- n_{1,a} n_{2,b} \rangle]^c\sim \left[\fourbubble{1}{x}{b}{y}{a}\rbrack\right].
\end{equation}

This correspondence can be carried further to any combination of operators. First consider,
\begin{align}
\bE\left[\langle S_{x_{1}}^{+}S_{y_{1}}^{-}...S_{x_{n}}^{+}S_{y_{n}}^{-}\rangle\right]^{c} & \sim-\sum_{\sigma\in S_{n},\tau\in S_{n-1}}\left[\nbubbleFour{1}{2n}{y_{\sigma(n)}}{x_{n}}{y_{\sigma(1)}}{x_{\tau(1)}}\right],
\end{align}
where $S_{n}$ is the permutation group of $n$ elements. 

Next, any correlation function of the form $\bE\left[\langle n_{1,a_{1}}\cdots n_{1,a_{l}} n_{2,b_{1}}\cdots n_{2,b_{k}} \rangle\right]^c$ is in fact a correlation function of the product of 1-replica operators, $\bE\left[\langle n_{a_{1}} \cdots n_{a_{l}} \rangle\langle n_{b_{1}}\cdots n_{b_{k}} \rangle\right]^c$. Therefore, the leading contribution of its connected part is not given by a single $l+k$ bubble, but rather by the connected part of the product of two bubbles with $l$ and $k$ nodes. These diagrams scale with even higher (negative) power of $L$ than a $k+l$-bubble, and their time evolution has not been considered in the last section. Still we can give the correspondence for 1-replica correlation functions of the form
\begin{equation}
\bE\left[\langle n_{a_{1}}...n_{a_{l}}\rangle\right]^{c}\sim(-1)^{l+1}\sum_{\sigma\in S_{l-1}}\left[\nbubbleFour{1}{l}{a_{\sigma(l-1)}}{a_{l}}{a_{\sigma(1)}}{a_{\sigma(2)}}\right].
\end{equation}
Finally, the most general correlation function $\bE\left[\langle S_{x_{1}}^{+}S_{y_{1}}^{-}...S_{x_{n}}^{+}S_{y_{n}}^{-}n_{1,a_{1}}...n_{1,a_{l}}n_{2,b_{1}}...n_{2,b_{k}}\rangle\right]^{c}$
has as its leading contribution bubbles with $2n+l+k$ nodes that must be arranged on a circle according to the following rules:
\begin{itemize}
    \setlength\itemsep{0.2em}
    \item every $x$ is followed by either $b$ or $y$;
    \item every $y$ is followed by either $a$ or $x$;
    \item every $a$ is followed by either $a$ or $x$;
    \item ever $b$ is followed by either $b$ or $y$.
\end{itemize}
One can convince oneself of these rules by looking at the example given above.

\section{Application: Operator spreading}
\label{sec:op-entangled}

In this section, we will consider the large scale dynamics of operator spreading, which attracted a significant amount of attention in recent years as an indicator of quantum chaos \cite{larkin1969quasiclassical,maldacena2016bound}. Research, thus far, has mostly focused on operator spreading in isolated quantum systems, or in random unitary circuits \cite{khemani2018velocity,gopalakrishnan2018operator,gopalakrishnan2018hydrodynamics,nahum2018operator,von2018operator,Chan2018Solution,gopalakrishnan2018operator,gopalakrishnan2018hydrodynamics,DeNardis2018Hydrodynamic}. It is, however, important to understand operator spreading in continuous (in time) systems, which are coupled to an environment. The Q-SSEP represents a perfect test-bed to address these questions. 

We will focus on two aspects of operator spreading, the first one being the out-of-time ordered correlators (OTOC), which can in some cases be related to Lyapunov exponents \cite{larkin1969quasiclassical,maldacena2016bound}, and secondly on the hydrodynamics of the operator entanglement spreading \cite{zanardi2001entanglement,prosen2007operator,znidaric2008complexity,pizorn2009operator, hartmann2009density,muth2011dynamical,dubail2017entanglement,zhou2017operator,jonay2018coarse,xu2018accessing,van2018entanglement,pal2018entangling,takayanagi2018holographic,nie2018signature,znidaric2020entanglement,alba2020diffusion}, which has been conjectured to be able to distinguish between integrable and non-integrable isolated systems \cite{prosen2007efficiency,Alba2019Operator}, even in cases when OTOCs fail to distinguish between different types of dynamics  \cite{Alba2019Operator}. Operator entanglement is also interesting due to its relation to the computational complexity in the matrix product ansatz based  algorithms.

In what follows we will consider the spreading of the single site fermion creation and annihilation operators $c_i$, $c^\dagger_i$. Without loss of generality, we will focus on the time evolution of operator $c_0$.
For every realization of the noise, the time evolution of these operators is free, meaning that 
\begin{equation}
	\label{opc0}
	c_{0}(t)=\sum_i A_i(t) c_i.
\end{equation}
Nevertheless, we have to keep in mind that coefficients 
\begin{equation}
	\label{weight}
	A_i(t)=\Tr(c_i^\dagger c_0(t))
\end{equation}
are fluctuating objects.
\subsection{Out-of-time ordered correlation functions}
In general, the weight of an operator $O(t)$ that has spread to site $x$ at time $t$ can be quantified by summing up the contributions of the commutators between the operator $O(t)$ and the basis elements of the local operator algebra $e_x^{(r)}\in \{c_x^\dagger, c_x, n_x,\mathbb{I} - n_x\}$
\begin{equation}
	C(x,t)=\sum_r\frac{1}{{\rm Tr}(\one)}{\rm Tr}\Big([O(t),e_x^{(r)}]^\dagger[O(t),e_x^{(r)}]\Big).
\end{equation}
For the operator $O(0)=c_0$, the OTOC $C(x,t)$ can be related to the amplitudes $A_x(t)$ defined in \eqref{weight}
\begin{equation}
	\label{otoc_amp}
	C(x,t)= 4-|A_x(t)|^2.
\end{equation}
Note the constant $4$, which arises due to the non-commutativity of the basis elements associated with different sites of the chain.

The OTOC amplitude \eqref{otoc_amp} can also be accessed by considering the dynamics of the two point function, if the system is initially prepared in the state with a single fermion at position $x$. Namely, the two point function
\begin{equation}
	G^{[x]}_{0,0}(t)=\Tr\big(c_0^\dagger c_0\rho^{[x]}(t)\big)=\Tr\big(c_0^\dagger(t) c_0(t)\rho^{[x]}\big)=A_{i}^\dagger(t)A_j(t) \Tr(c_i^\dagger c_j\rho^{[x]}).
\end{equation}
gives us
\begin{equation}
	C(x,t)=4-G^{[x]}_{0,0}(t),
\end{equation}
if we choose $\rho^{[x]}$ such that  $\Tr (c_{i'}^\dagger c_i\rho^{[x]})=\delta_{i,x}\delta_{i',x}$.

It is clear that at the level of averages the spreading of OTOCs is purely diffusive due to translation invariance
\begin{equation}
	\bE[C(x,t)]=4-\bE[G^{[0]}_{-x,-x}(t)].
\end{equation}
In the scaling limit $L\to\infty$ and for large times $t$, we have
\begin{equation}
	\bE[C(x,t)]=4- \frac{1}{\sqrt{4\pi t}} \exp(-x^2/4t)+\mathfrak{o}(t^{-1/2}).
\end{equation}
\subsection{Operator entanglement}
A complete operator basis of the system can be obtained by multiplying the basis operators associated with the local algebra $e_x^{(r)}$. Considering a bi-partition into the subsystem $A$ and the subsystem $B$ any operator can be represented as
\begin{equation}
	O/\sqrt{\Tr(O^\dagger O)}=\sum_{i,j} M_{i,j} O_{A,i}O_{B,j},
\end{equation}
where $O_{A,i}$ and $O_{B,i}$ are orthonormalized basis elements of the operator spaces associated with the subspaces $A$ and $B$ respectively
\be
\Tr_{A/B}\big(O_{A/B,i}^\dagger O_{A/B,j}\big)=\delta_{ij}\ .
\ee
Operator entanglement of the local operator $O$ can then be obtained by performing the singular value decomposition of the matrix $[M]_{i,j}=M_{i,j}$, which provides the Schmidt decomposition of the operator $O$
\begin{equation}
	O/\sqrt{\Tr(O^\dagger O)}=\sum_{i}\sqrt{\lambda_i}O_{A,i} O_{B,i}.
\end{equation}
Schmidt coefficients $\lambda_i$ are normalized,  $\sum_i\lambda_i=1$, and $\alpha$-th R\'{e}nyi operator entanglement entropy reads
\begin{equation}
	S_\alpha=\frac{1}{1-\alpha}\log\left(\sum_i\lambda_i^\alpha\right).
\end{equation}

We will focus on the operator entanglement $S_\alpha(x,t)$ of the operator \eqref{opc0}, for two connected bipartitions $A=[1,x]$ and $B=(x,L]$, which can be deduced from the decomposition
\begin{equation}
	O/\sqrt{\Tr(O^\dagger O)}=\sum_{i,j} M_{(r_1,r_2,...,r_x),(r_{x+1},...,r_{L})} \left(\prod_{k=1}^{x}e_k^{(r_k)}\right)\left(\prod_{h=x+1}^L e_h^{(r_h)}\right).
\end{equation}
Note that $S_\alpha(x,t)$ is defined for a fixed realization of the noise, and in the following we will be concerned with the averaged quantity.
The terms comprising the operator can be grouped in two parts
\beqa
	\label{opc1}
	c_{0}(t)&=& \sum_{i\in A} A_i(t) c_i+\sum_{i\in B} A_i(t) c_i \\
	&=& \left(\sum_{i\in A} A_i(t) c_i\right)\left(\prod_{i\in B}\one_i\right)+\left(\prod_{i\in A}\one_i\right)\left(\sum_{i\in B} A_i(t) c_i\right). \nonumber
\eeqa
Clearly the two parts satisfy the orthogonality condition, and the Schmidt coefficients can be obtained simply by computing the norms of associated operators:
\begin{equation}
	\lambda_{1}(x,t)=\sum_{i=-L/2+1}^x |A_i(t)|^2;\quad\lambda_2(x,t)=\sum_{i=x+1}^{L/2} |A_i(t)|^2.
\end{equation}
Operator entanglement is upper bounded by the logarithm of the number of non-zero Schmidt coefficients $S_{\alpha}(x,t)\leq \log(2)$, which can be understood as a direct consequence of the free dynamics of fermions for any realization of the noise. Similar upper bounds can be obtained for any composite operator \cite{dubail2017entanglement}.

Also in this case it proves useful to relate the dynamics of operator entanglement to the two point function $G_{i,j}^{[x^-]}(t)=\Tr(c_i^\dagger c_j \rho^-_x(t))$
for the domain wall initial conditions
\begin{equation}
	\Tr(c_{l'}^\dagger c_{k'}\rho^-_x(0))=\delta_{l'k'}\, \theta(x+1/2-l'),\quad x,l',k'\in [1,L],
\end{equation}
as
\begin{equation}
	\lambda_1(x,t)=G_{L/2,L/2}^{[x^-]}(t),
\end{equation}
The second Schmidt coefficients can be related to the quench with a complementary domain wall initial conditions
\begin{equation}
	\Tr(c_{l'}^\dagger c_{k'}\rho^+_x(0))=\delta_{l'k'}\, \vartheta(l'-x-1/2)\quad x,l',k'\in [1,L].
\end{equation}
as
\begin{equation}
	\lambda_2(x,t)=G_{L/2,L/2}^{[x^+]}(t),
\end{equation}
The operator entanglement therefore takes the following suggestive form
\begin{equation}
	S_{\alpha}(x,t)=\frac{1}{1-\alpha}\log\big(G_{L/2,L/2}^{[x^-]}(t)^\alpha+G_{L/2,L/2}^{[x^+]}(t)^\alpha\big).
\end{equation}

In what follows, we will be concerned with the large system size limit $L\to\infty$, while, for the moment, keeping $x$ and $t$ finite. Firstly, this allows us to relate operator entanglement to the spatio-temporal profile of two point functions, starting from the single initial domain wall condition, with the cusp at the origin
\begin{equation}
	S_{\alpha}(x,t)=\frac{1}{1-\alpha}\log\big(G_{-x,-x}(t)^\alpha+G_{x+1,x+1}(t)^\alpha\big).
\end{equation}
Here we used a simplified notation $G_{x,x}(t)\equiv G_{x,x}^{[\rho]}(t)$, for the domain wall at the origin
\begin{equation}
	\Tr(c_{l'}^\dagger c_{k'}\rho_0(0))=\delta_{l'k'}\, \theta(1/2+l').
\end{equation}

The intuition of the large $L$ limit, which we built in the preceding section, lies at the heart of our conjecture that in the scaling limit $t,x\to \infty$, while $\xi=\frac{x}{\sqrt{t}}$ is kept fixed, the identification \eqref{conjecture}
holds.

Equation \eqref{conjecture}, which asserts that products of Green functions are self-averaging at leading order in $1/L$, implies that the averaged operator entanglement 
\begin{equation}
S_{\alpha}(\xi):=\lim_{t\to\infty}S_\alpha(\xi \sqrt{t},t)
\end{equation}
can be, in the infinite size limit limit\footnote{To leading order in the system size.}, reduced to the entanglement of the averaged behaviour
\begin{equation}
	\bE[S_{\alpha}(\xi)]=\frac{1}{1-\alpha}\log(\bE[G(\xi)]^\alpha+\bE[G(-\xi)]^\alpha).
\end{equation}
In the continuum limit $\bE[G(\xi)]$ satisfies the diffusion equation, which implies that
\begin{equation}
	\label{asy}
	\bE[G(\xi)]=\tfrac{1}{2}(\text{Erf}\left(\xi/2\right)+1).
\end{equation}
The hydrodynamics of the operator entanglement is therefore described by
\begin{equation}
	\mathbb{E}[S_\alpha(\xi)]=\frac{1}{1-\alpha}\log \left[\left(\frac{\text{Erf}\left(\xi/2\right)+1}{2}\right)^\alpha+\left(\frac{\text{Erf}\left(-\xi/2\right)+1}{2}\right)^\alpha\right].
\end{equation}

\begin{figure}[ht]
	\begin{center}
		\vspace{0mm}
		\includegraphics[width=0.5\textwidth]{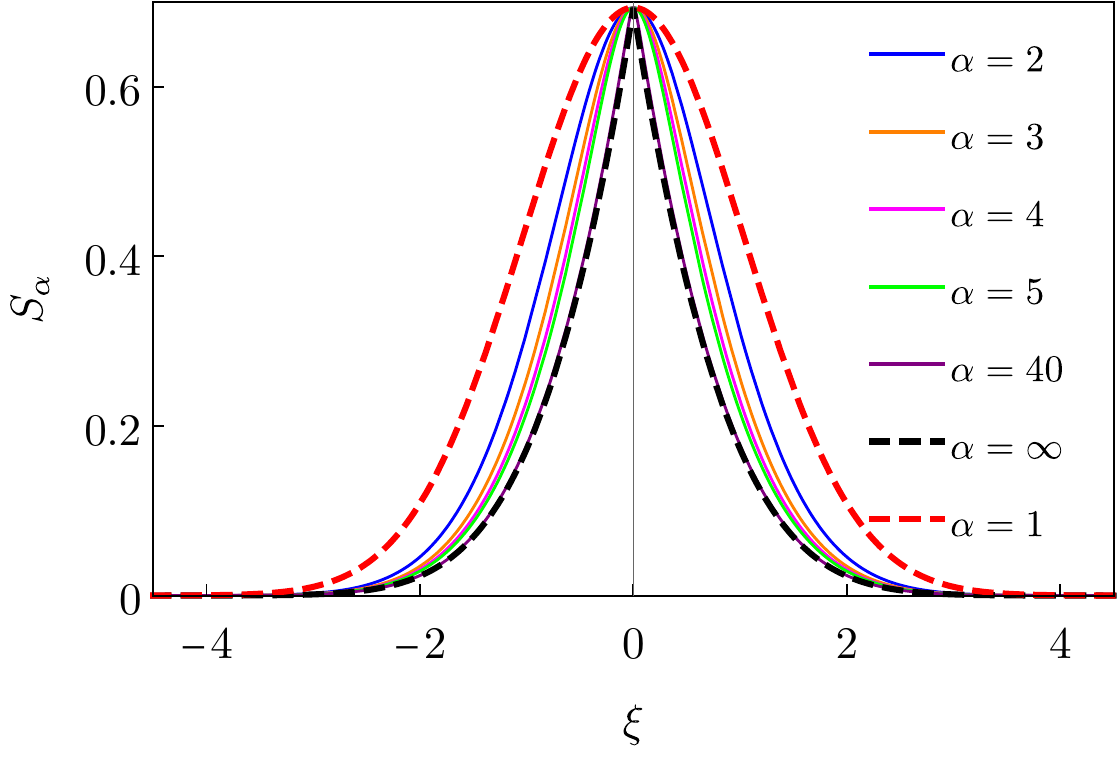}
		\caption{\label{op_ent2} Asymptotic profile of operator entanglement for different values of R\'{e}nyi indexes $\alpha$. The dashed curves correspond to the extremal values of parameters.}
	\end{center}
\end{figure}

Clearly the operator entanglement at the origin is maximal, $\bE[S_\alpha(0)]=\log(2)$, for any value of $\alpha$. There are two particularly interesting limits, which we might wish to consider. The first one is the Von Neumann entanglement entropy, which is obtained in the limit $\alpha \to 1$, and the $\infty$ entanglement entropy $\alpha\to\infty$, which corresponds to the logarithm of the largest Schmidt coefficient
\begin{eqnarray}
	\mathbb{E}[S_1(\xi)]&=&\text{erfc}\left(\frac{\xi}{\sqrt{2}}\right) \tanh ^{-1}\left[\text{erf}\left(\frac{\xi}{\sqrt{2}}\right)\right]-\log
	\left[\text{erf}\left(\frac{\xi}{\sqrt{2}}\right)+1\right]+\log 2,\\ \mathbb{E}[S_{\infty}(\xi)]&=&\log(2) - \log(1 + \text{erf}(|\xi|/\sqrt{2})).
\end{eqnarray}
In Fig.~\ref{op_ent2} we plot the entanglement entropy for different R\'{e}nyi indexes $\alpha$. 

In this section we have demonstrated that spreading of operators in Q-SSEP model falls in a different "dynamical universality class" from isolated quantum systems or random unitary circuits. While in the latter two cases operators spread ballistically with diffusive corrections, we observed that in Q-SSEP operator spreading is purely diffusive. This observation is important as it begs the question whether in open quantum systems, where the bath degrees of freedom are much faster than the dynamics in the quantum system under consideration, interactions with bath can in general reduce operator spreading. While this is a fundamental question in its own right, it is also related to the possibility of using weak dissipation in order to facilitate classical simulations of quantum dynamics. Similar behaviour was already exploited in a number of numerical studies of quantum transport employing the dissipative boundary driving, e.g. \cite{znidaric2011spin,znidaric2011transport,znidaric2016diffusive}.

\section{Summary and conclusions}
\label{sec:conclusions}

In this work we have conducted a detailed analysis of the dynamics of fluctuations in the quantum asymmetric simple exclusion process (Q-ASEP) with periodic boundary conditions. We have demonstrated that fluctuations of the fermionic degrees of freedom obey evolution equations of Lindblad type, and shown that that the corresponding Lindbladians can be represented as non-Hermitian quantum spin chains that can be expressed in terms of the generators of a $gl(2R)$ algebra. In case of the Q-SSEP this algebra is a symmetry of the Lindbladian. The operator space in our model fragments into exponentially many (in system size) sectors that are invariant under time evolution. This extends recent findings for the average time evolution to fluctuations. 
Focusing on the five sectors that describe the late time dynamics of the Q-SSEP we showed that two of them ($\cvec_{\pm1}$) correspond to a Yang-Baxter integrable model. Numerical checks of the $\cvec_0$ block revealed signatures consistent with integrability or weak integrability breaking in the level-spacing statistics. However, they are not conclusive given the limitations on the system sizes we were able to simulate.
In the particular case of Q-SSEP we determined the algebraic structure underlying the steady states and slow modes that govern the late time behaviour. We showed that the former can be understood in terms of "ferromagnetic" $gl(2R)$ multiplets, while the latter can be viewed as diffusive magnon-like excitations.
We then showed that the dynamics of fluctuations of observables in the Q-SSEP is described by a closed sets of coupled linear differential-difference equations. The behaviour of the solutions to these equations is essentially diffusive but with relevant deviations, that at sufficiently late times and large distances can be described in terms of a continuum scaling limit which we constructed. We established the applicability of this scaling limit over a significant range of time and space scales by comparing it to numerical solutions of the corresponding equations of motion for the lattice model. We finally applied this continuum description to the study of operator spreading at large scales, focusing on out-of-time ordered correlators and operator entanglement. In contrast to operator spreading in random unitary circuits and isolated many-particle Hamiltonian systems, where operators spread ballistically with diffusive corrections, we observe purely diffusive spreading in the Q-SSEP.

Our work raises a number of interesting questions that warrant further enquiry. First, it would be interesting to extend the results reported here for the Q-SSEP to the Q-ASEP. The determination of the steady state manifold will be addressed in a forthcoming publication, but the nature of excited states, dynamics of correlations and dynamics of operator spreading are significantly harder to address. Second, there should be a field theory that gives rise to the continuum description of the equations of motion for correlation functions and it is an open problem to construct it. Third, it would be interesting to extend our analysis to the case of open boundaries with particle injection and extraction. Fourth, one ought to go beyond purely dissipative dynamics and investigate the effects of a Hamiltonian part of the Lindbladian, at least in some limiting cases. Finally, in order to analyze the level-spacing statistics we introduced a conjecture of how to treat additional degeneracies arising from the presence of a higher-rank symmetry. It would be interesting to further test the validity of this conjecture by considering larger system sizes.


\vskip  0.5 truecm

\noindent {\bf Acknowledgements}: This work was initiated during the 
thematic trimester program \emph{Systems out of equilibrium} at the Institut Henri Poincar\'e. We are grateful to the IHP for hospitality and to Sorbonne University for support. D.B. acknowledges Michel Bauer and Jean-Bernard Zuber for regular discussions. M.M. thanks Tony Jin for initial collaboration on the project and, in particular, for his contributions to analysing the spectrum in the two-particle sector.
This work was in part supported by CNRS, by the ENS, by the ANR project ``ESQuisses'', contract number ANR-20-CE47-0014-01 and by the EPSRC under grant EP/S020527/1. 

\vskip 1.0 truecm

\appendix

\section{Dictionary: Super-operator versus Hilbert space doubling approach}
\label{...}

\subsection{One replica}
The basis of states \fr{states_1rep} in the Hilbert space doubling approach correspond to physical operators as follows
\begin{align}
|\!|1\rangle_j=n_j\ ,\quad
|\!|2\rangle_j=c_j\ ,\quad
|\!|3\rangle_j=c^\dagger_j\ ,\quad
|\!|4\rangle_j=1-n_j.
\end{align}
where $n_j=c^\dagger_j c_j$. In terms of the Hubbard operators \fr{Hubbardops} we have
\begin{align}
c^\dagger_j&=c^\dagger_j\otimes\mathds{1}=E^{12}_j+E^{34}_j\ ,\nn
\tilde{c}^\dagger_j&=\mathds{1}\otimes c^\dagger_j=-E^{13}_j+E^{24}_j\ .
\end{align}
Here the "doubled" operator $\tilde{c}^\dagger_j$ describes the right action of the annihilation operator $c_j$, so that the correspondence with the fermions $f_j^{l/r}$ defined in eqn \fr{ffermions} in the super-operator formalism is
\be
c^\dagger_j\leftrightarrow (f_j^l)^\dagger\ ,\quad
\tilde{c}^\dagger_j\leftrightarrow f_j^r\ .
\ee
The two states $|\!|1\rangle_j$ and $|\!|4\rangle_j$ form an $su(2)$ doublet with generators $J_j^+=E^{14}$, $J_j^z=E^{41}$ and $J_j^z=E^{11}-E^{44}$. The identity operator $|\!|1\rangle_j+|\!|4\rangle_j$ on this subspace is not an eigenstate of $J_j^z$ but of $J_j^x=J_j^+ + J_j^-= E^{14}+E^{41}$.

\subsection{Two replicas}
In the two-replica case the correspondence between the sixteen basis states \fr{states_2rep} and physical operators is as follows
\begin{center}
\begin{tabular}{|c|c|c|c|c|c|c|c|}
\hline
$|\!|0\rangle$ &
$|\!|1\rangle$ &
$|\!|2\rangle$ &
$|\!|3\rangle$&
$|\!|4\rangle$ &
$|\!|5\rangle$ &
$|\!|6\rangle$ &
$|\!|7\rangle$ \\
\hline
$c^\dagger_1c_2$
&$n_1n_2$
&$n_1(1-n_2)$
&$(1-n_1)n_2$
&$(1-n_1)(1-n_2)$
&$c_1c^\dagger_2$
&$n_1c_2$
&$c_1n_2$\\
\hline
\end{tabular}
\end{center}

\begin{center}
\begin{tabular}{|c|c|c|c|c|c|c|c|}
\hline
$|\!|8\rangle$ &
$|\!|9\rangle$ &
$|\!|10\rangle$ &
$|\!|11\rangle$ &
$|\!|12\rangle$ &
$|\!|13\rangle$ &
$|\!|14\rangle$ &
$|\!|15\rangle$ \\
\hline
$(1-n_1)c_2$
&$ c_1(1-n_2)$
&$n_1c_2^\dagger$
&$c_1^\dagger n_2$
&$(1-n_1)c_2^\dagger$
&$ c_1^\dagger(1-n_2)$
&$c_1c_2$
&$c_1^\dagger c_2^\dagger$\\
\hline
\end{tabular}
\end{center}
In terms of the Hubbard operators \fr{Hubbardops2} we have
\begin{align}
c^\dagger_{1,j}&={\cal E}_j^{0,8}
+{\cal E}_j^{1,7}
+{\cal E}_j^{2,9}
+{\cal E}_j^{6,14}
+{\cal E}_j^{10,5}
+{\cal E}_j^{11,3}
+{\cal E}_j^{13,4}
+{\cal E}_j^{15,12}\ ,\nn
c^\dagger_{2,j}&={\cal E}_j^{2,6}
+{\cal E}_j^{3,8}
-{\cal E}_j^{5,9}
-{\cal E}_j^{7,14}
+{\cal E}_j^{10,2}
-{\cal E}_j^{11,0}
+{\cal E}_j^{12,4}
-{\cal E}_j^{15,13}\ ,\nn
\tilde{c}^\dagger_{1,j}&=-{\cal E}_j^{1,11}
-{\cal E}_j^{2,13}
+{\cal E}_j^{5,12}
-{\cal E}_j^{6,0}
+{\cal E}_j^{7,3}
+{\cal E}_j^{9,4}
-{\cal E}_j^{10,15}
+{\cal E}_j^{14,8}\ ,\nn
\tilde{c}^\dagger_{2,j}&=-{\cal E}_j^{0,13}
-{\cal E}_j^{1,10}
-{\cal E}_j^{3,12}
+{\cal E}_j^{6,2}
+{\cal E}_j^{7,5}
+{\cal E}_j^{8,4}
+{\cal E}_j^{11,15}
-{\cal E}_j^{14,9}\ .
\end{align}
The correspondence with the fermions $f_j^{a;l/r}$ defined in eqn \fr{ffermions} in the super-operator formalism is
\be
c^\dagger_{a,j}\leftrightarrow (f_j^{a;l})^\dagger\ ,\quad
\tilde{c}^\dagger_{a,j}\leftrightarrow f_j^{a;r}\ .
\ee

\section{Steady states, low-lying states and gaps}
\label{Appendix:zero-mode}

To simplify notation we set $J=1$ in this Appendix.

Via eqn \eqref{eq:L-tensorC-GL} or  eqn \eqref{eq:L-tensorC-SL}, the (local) Q-SSEP Lindbladian is expressed in terms of the so-called tensor Casimir $\mathbb{C}$ for the $gl(2R)$ or $sl(2R)$ algebras, up to simple terms depending on the $C$ charges.  Namely, eqn \eqref{eq:L-tensorC-GL} can be written as (we set $D_0=1$ to simplify the notation)
\beq \label{eq:L-Casimir}
\mathcal{L}^*_{j+1;j} = \mathbb{C}_{j;j+1}  -  \frac{1}{2}(C_{j+1}+C_j) - R ,
\eeq
with $\mathbb{C}_{j;j+1}:=\sum_{AB}G_{j+1}^{AB}G_j^{BA}$ the $gl(2R)$ tensor Casimir. 

Imagine considering the tensor product of two representations $W_1\otimes W_2$ with highest weights $\mu_1$ and $\mu_2$ of some Lie algebra (and in particular $gl(2R)$ or $sl(2R)$). The tensor Casimir $\mathbb{C}_{12}$ is defined by
\beq \label{eq:tensor-C}
\mathbb{C}_{12} := \frac{1}{2}\big( C_{W_1\otimes W_2} - C_{W_1}\otimes\mathbb{I} - \mathbb{I}\otimes C_{W_2} \big) ,
\eeq
with $C_{W}$ the Casimir in the representation $W$. Now, if $\ket{\mu_{1,2}}$ are the highest weight vectors in $W_{1,2}$ then $\ket{\mu_{1}}\otimes \ket{\mu_{2}}$ is a highest weight vector in $W_1\otimes W_2$ with highest weights $\mu_1+\mu_2$. Thus, it is also an eigenvector of the tensor Casimir
$\mathbb{C}_{12}\, \ket{\mu_{1}}\otimes \ket{\mu_{2}} = \kappa_{\mu_1;\mu_2}\, \ket{\mu_{1}}\otimes \ket{\mu_{2}}$, 
with $ \kappa_{\mu_1;\mu_2}=\frac{1}{2}\big( \mathrm{Cas}_{\mu_1+\mu_2} - \mathrm{Cas}_{\mu_1} - \mathrm{Cas}_{\mu_2}\big)$, with $\mathrm{Cas}_{\mu}$ the value of the Casimir in the representation of weight $\mu$. 
Up to a normalisation factor (to be determined according the chosen convention), $\mathrm{Cas}_{\mu}=(\mu,\mu+2\rho)$ with $\rho$ half of the sum over the positive roots. Here $(\cdot,\cdot)$ denotes the scalar product on the weight space induced by the Killing form. Hence,  $\kappa_{\mu_1;\mu_2}=(\mu_1,\mu_2)$ and 
\beq
\mathbb{C}_{12}\, \ket{\mu_{1}}\otimes \ket{\mu_{2}} = (\mu_1,\mu_2)\, \ket{\mu_{1}}\otimes \ket{\mu_{2}} .
\eeq

To apply this to our problem, we have to check the normalisation of the Casimir in the fermionic  representation of $gl(2R)$ we are using. We have $G^{AB}:= f^{A\,\dag} f^{B}$ with $\hat C:=\sum_A G^{AA}$ the total number of $f$ fermions. Recall that $\hat C$ is related the $C$ charge via $\hat C= C+R$. Using the fermionic anti-commutation rules, we compute:
\[ \sum_{A,B}G^{AB}G^{BA} =  \sum_{A,B}f^{A\,\dag} f^{B}f^{B\,\dag} f^{A} = \hat C (2R+1-\hat C) .\]
This coincides with $(\mu_p,\mu_p+2\rho)$ for $\mu_p=\epsilon_1+\cdots+\epsilon_p$, with $\epsilon_i$ ortho-normalized, the highest weight of the $p$ antisymmetric tensor and $\hat C=p$ or equivalently $C=p-R$. In particular $(\mu_p,\mu_q)=\mathrm{min}(p,q)$ and this yields the eigien-value of the tensor Casimir in the corresponding tensor product of $gl(2R)$ representations.

Consider now the fully polarized state in the sector with $C$ charge $p-R$ with $0\leq p\leq 2R$. This state is an eigenstate of the Q-SSEP Lindbladian. Since $\mathcal{L}^*_{j+1;j} = \mathbb{C}_{j;j+1} -  \frac{1}{2}(C_{j+1}+C_j) - R$, its eigenvalue is $p - (p-R) - R=0$. Thus, as claimed in the main text, the state $|\!|\Omega_{\mu_p}\rangle:=|\mu_{p}\rangle\otimes\cdots\otimes|\mu_{p}\rangle$ are zero modes of the Lindbladian $\mathcal{L}^*\, |\!|\Omega_{\mu_p}\rangle = 0$. By action of $GL(2R)$ this generates $|\!|\Omega_x\rangle=x_1 x_2\cdots x_L$ with $x\in [\mu_p]$ and $\mathcal{L}^*\, |\!|\Omega_{x}\rangle = 0$.

Furthermore, for the tensor product states $\ket{\mu_p}\otimes\ket{\mu_q}$, with $p\not= q$, the eigen-value of the tensor Casimir is $\mathrm{min}(p,q)$. As a consequence of domain wall states $\ket{\mu_p}^{\otimes M}\otimes\ket{\mu_q}^{\otimes(L-M)}$ between sectors with $C$ charges $p-R$ and $q-R$ are eigenstates of the Q-SSEP Lindbladian with eigenvalues $-\frac{1}{2}|p-q|$ (because the eigenvalue is $(\mu_p,\mu_q) - \frac{1}{2}(p+q)$ and $(\mu_p,\mu_q)=\mathrm{min}(p,q)$).

We can use this observation to prove that the spectrum of the Lindbladian is gapped in any "non-homogeneous" sector (i.e. any $\cvec$ such that $\exists j,k$ s.t. $c_j\neq c_k$). Any such sector is made of a series of homogeneous portions separated by `domain wall edges' with different $C$-charges on its two ends. Let us denote these domain wall edges by $w_\alpha$, $\alpha=1, \cdots, d$, where $d$ is the number of homogeneous segments, and denote by $c^\pm_\alpha$ the two distinct $C$-charges on either side of the corresponding edge. Since the total Q-SSEP Lindbladian is the sum of Lindbladian attached to each of the edges, and since any Lindbladian is a non-positive operator, we have the operator inequality
\beq
\mathcal{L}^*_{\mathrm{ssep}} \leq \sum_\alpha \mathcal{L}^*_{j_\alpha,j_\alpha+1} ~,
\eeq
where $(j_\alpha,j_\alpha+1)$ are the two vertices connected by the edge $w_\alpha$. The two representations $[\mu_{R+c^+_\alpha}]$ and $[\mu_{R+c^-_\alpha}]$ are attached to these vertices, so that $\mathcal{L}^*_{j_\alpha,j_\alpha+1}$ is acting on their tensor product $[\mu_{R+c^+_\alpha}]\otimes[\mu_{R+c^-_\alpha}]$, as in eqn \eqref{eq:L-Casimir}. As a consequence, its maximum eigenvalue is that of the tensor Casimir on $[\mu_{R+c^+_\alpha}]\otimes[\mu_{R+c^-_\alpha}]$. The latter is diagonalized by decomposing the tensor product $[\mu_{R+c^+_\alpha}]\otimes[\mu_{R+c^-_\alpha}]$ into $gl(2R)$ irreducible representations, $[\mu_{R+c^+_\alpha}]\otimes[\mu_{R+c^-_\alpha}]=\oplus_\lambda[\mu_\lambda]$. From eqn \eqref{eq:tensor-C}, its maximal eigenvalue is given by the representation $\mu_\lambda$ maximising its Casimir. Since $\mathrm{Cas}_\mu=(\mu,\mu+2\rho)=|\mu+\rho|^2-|\rho|^2$ the maximum of the Casimir among the representation $[\mu_\lambda]$ occurring in the decomposition of the tensor product $[\mu_{R+c^+_\alpha}]\otimes[\mu_{R+c^-_\alpha}]$ is for $\mu_\lambda=\mu_{R+c^+_\alpha}+\mu_{R+c^-_\alpha}$. All other weights $\mu_\lambda+\rho$ in this decomposition have smaller norm. Hence, from the computation of the previous paragraph, we have
\beq
\mathcal{L}^*_{j_\alpha,j_\alpha+1}  \leq -\frac{1}{2}|c^+_\alpha-c^-_\alpha| ~.
\eeq
This proves that all inhomogeneous sectors are gapped with a gap given by the sum of the gaps associated with each domain wall edge.

Let us present here the proof that the one-magnons states $|\!|p;\gamma,\mu_n\rangle$ defined in eqn \eqref{eq:1magnon-R} are indeed eigenstates of the Q-SSEP Lindbladian $\mathcal{L}=\mathcal{L}^*$. Since the local charges $G^{-\gamma}_j$ is one of the conserved currents, commuting it with the Lindbladian yields, 
\beq
[\mathcal{L}^*, G^{-\gamma}_j] = {V}^{-\gamma}_{j}-{V}^{-\gamma}_{j-1} ~,
\eeq
 with 
 \beqs
 {V}^{-\gamma}_{j} &:= & -(\vec{\gamma}\cdot\vec{H}_{j+1})G^{-\gamma}_j + G^{-\gamma}_{j+1}(\vec{\gamma}\cdot\vec{H}_{j})
 + 2\sum_{\alpha \,:\, \alpha-\gamma\in\mathrm{roots}} G_{j+1}^{-\alpha}G_j^{\alpha-\gamma}.
\eeqs
For $|\ket{p;\gamma,\mu_n}$ to be an eigenstate of $\mathcal{L}^*$, the last sum in the above expression has to annihilate it. This amounts to say that either $\alpha-\gamma$ or $-\alpha$ have to be a positive root for all $\alpha$ such that $\alpha-\gamma$ is a root. We now show that this is indeed true iff $\gamma$ is a simple root. For a specific choice of the Weyl chamber, the positive (resp. negative) roots of $sl(2R)$ can be written as $\epsilon_k-\epsilon_l$ with $k<l$ (resp. $k>l$) with $\epsilon_k$ orthonormal basis. Choose $\gamma=\epsilon_i-\epsilon_j$, $i<j$. Then, either $\alpha=\epsilon_k-\epsilon_j$ (and $\alpha-\gamma=\epsilon_k-\epsilon_i$) or $\alpha=\epsilon_i-\epsilon_l$ (and $\alpha-\gamma=\epsilon_j-\epsilon_l$). It is then easy to check (by analysing each four cases) that we fulfil the condition that either $\alpha-\gamma$ or $-\alpha$ is positive, for all $\alpha$, if and only if $j=i+1$, i.e. iff $\gamma=\epsilon_i-\epsilon_{i+1}$ is a simple root. Thus, for $\gamma=\alpha_i$ a simple root, we have
\[ {V}^{-\alpha_i}_{j} |\!|\Omega_{\mu}\rangle = (\alpha_i,\mu)\,\big( G_{j+1}^{-\alpha_i}-  G_{j}^{-\alpha_i}\big)  |\!|\Omega_{\mu}\rangle ,\]
for any highest weight vector $\mu$. Finally, for $\mu$ one of the fundamental weight of $sl(2R)$, the scalar product $(\alpha_i,\mu)$ is equal either to $0$ or to $1$.

Finally, we describe here the computations needed to compare the two-replica Lindbladian with the $so(6)$ integrable spin chain Hamiltonian in the vector representation. The $so(6)$ generators in the vector representation can be written as
\[ J_{ij}= E_{ij}-E_{j'i'} ~,\]
with $E_{j}=\ket{i}\bra{j}$ and $i,j=1,\cdots,6$. The permutation $P$ and trace $Q$ operators are
\[ P=\sum_{ij} E_{ij}\otimes E_{ji} ~,\quad Q=\sum_{ij} E_{ij}\otimes E_{i'j'} ~.\]
The tensor Casimir is $\mathbb{C}=\sum_{ij} J_{ij}\otimes J_{ji}$. Computing, we get
\beqs
\mathbb{C} &=& \sum_{ij}(E_{ij}-E_{j'i'})\otimes(E_{ji}-E_{i'j'}) \\
&=& E_{ij}\otimes E_{ji} - E_{ij}\otimes E_{i'j'} - E_{j'i'}\otimes E_{ji} + E_{j'i'}\otimes E_{i'j'}\\
&=& 2(P-Q) ~.
\eeqs
Hence the Lindbladian is proportional to $\sum_j (P_{j;j+1}-Q_{j;j+1})$, up to a constant, which differs from the known Yang-Baxter $so(6)$ integrable Hamiltonian $H\propto \sum_j (2P_{j;j+1}-Q_{j;j+1})$.

\section{\sfix{$GL(2R)$ and $U(L)$ interplay on steady states}}
\label{Appendix:interplay}

The analysis of the $R$ replica Q-SSEP zero modes presented in the main text tells us how they decompose into representations of $gl(2R)$: steady states in the sector with $C$ charge equals to $p-R$ form an irreducible $gl(2R)$ isomorphic to that associated with a rectangular Young tableau with $p$ boxes vertically and $L$ boxes horizontally. The purpose of this Appendix is to understand the nature of these steady states in terms of the physical degrees of freedom. 

The system Hilbert space, generated by the physical fermions $c_j^\dag$, decomposes into the direct sum of sub-spaces with fixed particle numbers:
\beqs
 \mathcal{H}_{\rm sys} = \Lambda_1 \oplus  \Lambda_2 \oplus \cdots \oplus \Lambda_L ~, 
\eeqs
where $L$ is the number of sites.
Each such subspace $\Lambda_m$, with $m$ particles, forms an irreducible $SU(L)$ representation isomorphic to arank $m$ antisymmetric tensor in dimension $L$. The two representations with $m=0$ and $m=L$ are isomorphic, and  isomorphic to the scalar representation, i.e. $\Lambda_0\cong \Lambda_L\cong \mathbb{C}$. The others are not isomorphic. Thus, as a $SU(L)$ module, the system Hilbert space decomposes as
\beqs
 \mathcal{H}_{\rm sys}
 &\equiv&  \Lambda_{0}\otimes \mathbb{C}^2 \oplus \Lambda_1 \oplus \Lambda_2 \cdots \oplus \Lambda_{L-1}, 
 \eeqs 
where the first factor encodes for the multiplicity of the trivial representation, i.e. $\Lambda_{0}\otimes \mathbb{C}^2\equiv \Lambda_0\oplus \Lambda_L$. We shall denote by $\mathbb{P}_m$ the projector on the subspace $\Lambda_m$ with $m$ particles.

Let us first recall a few properties of the Q-SSEP dynamics and invariant measure:
\begin{enumerate}
\item{} For each realisation of the noise, the dynamics preserves the total number of particles $\hat M= \sum_jc_j^ \dag c_j$ and the spectrum of the density matrix (in particular it preserves pure states).
\item{} The Q-SSEP invariant measure is $SU(L)$ invariant. On each sector with fixed number of particle, it is that induced by the $SU(L)$ Haar measure on that orbit, see Ref. \cite{Bauer2019Equilibrium}. 
\end{enumerate}
As a consequence the $R$ replica steady states have to be covariant under the $SU(L)$ action and this is the route that we shall follow to understand them. We shall only discuss in details the cases of $R=1$ or $R=2$ replicas.

$\bullet$ {\it For one replica ($R=1$)}.

Here the symmetry algebra is $gl(2)$.
In the $\cvec_0$ sector, the zero modes form a multiplet of dimension $L+1$ (spin $L/2$). The $\cvec_{\pm 1}$ sectors have each a single zero mode, scalar under the $gl(2)$ symmetry. The $SU(L)$ invariance of the steady measure implies that the steady mean density matrix $\bar \rho := \mathbb{E}_{\rm inv}[\rho]$ is invariant under the $SU(L)$ adjoint action. As a consequence, by Schur's lemma, it has to read:
\[ \bar \rho := \mathbb{E}_{\rm inv}[\rho] = q_{0} \mathbb{P}_{0}\otimes \bar \sigma_{2\times 2} + q_1 \mathbb{P}_1 + \cdots + q_{L-1} \mathbb{P}_{L-1} ,\]
with $\bar \sigma_{2\times 2}$ a $2\times 2$ matrix acting on the $\mathbb{C}^2$ factor of the component $\Lambda_0\otimes \mathbb{C}^2$. The diagonal entries corresponding to the $L+1$ projectors $\mathbb{P}_m$, $m=0,\cdots, L$ (with $\mathbb{P}_0$ and $\mathbb{P}_L$ the two diagonal entries in $\bar \sigma_{2\times 2}$) are zero modes belonging to the $\cvec_0$ sector. The off-diagonal ones in $\bar  \sigma_{2\times 2}$ belong in the $\cvec_{\pm 1}$ sector and couple the empty ($m=0$) and full ($m=L$) states. These off-diagonal terms are non-zero if and only if there were already present in the initial mean density matrix (because these terms are frozen under the Q-SSEP dynamics, cf. eqn \fr{L2frag}). 

The $gl(2)$ symmetry acts on the diagonal parts (the off-diagonal terms are scalar under $gl(2)$), i.e. it acts on the projectors $\mathbb{P}_m$, $m=0,\cdots, L$. It is easy to verify that,
\[ J^+\cdot \mathbb{P}_m = (m+1) \mathbb{P}_{m+1},\ 
J^-\cdot \mathbb{P}_m = (L+1-m) \mathbb{P}_{m-1}, \
J^z\cdot \mathbb{P}_m = (2m-L) \mathbb{P}_{m} ,\]
for $m=0,\cdots, L$. Thus, these projectors form a spin $L/2$ representation of $gl(2)$ as they should. Phrased differently, the $gl(2)$ action on one replica steady state induces moves between steady states with different particle numbers.

$\bullet$ {\it For two replicas ($R=2$)}.

By $SU(L)$ invariance, the second moment $\bar \rho^{(2)}:=\mathbb{E}_{\rm inv}[\rho\otimes\rho]$ is also $SU(L)$ invariant under the {\it diagonal} adjoint action. Hence, by Schur's lemma, it is block diagonal on irreducible $sl(2R)$ sub-representations of  $\mathcal{H}_{\rm sys}\otimes \mathcal{H}_{\rm sys}$. That is: it decomposes as the sum over the projectors of the irreducible $SU(L)$ components in $\mathcal{H}_{\rm sys}\otimes \mathcal{H}_{\rm sys}$.

To decompose $\mathcal{H}_{\rm sys}\otimes \mathcal{H}_{\rm sys}$ as a sum of $SU(L)$ irreducible representations is a (nice) exercise in group theory~\footnote{We thank Jean-Bernard Zuber for his help in clarifying this problem.}. The output is the following decomposition:
\beq \label{eq:decomposeH2}
 \mathcal{H}_{\rm sys}\otimes \mathcal{H}_{\rm sys} 
\equiv (L+3) \Lambda_0 + \bigoplus_{k=1}^{L-1} (L+2) \Lambda_k 
+\bigoplus_{p=1}^{L-1}\bigoplus_{q=0}^{L-p-1} (q+1) \Lambda_{p+q;p} .
\eeq
where $\Lambda_r$ is the $r$-th fundamental representation of $SU(L)$, with Young tableau made of one column with $r$ boxes, and $\Lambda_{r;s;\cdots}$ the representation with Young tableau with $r$ boxes on the first column, $s$ boxes on the second column, etc. The proof of the above formula \eqref{eq:decomposeH2} follows from using rules for decomposing of the tensor product of two fundamental $SU(L)$ representations: for $r\geq s$, $\Lambda_r\otimes\Lambda_s = \oplus_{k=0}^{{\rm min}(s,L-r)} \Lambda_{r+k;s-k}$.

Let us give a few examples (at least to check the general formula):\\
-- For $L=2$, $\mathcal{H}_{\rm sys}= \Lambda_0\oplus \Lambda_1\oplus\Lambda_2$, with $\Lambda_2\equiv \Lambda_0$, and
\[
 \mathcal{H}_{\rm sys}\otimes \mathcal{H}_{\rm sys} 
= 5 \Lambda_0 \oplus 4 \Lambda_1 \oplus \Lambda_{1;1} ~.
\]
-- For $L=3$, $\mathcal{H}_{\rm sys}= \Lambda_0\oplus \Lambda_1\oplus\Lambda_2\oplus\Lambda_3$, with $\Lambda_3\equiv \Lambda_0$, and
\beqs
 \mathcal{H}_{\rm sys}\otimes \mathcal{H}_{\rm sys} 
 &=&  6 \Lambda_0 \oplus 5 \Lambda_1 \oplus 5 \Lambda_2  \oplus \Lambda_{1;1} \oplus 2 \Lambda_{2;1} \oplus \Lambda_{2;2} ~.
 \eeqs
-- For $L=4$, $\mathcal{H}_{\rm sys}= \Lambda_0\oplus \Lambda_1\oplus\Lambda_2\oplus\Lambda_3\oplus\Lambda_4$, with $\Lambda_4\equiv \Lambda_0$, and
\beqs
 \mathcal{H}_{\rm sys}\otimes \mathcal{H}_{\rm sys} 
  &=&  7 \Lambda_0 \oplus 6 \Lambda_1 \oplus 6 \Lambda_2 \oplus 6 \Lambda_3 \\
  &~& \oplus   \Lambda_{1;1} \oplus 2 \Lambda_{2;1} \oplus 3 \Lambda_{3;1} \\
  &~& \oplus \Lambda_{2;2} \oplus 2 \Lambda_{3;2}\\
  &~& \oplus \Lambda_{3;3} ~.
 \eeqs
We see a simple (triangular) pattern emerging. Formula \eqref{eq:decomposeH2} is a compact way to encode it.

As a consequence, the quadratic fluctuations $\mathbb{E}_{\rm inv}[\rho\otimes\rho]$ in the steady measure decompose as:
\beq \label{eq:2replica-decompose}
\mathbb{E}_{\rm inv}[\rho\otimes\rho] = \mathbb{P}_0\otimes \bar \sigma^{(0)}_{(L+3)\times (L+3)} 
+ \sum_{m=1}^{L-1} \mathbb{P}_m\otimes \bar \sigma^{(m)}_{(L+2)\times (L+2)} 
+ \sum_{p=1}^{L-1}\sum_{q=0}^{L-p-1} \mathbb{P}_{p+q;q}\otimes \bar \sigma^{(p+q;q)}_{(q+1)\times (q+1)} ~,
\eeq
where $\mathbb{P}_{r;s;\cdots}$ are the projectors on $\Lambda_{r;s;\cdots}$. The square matrices $\bar \sigma^{(*)}_{n\times n}$ are $n\times n$ matrices parametrizing the two replica steady states. The total number of parameters (without taking the trace normalization into account) is 
\beqs 
\#~\mathrm{parameters} &=& (L+3)^2 + (L+2)^2(L-1) +\sum_{k=1}^{L-1} k^2(L-k) \\
&=&  \frac{(L+1)(L+2)^2(L+3)}{12} + 2\frac{(L+1)(L+2)(L+3)}{6} + 2.
\eeqs
It matches the total dimension of the zero mode spaces that we identify in the main text using $gl(4)$ arguments. This shows that we didn't miss any zero modes using these arguments and that, given their $C$ charge, these zero modes form irreducible $gl(4)$ multiplets.

The above construction and counting do not distinguish between the different $C$ charge sectors. To do that we have to refine the decomposition \eqref{eq:decomposeH2} of $\mathcal{H}_{\rm sys}\otimes \mathcal{H}_{\rm sys}$ by taking into account the $u(1)$ charges in $U(L)$ and not only the $SU(L)$ content. Indeed, some of the Young tableaux appearing in the decomposition of $\mathcal{H}_{\rm sys}\otimes \mathcal{H}_{\rm sys}$ are of the form $\Lambda_{L;s}$ with a first column full with $L$ boxes. As $SU(L)$ representation $\Lambda_{L;s}$ and $\Lambda_s$ are  isomorphic but not as $U(L)$ representations. Going from $\Lambda_s$ to $\Lambda_{L;s}$ amounts to add $+1$ to all occupation numbers $n_j$ (recall that the $U(L)$ Cartan generators are the occupation numbers, $H_j=n_j$, whereas those of $SU(L)$ are $h_j=H_j-H_{j-1}$). We denote $\Lambda^+_{s}: = \Lambda_{L;s}$ (for instance $\Lambda_0^+=\Lambda_L$) and $\Lambda^{++}_{s}: = \Lambda_{L;L;s}$.

Using $\Lambda_r\otimes\Lambda_s = \oplus_{k=0}^{\mathrm{min}(s,L-r)} \Lambda_{r+k;s-k}$, for $r\geq s$, the detailed analysis of the decomposition of $\mathcal{H}_{\rm sys}\otimes \mathcal{H}_{\rm sys}$ is:\\
-- $\Lambda_0$ comes either (a) from the tensor products $\Lambda_0\otimes\Lambda_0\leadsto \Lambda_0$, $\Lambda_0\otimes\Lambda_L\leadsto \Lambda_0^+$, $\Lambda_L\otimes\Lambda_0\leadsto \Lambda_0^+$ and $\Lambda_L\otimes\Lambda_L\leadsto \Lambda_0^{++}$, or (b) from $\Lambda_k\otimes\Lambda_{L-k}\leadsto \Lambda_0^+$, with $k=1,\cdots,L-1$;\\
-- $\Lambda_k$, for any $k=1,\cdots, L-1$, comes either (a) from the tensor products $\Lambda_r\otimes\Lambda_s\leadsto \Lambda_{r+s=k}$ with $r=0,\cdots,[\frac{k}{2}]$, $s=[\frac{k+1}{2}],\cdots, k$ or with the role of $r$ and $s$ exchanged, and there are $(k+1)$ such possibilities, or (b) from $\Lambda_r\otimes\Lambda_s\leadsto \Lambda_{r+s=L+k}$ with $r,s=k,\cdots, L$, and there are $(L+1-k)$ such possibilities.\\
-- $\Lambda_{p+q;q}$ comes from $\Lambda_r\otimes\Lambda_s\leadsto \Lambda_{p+q;q}$, with $r+s=2p+q$ and $r,s=p,\cdots,p+q$. They are all of the form $\Lambda_{p+q;q}$ and not $\Lambda_{p+q;q}^+$ because their Young tableaux have two columns (so that a full column could not have been produced from the tensor product $\mathcal{H}_{\rm sys}\otimes \mathcal{H}_{\rm sys}$).

Thus, the $U(L)$ decomposition of $\mathcal{H}_{\rm sys}\otimes \mathcal{H}_{\rm sys}$ is
\beqa \label{eq:decomposeH2U(L)}
 \mathcal{H}_{\rm sys}\otimes \mathcal{H}_{\rm sys} 
&\equiv& ( \Lambda_0 +(L+1)\Lambda_0^++ \Lambda_0^{++}) + \bigoplus_{m=1}^{L-1} ((m+1) \Lambda_m + (L+1-m) \Lambda_m^+) \nonumber\\
&& ~ +\bigoplus_{p=1}^{L-1}\bigoplus_{q=0}^{L-p-1} (q+1) \Lambda_{p+q;p} . 
\eeqa
In particular, $\Lambda_0$ is the empty state and $\Lambda_0^{++}$ the totally (on both replicas) full state.

The zero modes in the $\cvec_{+1}$ sector are the intertwiners from $\Lambda_r$ to $\Lambda_r^+$, or from $\Lambda_r^+$ to $\Lambda_r^{++}$ (i.e. they are operators intertwining representations whose Young tableaux differ by a full column). Explicitly, in the $\cvec_{+1}$ sector, we have the intertwiners:
\beqs
&& \Lambda_0 \to  \Lambda_0^+\ [(L+1)\mathrm{-times}],\quad \Lambda_0^+ \to  \Lambda_0^{++}\ [(L+1)\mathrm{-times}]\\
&& \Lambda_m \to  \Lambda_m^+\ [(m+1)(L+1-m)\mathrm{-times}],\quad m=1,\cdots, L-1.
\eeqs
The total number of $C=1$ intertwiners is:
\[ \#~\mathrm{off.diag} = 2(L+1) + \sum_{m=1}^{L-1} (m+1)(L+1-m) = \frac{(L+1)(L+2)(L+3)}{6} .\]
 It matches the dimension of the $\cvec_{+1}$ (or $\cvec_{-1}$) zero modes, as it should.
The $\cvec_{-1}$ intertwiners are the transposed. The $\cvec_{\pm2}$ intertwiners are the empty-to-full interwiners $\Lambda_0^{++} \to  \Lambda_0$ and $\Lambda_0 \to  \Lambda_0^{++}$, respectively.

As a consequence, the decomposition \eqref{eq:2replica-decompose} of $\mathbb{E}_{\rm inv}[\rho\otimes\rho]$ can be refined. The matrices $\bar \sigma^{(0)}_{(L+3)\times (L+3)}$ can be decomposed into blocks of sizes $1,\, L+1,\, 1$, according to the decomposition $\Lambda_0 +(L+1)\Lambda_0^++ \Lambda_0^{++}$ in eqn \eqref{eq:decomposeH2U(L)}. The matrices $\bar \sigma^{(m)}_{(L+1)\times (L+1)}$ can be decomposed into blocks of sizes $m+1,\, L+1-m,$ according to the decomposition $(m+1) \Lambda_m + (L+1-m) \Lambda_m^+$ in eqn \eqref{eq:decomposeH2U(L)}. The off-diagonal blocks correspond to the $\cvec_{\pm 1}$ sector (and $\cvec_{\pm 2}$). The $gl(4)$ algebra act (faithfully on each sector of given charge $C=0,\pm1,\pm2$) on these intertwiners.


\section{Unfolding and level-spacing for a Poisson process.}\label{app:unfolding}
Given $N$ i.i.d random variables $X_i$ (the eigenvalues of an integrable Hamiltonian) taking values in $\mathbb{R}$ with density $p_X(x)$, we perform a local change of variables such that the new variable $\hat x(x)$ describes the average number of old variables below $x$,
\begin{equation}
x\to \hat x(x)=N \int_{-\infty}^x p_X(x')dx'.
\end{equation}
This procedure is called "unfolding the spectrum" and it ensures that the density of the new variables $\hat p_X(\hat x) d \hat x = p_X(x) dx$ is indeed uniform, $\hat p_X(\hat x)=1/N$. 

The probability to find $N_s=k$ of the new random variables in the interval $[0,s]$ is now is given by the Poisson distribution
\begin{equation}
\mathbb{P}[N_s=k]= \frac{(\lambda s)^k}{k!}e^{-\lambda s}, 
\end{equation}
with average "even rate" $\lambda=1$. We can define the probability $p_S(s)$ to observe a spacing $s$ between adjacent eigenvalues by 
\begin{equation}
\mathbb{P}[N_s=0]=\int_s^\infty p_S(s')ds'.
\end{equation}
Deriving w.r.t $s$ provides us with the exponential distribution $p_S(s)=e^{-s}$.
\section{Two point functions in the Lindbladian formulation}
\label{app:2particlesector}

As the state $|\!|\mathds{1}_2\rangle$ belongs to the sector $\cvec_0$ we can project the equations of motion \fr{SEq} to this sector. This means that at each site we need to consider only the states $|\!\ket{0}_j$, $|\!\ket{1}_j$, $|\!\ket{2}_j$, $|\!\ket{3}_j$, $|\!\ket{4}_j$, $|\!\ket{5}_j$, which form a six dimensional $gl(4)$ multiplet with $C$ charge zero. A representation of the $gl(4)$ generators on the corresponding subspace is obtained by retaining only the Hubbard operators $\mathcal{E}_j^{k,l}$, with $k,l=1,\cdots,6$ in eqn \fr{notationsfermions}. The two $gl(2)$ sub-algebras associated with each of the replicas are spanned by generators leaving operators on the opposite replica invariant. The action of $J_1^\alpha$ (resp. $J_2^\alpha$) is as in the 1-replica case, i.e.
\begin{align}
J_{1,j}^+&=E^{14}_j\otimes\mathds{1}=G^{12}_j\ ,\quad
J_{2,j}^+=\mathds{1}\otimes E^{14}_j=G^{43}_j\ ,\nn
J^z_{1,j}&=(E_j^{11}-E_j^{44})\otimes\mathds{1}=G^{11}_j-G^{22}_j\ ,\nn
J^z_{2,j}&=\mathds{1}\otimes(E_j^{11}-E_j^{44})=G^{44}_j-G^{33}_j\ .
\end{align}
On this subspace we have $2n_{a,j}-1=J^z_{a,j}$, which allows us to rewrite \fr{SEq} as
\be
\frac{d}{dt}\langle\mathds{1}_2|\!|(J^z_{1,j}+1)(J^z_{2,k}+1)|\!|\rho^{(2)}(t)\rangle=
\langle\mathds{1}_2|\!|(J^z_{1,j}+1)(J^z_{2,k}+1)\hat{\cal L}_{2,\cvec_0}|\!|\rho^{(2)}(t)\rangle\ .
\label{Seq1}
\ee
We now observe that while the identity $\|\mathds{1}_2\rangle$ is not an eigenstate of $J_{a,j}^z$, it is in fact a simultaneous eigenstate of $J_{a,j}^x=J_{a,j}^++J_{a,j}^-$, $a=1,2$. To proceed we therefore "rotate the quantization axis" and work in a basis of eigenstates of $J^x_{a,j}$ rather than $J^z_{a,j}$. This basis is given by
\beqs
|\!\ket{\mathfrak{1}}_j = \frac{1}{2} ,\quad |\!\ket{\mathfrak{2}}_j = n_{1,j}-\frac{1}{2} &,& |\!\ket{\mathfrak{3}}_j = n_{2,j}-\frac{1}{2},\quad |\!\ket{\mathfrak{4}}_j = 2(n_{1,j}-\frac{1}{2})(n_{2,j}-\frac{1}{2}) ,\\
|\!\ket{\mathfrak{0}}_j = c_{1,j}^\dag c_{2,j} &,& |\!\ket{\mathfrak{5}}_j = c_{1,j}c_{2,j}^\dag \ .
\eeqs
 The eigenvalues of $J^x_{1,j},J^x_{2,j}$ on these states are $(1,1)$, $(-1,1)$, $(1,-1)$, $(-1,-1)$, $(0,0)$ and $(0,0)$ respectively. We note that $\|\mathds{1}\rangle=2^L\otimes_{j=1}^L|\mathfrak{1}\rangle_j$. It is now apparent that $J^z_{a,j}$ as well as $S^\pm_j$ defined in \fr{nspm} create single-particle excitations over the state $\|\mathds{1}\rangle$. As the Linbladian $\hat{\cal L}_{2,\cvec_0}$ commutes with $J^x_{1,2}$ the corresponding particle numbers are good quantum numbers and \fr{Seq1} as well as its analog for $g_-(j,k;t)$ can therefore be viewed as Schr\"odinger equations in the two-particle sector.

\section{Spectrum of the two-replica Lindbladian in the two-particle sector}
\label{Appendix:two-particle}
As a consequence of the free dynamics governing the behaviour of Q-SSEP model for every realization of the noise, the equations for the $n$-point functions wrt the noise close, as discussed in the main text. 
In this appendix we will focus on the spectral properties in the case of two point functions. Furthermore, we will specialize our discussion to the closed loop diagrams, on which, once again, the dynamics closes. Correlation functions associated with disconnected diagrams decay exponentially fast and are, as such, of lesser importance when considering the hydrodynamics of the model.

On the level of two-point functions it proves advantageous to define two sets of observables, comprised of diagonal and off-diagonal correlations
\begin{align}
	\label{cors}
	A_{jk}^\pm&:=\mathbb{E}[G_{jj}G_{kk}\pm G_{kj}G_{jk}].
\end{align}
Interestingly, the dynamics for correlations $\eqref{cors}$ decouples and is governed by equations
\begin{equation} \label{2pt_pm}
	\frac{d}{dt}A^\pm_{ij} = (\Delta_i+\Delta_j)A^\pm_{ij}\pm 2\delta_{i,j}(A^\pm_{i,i+1}+A^\pm_{i-1,i})\mp 2(\delta_{j,i+1}+\delta_{i,j+1})A^\pm_{i,j} .
\end{equation}

\begin{figure}[ht]
	\begin{center}
		\vspace{0mm}
		\includegraphics[width=1\textwidth]{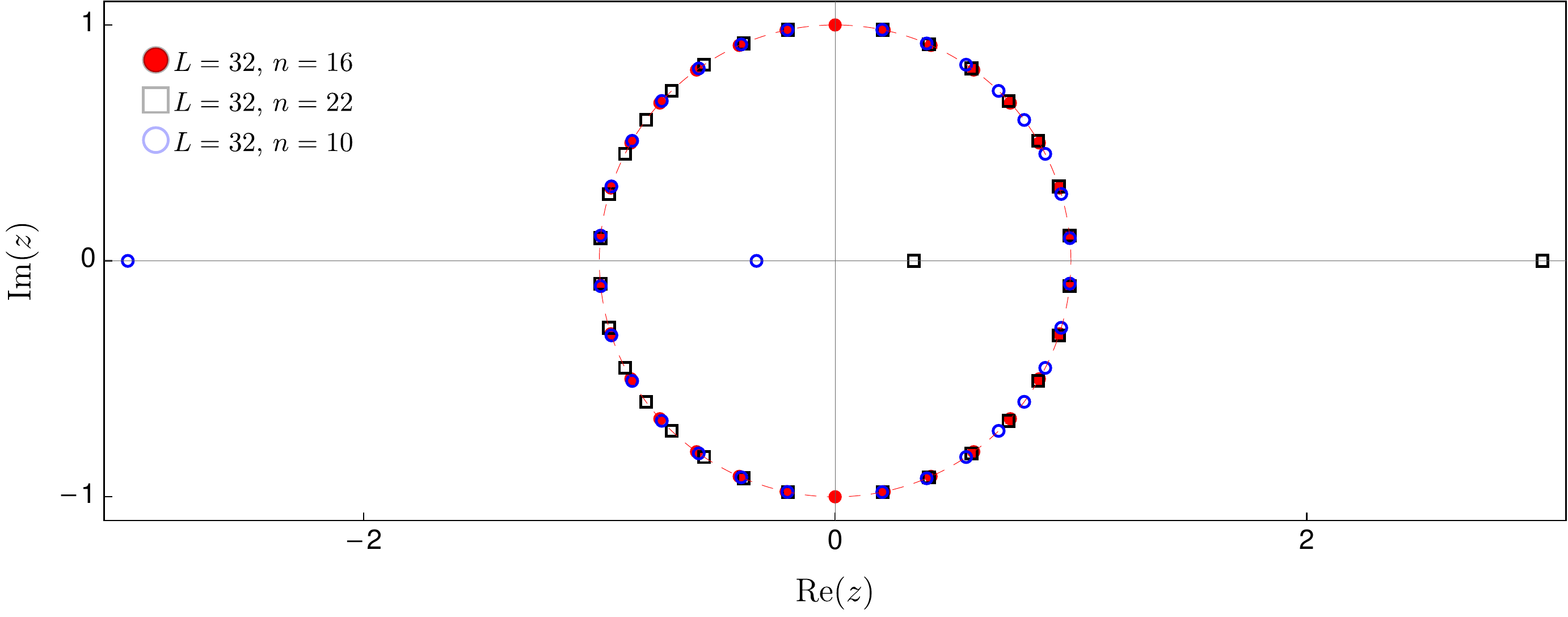}
		\caption{\label{eig_val} Solutions of the equation \eqref{eq:neven_Neven} for $N=12$ and $n=4,6,8$. All of the solutions lie on the unit circle (red dashed line) except for two solutions at any $n$, if $n\neq N/2$.}
	\end{center}
\end{figure}

We can consider equations \eqref{2pt_pm} as a two particle problem, where the positions of two particles is associated with indices $i$ and $j$, which can be solved through the scattering wave ansatz
\begin{align*}
	A_{j,k}^\pm(t) & =\big(e^{i(pj+qk)}+S^\pm(p,q)\,e^{i(qj+pk)}\big)\,e^{\omega t},\hspace{1em}\text{for}\ j<k ,\\
	A_{j,j}^\pm(t) & =C^\pm(p,q)\,e^{i(p+q)j+\omega t},\\
	A_{j,k}^\pm(t) & =\big(e^{i(pk+qj)}+S(p,q)\,e^{i(qk+pj)}\big)\,e^{\omega t},\hspace{1em}\text{for}\ j>k.
\end{align*}
The decay rate $\omega$ is governed by the non-interacting, i.e. purely diffusive, part of equations \eqref{2pt_pm} $(\Delta_i+\Delta_j)$, which yields
\begin{equation}
	\omega(p,q)=2(\cos p+\cos q-2).
\end{equation}
Coefficients $C^\pm(p,q)$ and $S^\pm(p,q)$ can be obtained by considering the equations for $A^\pm_{i,i+1}(p,q)$ and $A^\pm_{i,i}(p,q)$, yielding
\begin{eqnarray}
	S^+(p,q) &=& -e^{i(q-p)}\,\frac{(\cos p+\cos q)(e^{ip}+e^{-iq}+2)-4(1+\cos(p+q))}{(\cos p+\cos q)(e^{-ip}+e^{iq}+2)-4(1+\cos(p+q))},\\
	C^+(p,q) &=& \frac{4}{(\omega+4)}\big(e^{-ip}+e^{iq}+S^+(p,q)(e^{-iq}+e^{ip})\big),
\end{eqnarray}
and 
\begin{align}
	S^-(p,q) &=-\frac{e^{i(p+q)}-2e^{iq}+1}{e^{i(p+q)}-2e^{ip}+1},\\
	C^-(p,q) &= 0.
\end{align}
Note that the coefficients $S^-$ and $C^-$, as well as the equations for $A^-_{ij}(t)$ match perfectly the Bethe ansatz coefficients and equations for the two particle sector of the Heisenberg $XXX$ chain.
Finally, periodicity implies the quantization conditions
\begin{equation}
	S^\pm(p,q)\,e^{ipL} = S^\pm(p,q)\, e^{-iqL} = 1.
\end{equation}
Due to the translational invariance the total momentum takes the values $k=p+q=\frac{2\pi n}{L}, n\in \ZZ$, which means that we can reduce the eigenvalue problem at the fixed total momentum $k$ to a single transcendental equation.
The quantization condition can be represented in terms of a new variable $\xi$
\begin{equation}
	z:= e^{iu/2},\ \xi := z+z^{-1}=2 \cos(u/2),
\end{equation}
which parametrizes the difference of particle momenta $u=p-q$, and  gives a compact expression for the eigenvalue
\beq \label{eq:wz}
\omega_n(z) = 2(z+z^{-1})c_n - 4 = 2(c_n\, \xi - 2),
\eeq
where we introduced $c_n=\cos(\frac{2\pi n}{L})$, and eigenfunctions\footnote{up to the multiplicative constant}
\beqs
A_{ij}(t) &=& e^{ik(i+j)/2}\, \big[ z^{i-j+N} + (-)^n z^{-(i-j+N)}\big]\, e^{\omega t} ,\quad \mathrm{for}\ i<j,\\
A_{jj}(t) &=& e^{ikj}\,4\big[\frac{ z^{N-1}+(-)^nz^{1-N}}{z^{-1}+z}\big]\, e^{\omega t} .
\eeqs
For $k=\frac{2\pi n}{L}\not=\pi$ (i.e. for $\cos(k/2)\not=0$), the $S$-matrix, for a total moment $k=\frac{2\pi n}{L}$, reads\footnote{Note that for $A^+$ the special case with $k=\pi$ has to be analysed separately.}
\begin{align} \label{eq:sz}
	S_n^+(z) &= - \frac{ z^{-1} R_n(z)}{z R_n(z^{-1})}; \quad \ 
	R_n^+(z) := 4c_n - (z+z^{-1})(1+z c_n),\\
	S_{n}^-(z) & =-\frac{c_n-1/z}{c_n-z}.
\end{align}
and since $e^{ikL/2}=(-1)^n$, the quantization condition reduces to
\beq \label{eq:sLz}
S_n^\pm(z)\,  z^{L} = (-1)^{n}.
\eeq
The condition can be neatly represented in terms of Chebyshev-like polynomials
\begin{align}
	P^+_p(\xi)&:=z^{2p}+z^{-2p},\quad Q^+_p(\xi)\xi:= z^{2p-1}+z^{1-2p}\\ Q^-_p(\xi)(z-z^{-1})\, \xi&:=z^{2p}-z^{-2p},\quad  P^-_p(\xi)\,(z-z^{-1}):=z^{2p+1}-z^{-1-2p}  ,
\end{align}
Namely, for chains of even length $L=2N$, we have to consider the following cases for $A^+$:
\begin{itemize}
	\item $N=2M$ even, $n$ even
	\beq \label{eq:neven_Neven}
	(4c_n-\xi)\,\xi\, Q^+_{M}(\xi) = c_n\, \xi\, P^+_{M}(\xi), 
	\eeq
	\item $N=2M+1$ odd, $n$ even
	\beq \label{eq:neven_Nodd}
	(4c_n-\xi)\,P^+_{M}(\xi) = c_n\, \xi^2\, Q^+_{M+1}(\xi),
	\eeq
	\item $N=2M+2$ even, $n$ odd
	\beq \label{eq:nodd_Neven}
	(4c_n-\xi)\,P^-_M(\xi) = c_n\, \xi^2\, Q^-_{M+1}(\xi),
	\eeq
	\item $N=2M+1$ odd, $n$ odd
	\beq \label{eq:nodd_Nodd}
	(4c_n-\xi)\,\xi\, Q^-_M(\xi) = c_n\, \xi\, P^-_M(\xi),
	\eeq
\end{itemize}
And similarly for $A^-$:

\begin{itemize}
	\item $N=2M$ even
	\begin{equation}
		c_{n}P_{M}^+(\xi)=Q_{M}^+(\xi),
	\end{equation}
	\item $N=2M+1$ odd
	\begin{equation}
		c_{n}Q_{M-1}^+(\xi)=P_{M}^+(\xi).
	\end{equation}
\end{itemize}
A numerical analysis of these equations reveals that solutions $z$ always  lie on the unit circle $|z|=1$, except for two reciprocal values, see Fig.~\ref{eig_val}.

\section{Derivation of the time evolution of \sfix{$n$}-point bubbles}
\label{app:derivation_n_bubbles}
Here we show how to arrive at (\ref{eq:evolution_continuous_n_bubble_lead}) in the main
text that describes time evolution of the $n$-point bubble $g_{n}(l_{1},...,l_{n})=[G_{l_{1}l_{2}}G_{l_{2}l_{3}}...G_{l_{n}l_{1}}]$, where $[\dots]$ is shorthand for $\bE[\dots]$. The time evolution of the two-point functions $G_{ij}=Tr(\rho c_j^\dagger c_i)$ is given by the stochastic differential equation \cite{Bauer2019Equilibrium}
\beqa
\label{eq:dG}
dG_{ij}&=&(\delta_{ij}(G_{i+1,i+1}+G_{i-1,i-1})-2G_{i,j})Jdt \\
&& +i\left(G_{i,j-1}d\bar{W}^{j-1}+G_{i,j+1}dW^{j}-G_{i-1j}dW^{i-1}-G_{i+1,j}d\bar{W}^{i}\right), \nonumber
\eeqa
where we used the convention for the Brownian motion $dW$ from \eqref{eq:def-dW} with $p=q$ (Q-SSEP), which makes $dW$ dimensionless by introducing the rate parameter $J$. Using Ito convention we have
\begin{align}\label{eq:dGdG}
\frac{[dG_{ij}dG_{kl}]}{J dt}=& ~\delta_{jk}[G_{i,j\pm1}G_{k\pm1,l}]+\delta_{il}[G_{i\pm1,j}G_{k,l\pm1}]-(\delta_{i,k\pm1}+\delta_{j,l\pm1})[G_{il}G_{kj}],
\end{align}
where we introduced the shorthand notation $G_{i,j\pm1}G_{k\pm1,l}:=G_{i,j+1}G_{k+1,l}+G_{i,j-1}G_{k-1,l}$ and similarly $\delta_{i,k\pm1}:=\delta_{i,k+1}+\delta_{i_,k-1}$.
\subsection{Three-point bubble}
To illustrate the derivation, we will first explicitly derive the evolution of the three-point bubble,
\begin{align}\label{eq:dg_3}
dg_3(i,j,k)=
&[dG_{ij} G_{jk} G_{ki}] + [G_{ij} dG_{jk} G_{ki}] + [G_{ij} dG_{jk} G_{ki}] \\
&+[dG_{ij} dG_{jk} G_{ki}]+[G_{ij} dG_{jk} dG_{ki}]+[dG_{ij} G_{jk} dG_{ki}].\nonumber
\end{align}
In particular we need
\begin{align}
\frac{[dG_{ij}]}{J dt}&=\delta_{ij} [G_{i\pm 1,i\pm 1}]-2[G_{ij}]\\
\frac{[dG_{ij}dG_{jk} G_{ki}]}{J dt}&=[G_{i,j\pm 1}G_{j\pm 1,k} G_{ki}]+\delta_{ik}[G_{i\pm 1,j}G_{j,i\pm1} G_{ii}]-(\delta_{i,j\pm 1}+\delta_{j,k\pm 1})[G_{ik}G_{kk} G_{ki}]
\end{align}
The terms in (\ref{eq:dg_3}) without Kronecker deltas yield discrete Laplacians $(\Delta_i +\Delta_j +\Delta_k)g_3(i,j,k)$. The contact terms (terms with Kronecker delta) are built by products of two bubbles which appear in a symmetric manner. We therefore introduce the notation
\begin{align}\label{eq:notation_f_g}
g_3(i|j,k):&=[G_{ii}G_{jk}G_{kj}]=\left[\onebubble{1}{i}\twobubble{1}{k}{j}\right] \\
f_3(i|j):&=g_3(i|j,k)+g_3(j|i,k)=\left[\onebubble{1}{i}\twobubble{1}{k}{j}\right]+\left[\onebubble{1}{j}\twobubble{1}{k}{i}\right].
\end{align}
The idea behind this notation is that the vertical bar separates bubbles into disconnected pieces. In writing $f_n(\text{var.})$ we mean the symmetrization of $g_n(\text{var.}+\text{other var.})$ in the displayed variables. Of course, $f_n(\text{var.})$ also depends on the other variables (in total there must be $n$ variables), but for simplicity they are suppressed in this notation.
Denoting $\mathcal{D}^{ij}$ the operator that produces the contact terms between $i$ and $j$, we find
\begin{equation}\label{eq:discrete_contact_terms}
\begin{split}
\mathcal{D}^{ij}g_3(i,j,k)
&=\delta_{ij}\left(f_3(i+1|j)+f_3(i-1|j)\right)-(\delta_{i+1,j}+\delta_{i-1,j})f_3(i|j) \\
&=\delta_{ij}(\Delta_i f_3(i|j))-(\Delta_i \delta_{ij})f_3(i|j).
\end{split}
\end{equation}
To summarize the discrete evolution, we have found that 
\begin{align}\label{eq:evolution_discrete_3_bubble}
\frac{dg_3(i,j,k)}{Jdt}=\left( \sum_{a\in\{i,j,k\}} \Delta_a  +\sum_{\substack{a,b\in\{i,j,k\} \\ \text{unorderd pairs}}}\mathcal{D}^{a,b}\right)g_3(i,j,k)
\end{align}
This sum over nodes and unordered pairs of nodes can be easily generalized to $n$-point bubbles with an appropriate generalization of $f_3(i|j)$, see (\ref{eq:f_n}) to understand the spirit of this equation. 

The scaling limit of (\ref{eq:evolution_discrete_3_bubble}) is taken by introducing continuous variables $$x=i/L, y=j/L, z=k/L,$$ while sending $J=L^2$ to infinity (note that we set $\mathfrak{L}=\mathfrak{D}=1$ for simplicity). For simplicity, will denote the 3-point bubbles in the scaling limit by the same name
\footnote{
There is a small subtlety in taking the scaling limit. Formally we should contract the discrete equation with a test function (here done for a one-point function), $(h,g_1)=\sum_{i=1}^L h(i)g_1(i)$, and then approximate the sum as an integral, $(h,g_1)\to\int_0^1 dx \tilde{h}(x)\tilde{g}_1(x)$ (we denote the continuous function by a tilde in this footnote). However, the approximation leads to an error that also scales with some power of $L$. It can be quantified to scale as $\mathcal{O}(L^{-2})$ using the "mid-point rule",
\[
\left|\sum_{i=1}^{L}\frac{1}{L}f({x_{i}})-\int_{0}^{1}f(x)dx\right|\le\frac{1}{24L^{2}}\max_{x\in[0,1]}f''(x),
\]
where $x_i=(i-1/2)/L$. To have an error of at most $\mathcal{O}(L^{-2} \tilde{g}_3)$ in (\ref{eq:scaling_laplac}) and (\ref{eq:scaling_source}), one should therefore define the continuous 3-point function as $\tilde{g}_3((i-1/2)/L,...)=g_3(i,...)$. When $L\to \infty$ this becomes the same as the definition in the main text.
}.
Expanding in inverse powers of $L$, the Laplacian becomes (the factor $L^2$ on the rhs.\ is there to cancel with $1/J$) 
\begin{align}\label{eq:scaling_laplac}
L^2 \Delta_i g_3(i,j,k) \to  \Delta_x g_3(x,y,z) + \mathcal{O}(L^{-2}g_3),
\end{align}
while a Kronecker delta scales as
\begin{align}
\delta_{ij} \to L^{-1} \delta(x-y)+\mathcal{O}(L^{-2}).
\end{align}
Therefore the contact term becomes
\begin{equation}\label{eq:scaling_source}
\begin{split}
L^2 \mathcal{D}^{ij}g_3(i,j,k)
\to &L^{-1}\left(\delta(x-y)\Delta_x f_3(x|y)-\delta''(x-y)f_3(x|y)\right)+\mathcal{O}(L^{-2} g_3) \\
&=L^{-1}\partial_x\partial_y (\delta(x-y) f_3(x|y))+\mathcal{O}(L^{-2} g_3).
\end{split}
\end{equation}
The last step is obtained by partial integration w.r.t $x$, which is allowed, since all of these equations should be thought of evaluated against a test function, i.e.
\begin{align}
\int dx dy \varphi(x,y) \partial_x \partial_y (\delta(x-y) f_3(x|y)).
\end{align}

Then, we can summarize the continuous evolution equation as
\begin{align}\label{eq:evolution_continuous_3_bubble}
\frac{dg_3(x,y,z)}{dt}=\sum_{a\in\{x,y,z\}} \Delta_a g_3(x,y,z) + L^{-1}\sum_{\substack{a,b\in \{x,y,z\} \\ \text{unordered pairs}}} \partial_x\partial_y(\delta(a-b) f_3(a|b)).
\end{align}

We now discuss the scaling of the 3-point bubble in inverse powers of $L$ that, a priori, could have an expansion according to
$$g_3(x,y,z)=g_3^{(0)}+L^{-1}g_3^{(1)}+L^{-2}g_3^{(2)}+...$$
If we assume that the initial density matrix $\rho_0$ does not involve long range correlations, then all $n$-point bubbles are originally zero, except for the average fermion density $g_1(x)$. At a later time, $g_3(x,y,z)$ can only be non-zero if the source term $L^{-1} f_3(x|y)$ is non-zero. But $f_3(x|y)$ also satisfies a diffusion equation with new source terms\footnote{
Explicitly, one finds
\begin{align*}
\frac{d}{dt}g_3(x|y,z)=(\Delta_x +\Delta_y + \Delta_z)g_3(x|y,z)
&+L^{-1}2\partial_y\partial_z(\delta(y-z)g_3(x|y|z))\\
&+L^{-1}\left(\partial_x\partial_y(\delta(x-y)\ \bullet) +\partial_x\partial_z(\delta(x-z)\ \bullet)\right)(f_3(x,y,z)).
\end{align*}
Here we made full use of the notation introduced in (\ref{eq:notation_f_g}): $g_3(x|y|z)$ is a product of three one-bubbles and $f_3(x,y,z)=g_3(x,y,z)+g_3(x,z,y)$ is the symmetrization of a a single three-bubble in all variables.
} of which one is $L^{-1} g_3(x|y|z)$. At $0$\textsuperscript{th} order in $1/L$, $g_3(x|y|z)$ satisfies a pure diffusion equation with non-zero initial condition, because at time zero (and therefore at all times) it factorizes, $g_3(x|y|z) =g_1(x)g_1(y)g_1(z)$, and $g_1(x)$ is non-zero as argued above. In the original equation for $g_3(x,y,z)$, this gives rise to a non-zero source term only at order $L^{-2}$ (consecutively "descending" the source terms we encountered two factors of $L^{-1}$) and the lower orders are actually zero, $g_3^{(0)}=g_3^{(1)}=0$. More general, the leading order of an $n$-bubble is $n-1$ and we denote it by the superscript $\#$ in the following. This argument also shows that at leading order the expectation value of products of bubbles factorizes. Rewriting (\ref{eq:evolution_continuous_3_bubble}) at leading order in terms of diagrams gives rise to (\ref{eq:evolution_continuous_3_bubble_lead}) in the main text.

\subsection{n-point bubbles}
We denote the discrete indices by $l_1,...,l_n \in \{1,...,L\}$ and the continuous variables by $x_1,...,x_n \in [0,1]$. The generalization of (\ref{eq:evolution_discrete_3_bubble}) is
\begin{align}
\frac{dg_n(l_1,...,l_n)}{Jdt}=\left( \sum_{a=1}^L \Delta_{l_a}  + \sum_{\substack{a,b=1 \\ a<b}}^L\mathcal{D}^{a,b}\right)g_n(l_1,...,l_n)
\end{align}
where
\begin{align}
\mathcal{D}^{a,b}=\delta_{l_a,l_b}(\Delta_{l_a} f_n(l_a|l_b))-(\Delta_{l_a} \delta_{l_a,l_b})f_n(l_a|l_b)
\end{align}
and for $a<b$ we generalize
\begin{align}\label{eq:f_n}
f_n(l_a|l_b)
&=g_n(l_a,l_{a+1},...,l_{b-1}|l_b,l_{b+1},...,l_{a-1})+g_n(l_b,l_{a+1},...,l_{b-1}|l_a,l_{b+1},...,l_{a-1})\\
&=\left[\nbubbleThree{1.2}{b-a}{l_{b-1}}{l_{a}}{l_{a+1}}\nbubbleThree{1.2}{n-b+a}{l_{a-1}}{l_{b}}{l_{b+1}}\right]
+\left[\nbubbleThree{1.2}{b-a}{l_{b-1}}{l_{b}}{l_{a+1}}\nbubbleThree{1.2}{n-b+a}{l_{a-1}}{l_{a}}{l_{b+1}}\right]
\end{align}
In terms of diagrams the contact term $f_n(l_a|l_b)$ has an easy visualisation: The original $n$-bubbles is squeezed together such that the nodes $l_a$ and $l_b$ touch each other and the diagram forms an eight. Then the eight is split apart into two disconnected bubbles and we sum over the two possible ways how one can attribute $l_a$ and $l_b$ to the two nodes at the splitting junction.

Then it is easy to see, that the generalization of (\ref{eq:evolution_continuous_3_bubble}) is just
\begin{align}\label{eq:evolution_continuous_n_bubble}
\frac{dg_n(x_1,...,x_n)}{dt}=\sum_{a=1}^L \Delta_{x_a} g_n(x_1,...,x_n) + L^{-1}\sum_{\substack{a,b=1 \\ a<b}}^L \partial_{x_a}\partial_{x_b}(\delta(x_a-x_b) f_n(x_a|x_b)).
\end{align}
In particular, at leading order in inverse powers of $L$, we recover (\ref{eq:evolution_continuous_n_bubble_lead}) in the main text.

\bibliography{references}

\end{document}